\documentclass[12pt]{article}
\usepackage{epsf,epsfig,cite}

\newcount\timecount
\newcount\hours \newcount\minutes  \newcount\temp \newcount\pmhours

\hours = \time
\divide\hours by 60
\temp = \hours
\multiply\temp by 60
\minutes = \time
\advance\minutes by -\temp
\def\hour{\the\hours}
\def\minute{\ifnum\minutes<10 0\the\minutes
            \else\the\minutes\fi}
\def\clock{
\ifnum\hours=0 12:\minute\ AM
\else\ifnum\hours<12 \hour:\minute\ AM
      \else\ifnum\hours=12 12:\minute\ PM
            \else\ifnum\hours>12
                 \pmhours=\hours
                 \advance\pmhours by -12
                 \the\pmhours:\minute\ PM
                 \fi
            \fi
      \fi
\fi
}

\def\monthname{\relax\ifcase\month 0/\or January\or February\or
   March\or April\or May\or June\or July\or August\or September\or
   October\or November\or December\else\number\month/\fi}

\def\bold#1{\setbox0=\hbox{$#1$}%
     \kern-.025em\copy0\kern-\wd0
     \kern.05em\copy0\kern-\wd0
     \kern-.025em\raise.0433em\box0 }

\def\gappeq{\mathrel{\rlap {\raise.5ex\hbox{$>$}}
{\lower.5ex\hbox{$\sim$}}}}

\def\lappeq{\mathrel{\rlap{\raise.5ex\hbox{$<$}}
{\lower.5ex\hbox{$\sim$}}}}

\def\beq{\begin{equation}}
\def\eeq{\end{equation}}

\def\m12{m_{1\!/2}}

\input paperdef

\begin{document}
\begin{titlepage}
\pagestyle{empty}
\baselineskip=21pt
\begin{flushright}
hep-ph/0411216 \hfill
CERN--PH--TH/2004-220\\
DCPT/04/146, IPPP/04/73 \hfill
UMN--TH--2326/04, FTPI--MINN--04/40
\end{flushright}
\vskip 0.10in
\begin{center}
{\large{\bf Indirect Sensitivities to the Scale of Supersymmetry}}
\end{center}
\begin{center}
\vskip 0.05in
{{\bf John Ellis}$^1$, 
{\bf Sven Heinemeyer}$^1$,
{\bf Keith A.~Olive}$^{2}$
and {\bf Georg Weiglein}$^{3}$}\\
\vskip 0.05in
{\it
$^1${TH Division, Physics Department, CERN, Geneva, Switzerland}\\
$^2${William I.\ Fine Theoretical Physics Institute,\\
University of Minnesota, Minneapolis, MN~55455, USA}\\
$^3${Institute for Particle Physics Phenomenology, University of Durham,\\
Durham DH1~3LE, UK}\\
}
\vskip 0.1in
{\bf Abstract}
\end{center}
\baselineskip=18pt \noindent

Precision measurements, now and at a future linear electron-positron
collider (ILC), can provide indirect information about the possible scale
of supersymmetry. We illustrate the present-day and possible future ILC
sensitivities within the constrained minimal supersymmetric extension of
the Standard Model (CMSSM), in which there are three independent soft
supersymmetry-breaking parameters $m_{1/2}, m_0$ and $A_0$.  We analyze
the present and future sensitivities separately for $\MW$, $\sweff$,
$(g-2)_\mu$, $\br(b \to s \ga)$, $\br(B_s \to \mu^+ \mu^-)$, $\Mh$ and
Higgs branching ratios. We display the observables as functions of
$m_{1/2}$, fixing $m_0$ so as to obtain the cold dark matter density
allowed by WMAP and other cosmological data for specific values of $A_0$,
$\tb$ and $\mu > 0$. In a second step, we investigate the combined
sensitivity of the currently available precision observables, $\MW$,
$\sweff$, $(g-2)_\mu$ and $\br(b \to s \ga)$, by performing a $\chi^2$
analysis.  The current data are in very good agreement with the
CMSSM prediction for $\tb = 10$, with a clear preference for relatively
small values of $m_{1/2} \sim 300$~GeV. In this case, there would be good
prospects for observing supersymmetry directly at both the LHC and 
the ILC, and some chance already at the Tevatron collider. For $\tb = 50$, 
the quality of the fit is worse,
and somewhat larger $m_{1/2}$ values are favoured. With the prospective
ILC accuracies the sensitivity to indirect effects of supersymmetry
greatly improves. This may provide indirect access to supersymmetry even
at scales beyond the direct reach of the LHC or the ILC.

\vfill
\vskip 0.15in
\leftline{CERN--PH--TH/2004-220}
\leftline{November 2004}
\end{titlepage}
\baselineskip=18pt


\section{Introduction}

Measurements at low energies may provide interesting indirect information
about the masses of particles that are too heavy to be produced directly.
A prime example is the use of precision electroweak data from LEP, the
SLC, the Tevatron and elsewhere to predict (successfully) the mass of the
top quark and to provide an indication of the possible mass of the
hypothetical Higgs boson~\cite{lepewwg}. 
Predicting the masses of supersymmetric
particles is much more difficult than for the top quark or even the Higgs 
boson, because the renormalizability of the
Standard Model and the decoupling theorem imply that many low-energy
observables are insensitive to heavy sparticles. Nevertheless, present
data on observables such as $\MW$, $\sweff$, $(g-2)_\mu$ and
$\br(b \to s \ga)$ already provide interesting information on the scale of
supersymmetry (SUSY), as we discuss in this paper, and have a 
great potential in view of prospective improvements of experimental and
theoretical accuracies.

In the future, a linear $e^+ e^-$ collider (ILC) will be the best available
tool for making many precision measurements~\cite{lctdrs}. 
It is important to understand
what information ILC measurements may provide about supersymmetry, 
both for the part of the
spectrum directly accessible at the LHC or the ILC and for sparticles
that would be too heavy to be produced directly.
Comparing the
indirect indications with the direct measurements would be an important
consistency check on the theoretical framework of supersymmetry.

Improved and more complete calculations of the supersymmetric
contributions to a number of low-energy observables such as $\MW$ and
$\sweff$ have recently become available (see the discussion in
\refse{sec:ewpo} below). These, combined with
estimates of the experimental accuracies attainable at the ILC and
future theoretical uncertainties from unknown higher-order corrections, 
make now an
opportune moment to assess the likely sensitivities of ILC measurements.

There have been many previous studies of the sensitivity of low-energy
observables to the scale of supersymmetry, including, for example, the
precision electroweak 
observables~\cite{oldfits,oldfits2,oldfits3,gigaz,deboer1,recentfit,recent2}. 
Such analyses are bedevilled by the
large dimensionality of even the minimal supersymmetric extension of the
Standard Model (MSSM), once supersymmetry-breaking parameters are taken
into account. For this reason, simplifying assumptions that may be more or
less well motivated are often made, so as to reduce the parameter space to
a manageable dimensionality. Following many previous studies, we work here
in the framework of the constrained MSSM (CMSSM), in which the soft
supersymmetry-breaking scalar and gaugino masses are each assumed to be
equal at some GUT input scale. In this case, the new independent MSSM
parameters are just four in number: the universal gaugino mass $m_{1/2}$,
the scalar mass $m_0$, the trilinear soft supersymmetry-breaking parameter
$A_0$, and the ratio $\tb$ of Higgs vacuum expectation values. The 
pseudoscalar Higgs mass $\MA$ and the magnitude of the Higgs mixing 
parameter $\mu$ can be determined by using the electroweak vacuum 
conditions, leaving the sign of $\mu$ as a residual ambiguity.

The non-discoveries of supersymmetric particles and the Higgs boson at 
LEP
and other present-day colliders impose significant lower bounds on
$m_{1/2}$ and $m_0$. An important further constraint is provided by the
density of dark matter in the Universe, which is tightly constrained by
WMAP and other astrophysical and cosmological data~\cite{WMAP}. 
These have the effect within the CMSSM, assuming that the dark matter 
consists largely of neutralinos~\cite{EHNOS},
of restricting $m_0$ to very narrow allowed strips
for any specific choice of $A_0$, $\tb$ and the sign of 
$\mu$~\cite{WMAPstrips,wmapothers}.
Thus, the dimensionality of the supersymmetric parameter space is further
reduced, and one may explore supersymmetric phenomenology along these
`WMAP strips', as has already been done for the direct detection of
supersymmetric particles at the LHC and linear colliders of varying 
energies~\cite{oldbench,SPS,BDEGOP,otherAnalyses1,Ellis:2004bx,otherAnalyses2}.
A full likelihood analysis of the 
CMSSM planes incorporating uncertainties in the cosmological
relic density was performed in \citere{eoss4}.
The principal aim of this paper is to extend this analysis to indirect
effects of supersymmetry.

We consider the following observables: the $W$~boson mass, $\MW$, the
effective weak mixing angle at the $Z$~boson resonance, $\sweff$, the
anomalous magnetic moment of the muon, \mbox{$(g-2)_\mu$} and the rare
$b$ decays $\br(b \to s \ga)$ and $\br(B_s \to \mu^+ \mu^-)$, as well
as the mass of 
the lightest $\cp$-even Higgs boson, $\Mh$, and the Higgs branching ratios
$\br(\hbb) / \br(\hWW)$.  We first analyze the sensitivity of each
observable to indirect effects of supersymmetry, taking into account the
present and prospective future experimental and theoretical uncertainties.
We then investigate the combined sensitivity of those observables for
which experimental determinations exist at present, i.e., $\MW$, $\sweff$,
$(g-2)_\mu$ and $\br(b \to s \ga)$. We perform $\chi^2$ analyses both for
fixed values of $A_0$ and for scans in the \plane{m_{1/2}}{A_0} for $\tb =
10$ and 50 with $\mu > 0$. We find a remarkably high sensitivity of the
current data for the electroweak precision observables to the scale of
supersymmetry. In the case $\tb = 10$, we find a preference for
moderate values of $m_{1/2} \sim 300 \gev$, in which case sparticles
should be observable at both the LHC and the ILC. In the case $\tb = 50$,
the global fit is not so good, and low values of $m_{1/2}$ are not so
strongly preferred. In order to investigate the possible future
sensitivities we study the combined effect of all the above observables
(except $\br(B_s \to \mu^+ \mu^-)$, which is discussed separately). For
this purpose we choose certain values of $(m_{1/2}, A_0)$ as assumed
future `best-fit' values (corresponding to the central values of the
observables) and investigate the indirect constraints arising from the
precision observables for prospective experimental and theoretical
uncertainties.

In Section~2 of the paper we specify the WMAP strips and discuss their
dependences on $A_0$ and the top-quark mass. We discuss in Section~3 the
present and future sensitivities of the different precision observables to
the scale of supersymmetry, represented by $m_{1/2}$ as one moves along
different WMAP strips. In Section~4 we analyze the combined sensitivity of
the precision observables for the present situation, and Section~5
presents the prospective combined sensitivity assuming the accuracies
expected to become available at the ILC with its GigaZ option. Finally,
Section~6 gives our conclusions. In most of the scenarios studied, even if
it does not produce sparticles directly, the ILC will check the
consistency of the CMSSM at the loop level and thereby provide valuable
extra information beyond that obtainable with the LHC.


\section{Supersymmetric dark matter and WMAP strips}
\label{sec:CDMstrips}

It is well known that the lightest supersymmetric particle (LSP) is an
excellent candidate for cold dark matter (CDM)~\cite{EHNOS}, with a density 
that
falls naturally within the range $0.094 < \Omega_{\rm CDM} h^2 < 0.129$
favoured by a joint analysis of WMAP and other astrophysical and
cosmological data~\cite{WMAP}. Assuming that the cold dark matter is
composed predominantly of LSPs, the uncertainty in the determination of
$\Omega_{\rm CDM} h^2$ effectively reduces by one the dimensionality of the
MSSM parameter space. Specifically, if one assumes that the soft
supersymmetry-breaking gaugino masses $m_{1/2}$ and scalar masses $m_0$
are universal at some GUT input scale, as in the CMSSM studied here, the
\plane{m_{1/2}}{m_0}s usually studied for fixed $A_0$, $\tb$ and
sign of $\mu$ are effectively reduced to narrow strips of limited
thickness in $m_0$ for any given value of $m_{1/2}$~\cite{WMAPstrips} 
and the other parameters.

These strips have been delineated and parametrized when $A_0= 0$ for
several choices of $\tb$ for each sign of $\mu$, and the possible
LHC and ILC phenomenology along these lines has been
discussed~\cite{BDEGOP}. As preliminaries to studying indirect
sensitivities to the scale of supersymmetry along some of these WMAP
strips, we first address a couple of physics issues. One is that the
experimental central value of $\mt$ has changed since~\citere{BDEGOP}, 
from 174.3~GeV to
178.0~GeV~\cite{mtopexpnew}, and the other is the dependence of 
the WMAP strips 
on $A_0$. The change in $\mt$ has a significant effect on the
regions of CMSSM parameter space allowed, particularly in the 
focus-point
region where the range of $m_0$ allowed by cosmology now starts above 
4~TeV.
In view of the high values of $m_0$ and the sensitivity to $\mt$~\cite{rs}, 
we do
not study the focus-point region further in this paper. There are also
$\mt$- and $A_0$-dependent effects in the `funnels' where neutralinos
annihilate rapidly via the $H, A$ poles. These affect the dependence of
$m_0$ on $m_{1/2}$ along the WMAP lines, as we now discuss in more 
detail.
As explained below, because of the anomalous magnetic moment of the muon, 
we focus on cases with $\mu > 0$.

\begin{figure}
\begin{center}
\includegraphics[width=.48\textwidth]{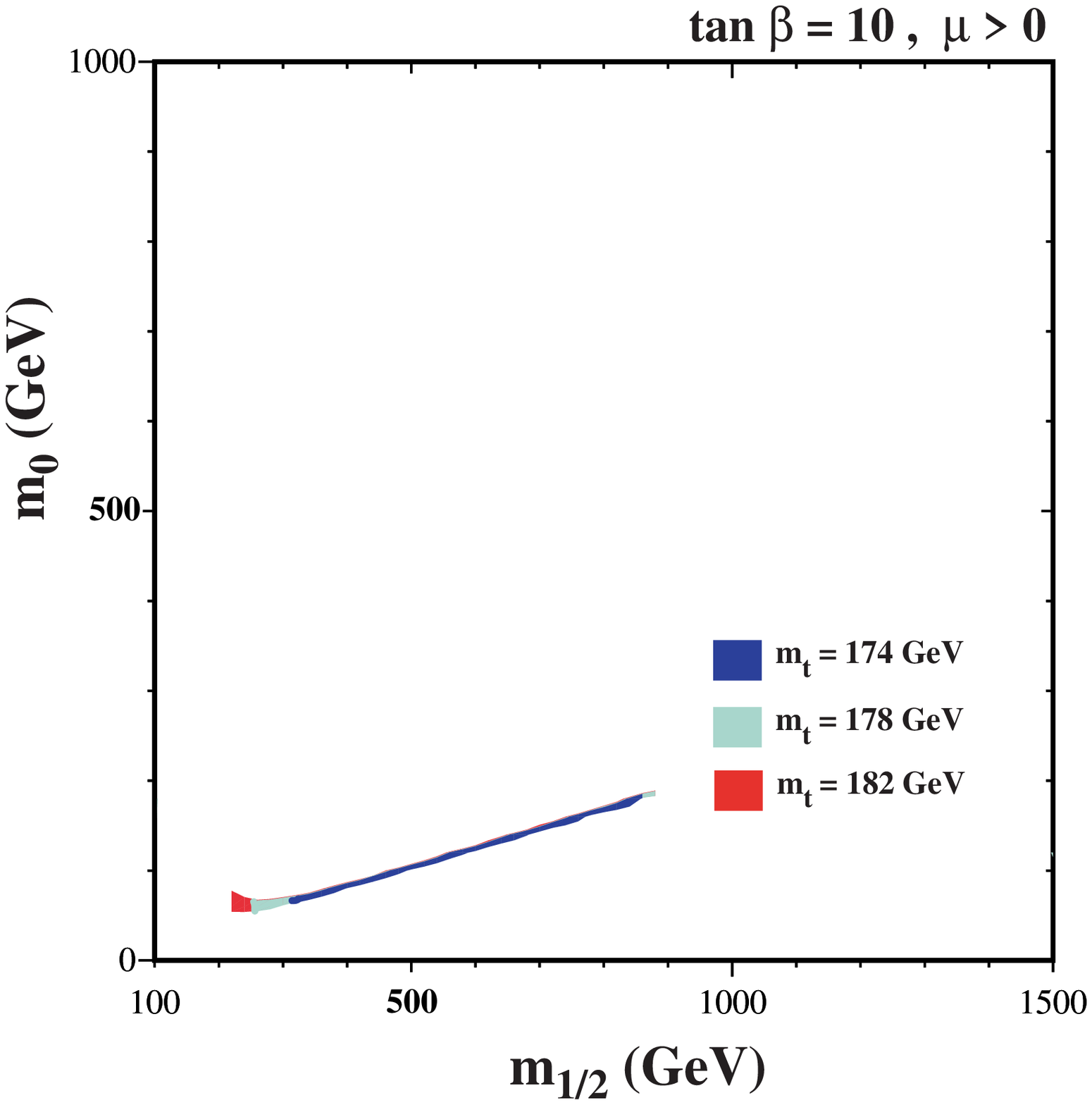}
\includegraphics[width=.48\textwidth]{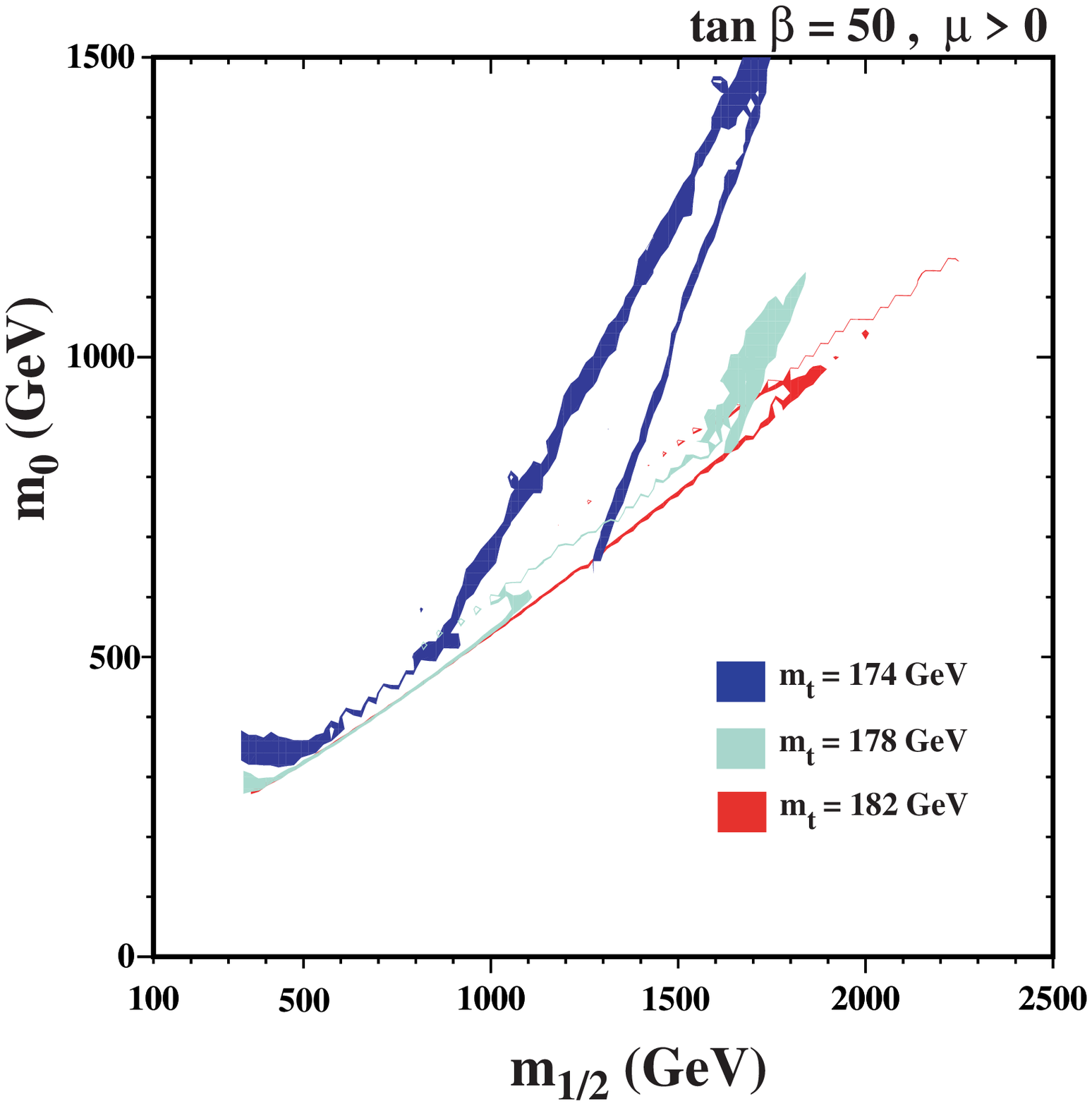}
\caption{The WMAP strips for $\mu > 0$, $A_0 = 0$
and (a) $\tb = 10$, (b)
$\tb = 50$, showing the dependence on the top-quark mass, for 
$\mt = 174, 178$ and 182~GeV.}
\label{fig:mt}
\end{center}
\end{figure}

Plotted in \reffi{fig:mt} is the region in the \plane{m_{1/2}}{m_0}
for fixed $\tb, A_0$ and $\mu > 0$ for which the relic density
is in the WMAP range (the results of~\cite{recent2} are in qualitative
agreement with \citere{recent3}).  We have applied cuts based on the lower
limit to the Higgs mass, $b \to s \gamma$, and require that the LSP be
a neutralino rather than the stau. The thin strips correspond to the 
relic density
being determined by either the coannihilation between nearly degenerate
$\tilde \tau$'s and $\chi$'s or, as seen at high $\tb$, by
rapid annihilation when $m_{\chi} \approx \MA/2$.
We see in \reffi{fig:mt}(a) that the WMAP strip for $\mu > 0$ and 
$\tan
\beta = 10$ does not change much as $\mt$ is varied, reflecting the fact
that the allowed strip is dominated by annihilation of the neutralino 
LSP
$\chi$ with the lighter stau slepton ${\tilde \tau_1}$. The main effect 
of
varying $\mt$ is that the truncation at low $m_{1/2}$, due to the Higgs
mass constraint, becomes more important at low $\mt$. This effect is not
visible in \reffi{fig:mt}(b) for $\tb = 50$, where the cutoff 
at
low $m_{1/2}$ is due to the $b \to s \gamma$ constraint, and rapid $\chi
\chi \to A, H$ annihilation is important at large $m_{1/2}$. The allowed
regions at larger $m_{1/2}$ vary significantly with $\mt$ when 
$\tb = 50$, because the $A, H$ masses and hence the rapid-annihilation 
regions
are very sensitive to $\mt$ through the renormalization group (RG)
running. Indeed, the rapid-annihilation region almost
disappears for $\mt = 182 \gev$ at this value of $\tb$.
In this case, in particular, we see a wisp
of allowed CMSSM parameter space running almost parallel to, but
significantly above, the familiar coannihilation strip, which is due to
rapid $\staustauH$ annihilation.  At 
higher values of
$\tb$ the rapid-annihilation region would reappear for $\mt = 182$ GeV.

\begin{figure}[htb!]
\vspace{1em}
\begin{center}
\includegraphics[width=.48\textwidth]{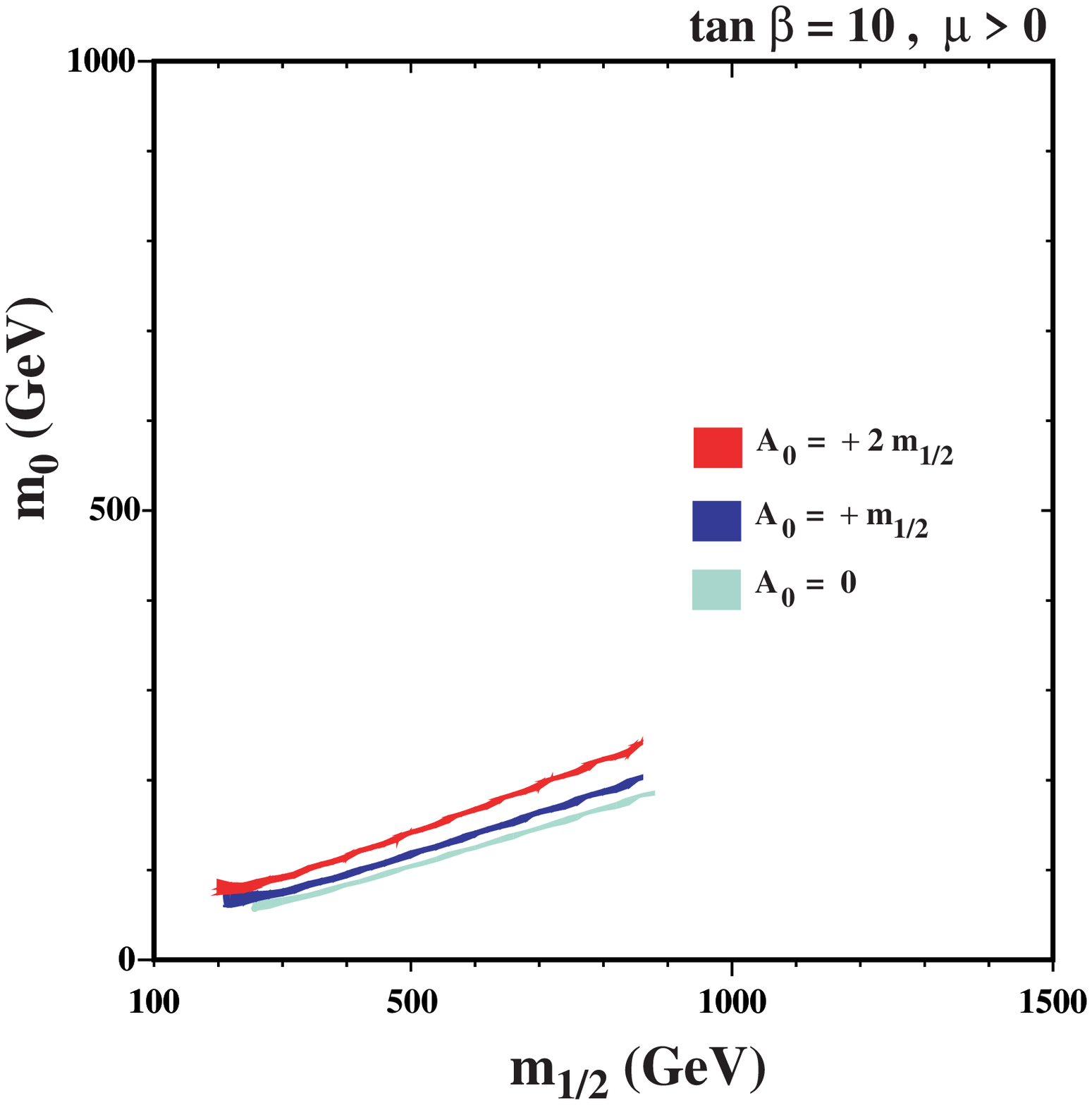}
\includegraphics[width=.48\textwidth]{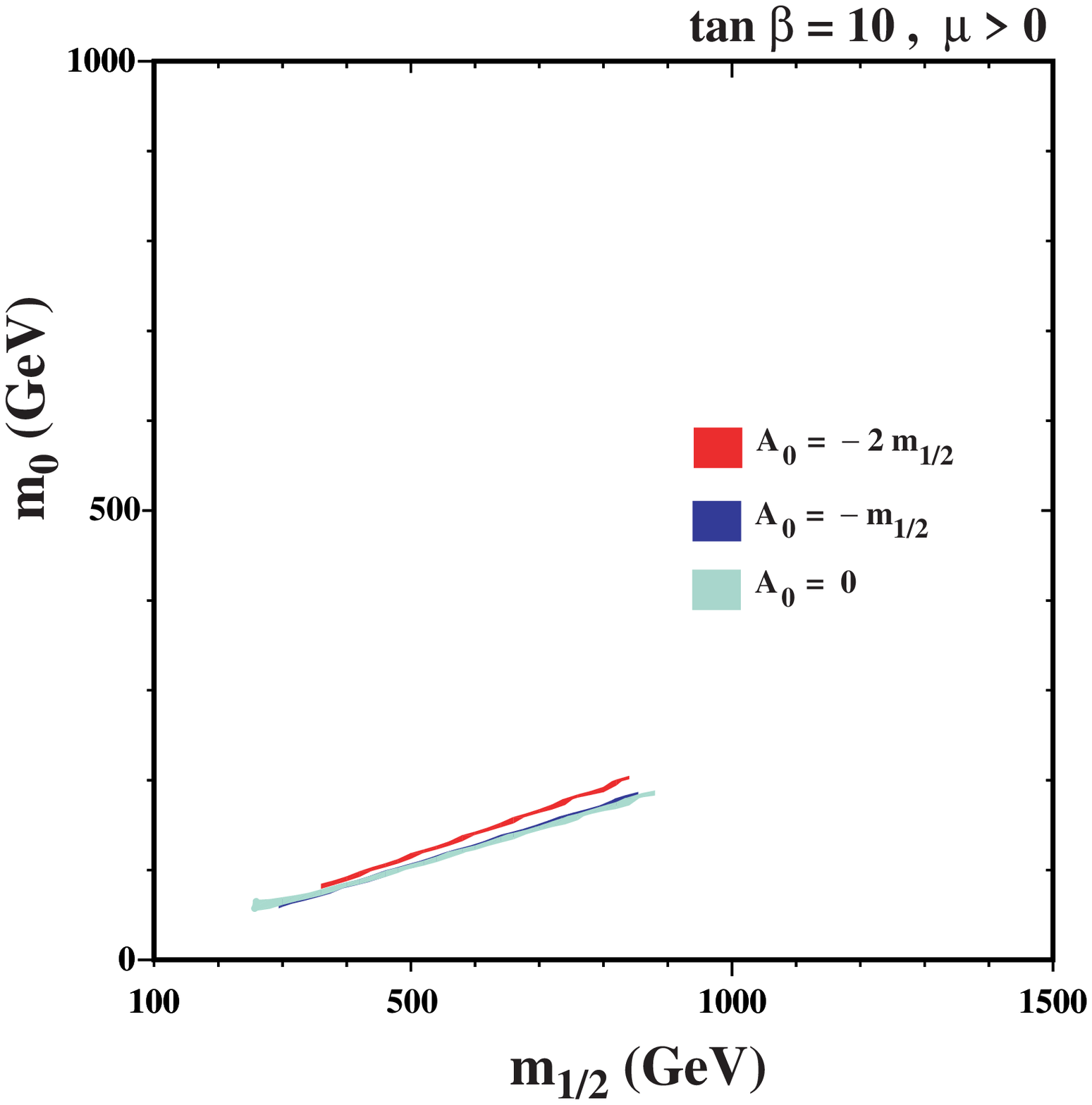}\\[3em]
\includegraphics[width=.48\textwidth]{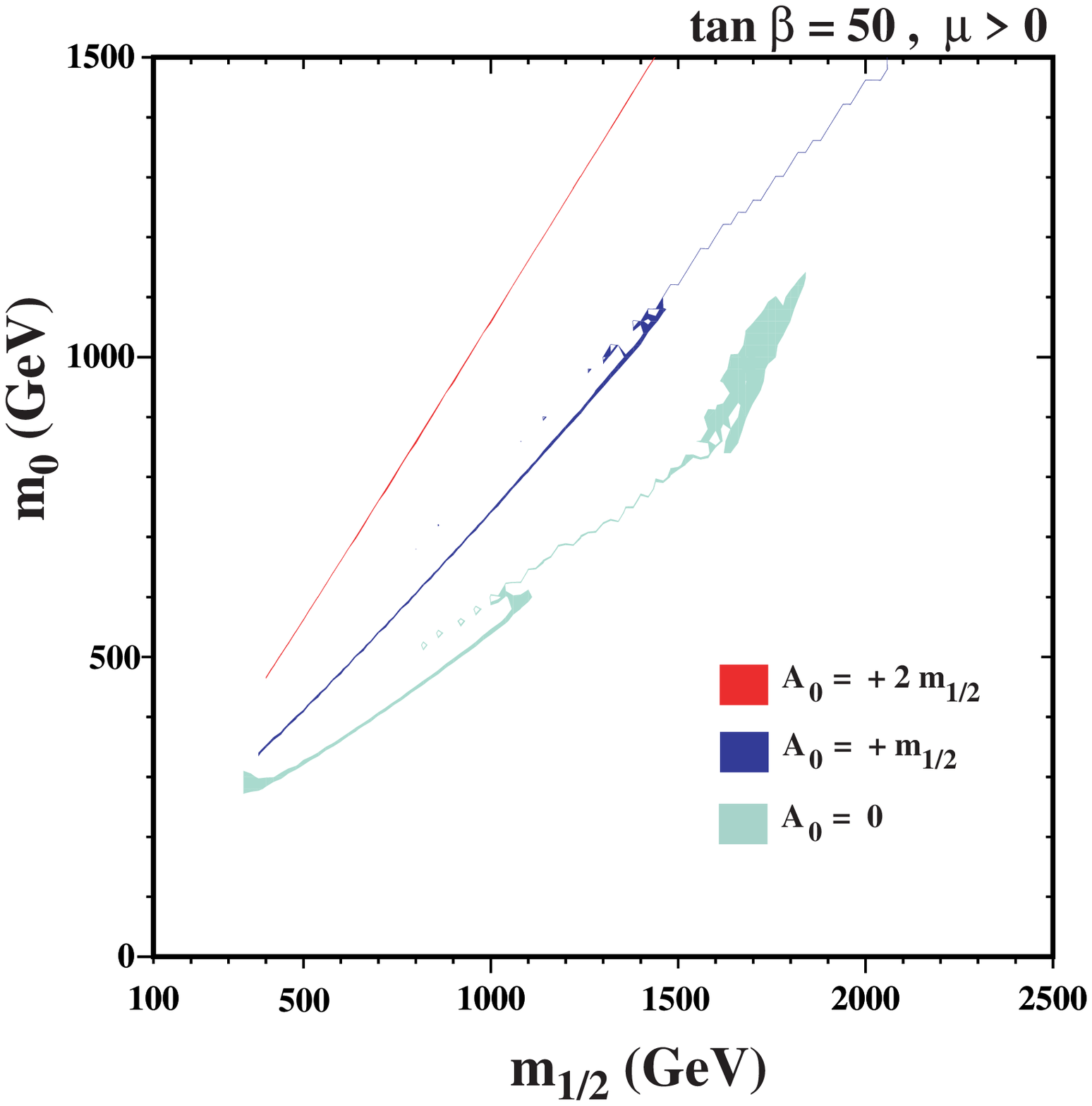}
\includegraphics[width=.48\textwidth]{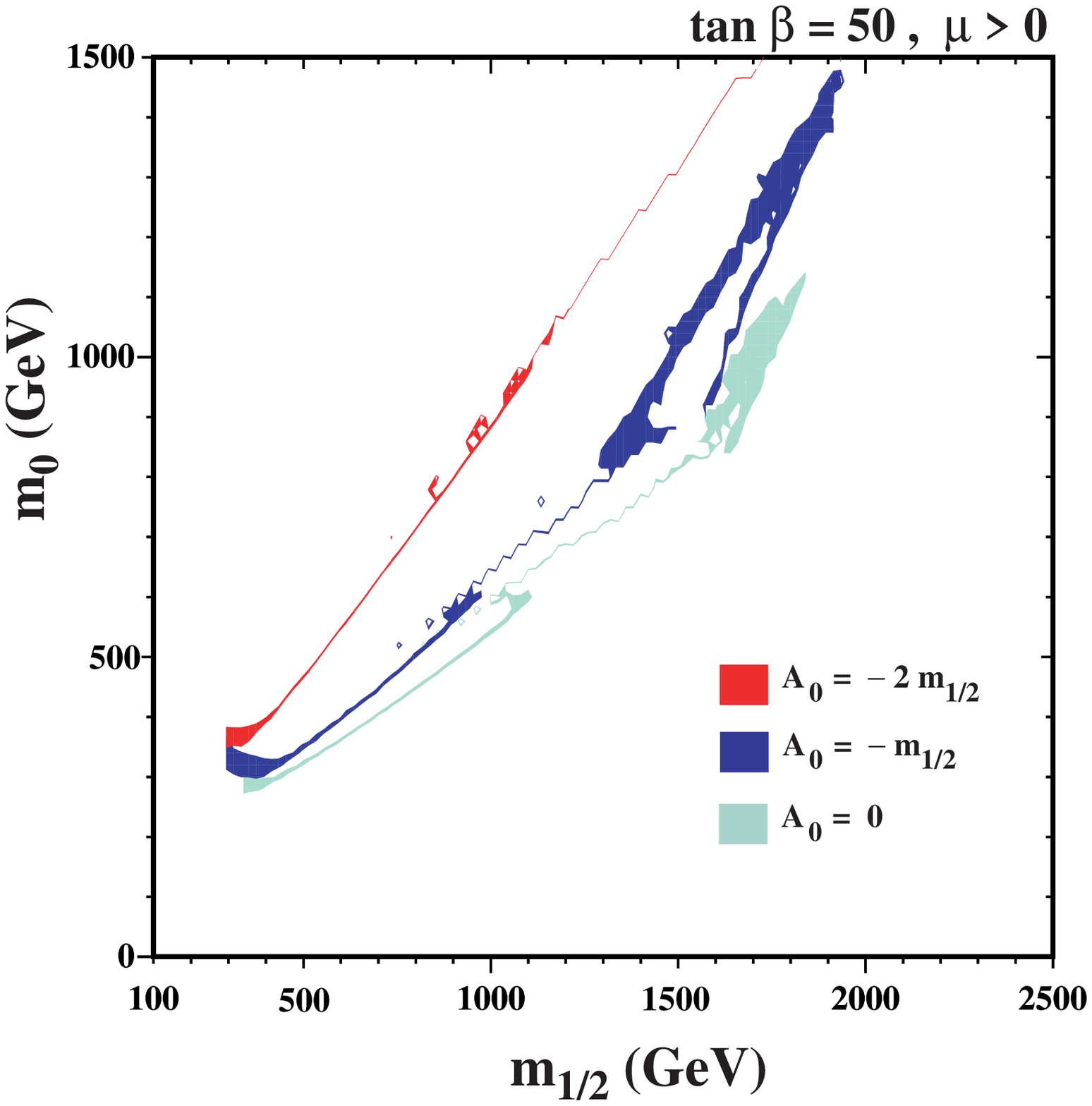}
\caption{The WMAP strips for $\mu > 0$, $\mt = 178 \gev$ 
and (a) $\tb = 10$, $A_0 \ge 0$ (upper left), 
(b) $\tb = 10$, $A_0 \le 0$ (upper right), 
(c) $\tb = 50$, $A_0 \ge 0$ (lower left), 
(d) $\tb = 50$, $A_0 \le 0$ (lower right)
showing the dependence on $A_0$ for $A_0 = 0, \pm m_{1/2}$
and $\pm 2 m_{1/2}$.
}
\label{fig:A}
\end{center}
\vspace{1em}
\end{figure}

We now turn to the variation of the WMAP strips for different $A_0$,
but with $\mt$ fixed to $\mt = 178 \gev$.
Since the WMAP strips are largely independent of the sign of $\mu$,
for clarity we show them in \reffi{fig:A} only for $\mu > 0$.
We see in \reffi{fig:A}(a,b) that the WMAP strip for $\tb = 10$
also does not change much as $A_0$ is varied: the main effect
is for the strip to move to larger $m_0$ as $|A_0|$ is increased. This is
because the main effect of $A_0$ is on the running of the diagonal
stau masses, whose RG equations depend only on $A_0^2$. The splitting
of the two stau masses depends 
on the sign of $A_0$ via the off-diagonal entries in the stau mass matrix,
but the impact of this effect on the final stau masses is relatively small.
Hence the WMAP strips rise for both signs of $A_0$. For a given value of
$m_{1/2}, m_0$ and $\tb$, the low-energy value of $A_\tau$ is
shifted from its high-energy value, $A_0$, by an amount $\Delta A$ that is
relatively independent of $A_0$.  Therefore, for $|A_0|$ much larger than
$\Delta A$, the low-energy value of $A_\tau$ will be larger than that for
$A_0 = 0$, causing the right-handed stau soft mass to drop. This in turn
increases the value of $m_0$ corresponding to the coannihilation strip.  
Only when the low-energy value of $|A_\tau|$ is less than and of opposite
sign to $\Delta A$ does the light stau mass increase.  In the
specific examples shown in \reffi{fig:A}(a,b), $\Delta A$ ranges from
about 130 GeV at low $m_{1/2}$ to about 550 GeV at high $m_{1/2}$. Since
the shifts are always positive, the coannihilation strip rises less
for negative values of $A_0$ (\reffi{fig:A}(b)) than for positive
values (\reffi{fig:A}(a)).

The WMAP regions for $\tb = 50$ vary much more rapidly with $|A_0|$,
because of the sensitivity of the $A, H$ masses and hence the
rapid-annihilation regions. In \reffi{fig:A}(c) the case for $A_0 \ge 0$
can be seen, whereas \reffi{fig:A}(d) shows $A_0 \le 0$.
We again see wisps of allowed CMSSM parameter space due to rapid 
$\staustauH$ annihilation.  In this case, as described above, the
right-handed stau mass is sensitive to the value of $A_0$.  Therefore, for
$A_0 \ne 0$ (\reffi{fig:A}(c,d)), the cosmologically preferred region
shifts to larger $m_0$ for both signs of $A_0$.
In addition, the value of the heavy Higgs 
scalar and pseudoscalar masses depends on $A_0$ (not only $A_0^2$) and the
position of the rapid-annihilation funnels therefore depends sensitively 
on~$A_0$.

In the following, we mainly present our results along the WMAP strips for
$\mt = 178 \gev$, the present experimental central value~\cite{mtopexpnew},
but we do show results for different values of $|A_0|$. This is because the
variation with $\mt$ is less important for $\tb = 10$, and
comparable with that due to varying $|A_0|$ when $\tb = 50$.
Additionally, we present scans of the $(m_{1/2}, A_0)$ planes for $\tb = 
10$ and 50.


\section{Present and future sensitivities to the scale of supersymmetry 
from
low-energy observables}
\label{sec:ewpo}

In this section, we briefly describe the low-energy observables used in
our analysis. We discuss the current and prospective future precision of
the experimental results and the theoretical predictions. 
In the following, we refer to the theoretical uncertainties from unknown
higher-order corrections as `intrinsic' theoretical uncertainties and
to the uncertainties induced by the experimental errors of the input
parameters as `parametric' theoretical uncertainties.  We also give 
relevant details of the higher-order perturbative corrections that we include.
We do not discuss theoretical
uncertainties from the RG running between the high-scale parameters
and the weak scale (see \citere{otherAnalyses2} for a recent discussion
in the context of predicting the CDM density). At present, 
these uncertainties are expected to be 
less important than the experimental and theoretical uncertainties of
the precision observables. In the future, both the uncertainties from 
unknown higher-order terms in the RG running and from the parameters 
entering
the running will considerably improve.

Results for these observables are shown as a function of $m_{1/2}$ with
$A_0$ varied, $m_0$ determined by the WMAP constraint (see
\refse{sec:CDMstrips}), and $\tb = 10, 50$. In this way the indirect
sensitivities of the low-energy observables to the scale of supersymmetry
are investigated.


\subsection{The $W$ boson mass}
\label{subsec:mw}

The $W$~boson mass can be evaluated from
\BE
\MW^2 \KL 1 - \frac{\MW^2}{\MZ^2}\right) = 
\frac{\pi \al}{\sqrt{2} \GF} \left(1 + \De r\KR,
\label{eq:delr}
\EE
where $\al$ is the fine structure constant and $\GF$ the Fermi constant.
The radiative corrections are summarized 
in the quantity $\De r$~\cite{sirlin}.
The prediction for $\MW$ within the Standard Model (SM)
or the MSSM is obtained from 
evaluating $\De r$ in these models and solving \refeq{eq:delr} in an
iterative way.

The one-loop contributions to $\De r$ can be written as
\BE
\De r = \De\al - \frac{\cw^2}{\sw^2}\De\rho 
        +(\De r)_{\rm rem},
\label{defdeltar2}
\EE
where $\De\al$ is the shift in the fine structure constant due to the 
light fermions of the SM, $\De\al \propto \log m_f$, and $\De\rho$ is the
leading contribution to the $\rho$ parameter. It is given by fermion and
sfermion loop contributions to 
the transverse parts of the 
gauge boson self-energies at zero external momentum, 
\BE
\De\rho = \frac{\Si_Z(0)}{\MZ^2} - \frac{\Si_W(0)}{\MW^2}~.
\EE
The remainder part, 
$(\De r)_{\rm rem}$, contains in particular the contributions from the
Higgs sector.

We include the complete \onel\ result in the
MSSM~\cite{deltarMSSM1lA,deltarMSSM1lB} as well as higher-order QCD
corrections of SM type of \order{\al\als}~\cite{drSMgfals,deltarSMgfals}
and \order{\al\als^2}~\cite{drSMgfals2,drSMgfals2LF}. Furthermore, we
incorporate 
supersymmetric corrections of \order{\al\als}~\cite{dr2lA} and of
\order{\al_t^2}~\cite{drMSSMgf2B} to $\De\rho$. 

The remaining intrinsic theoretical uncertainty in the prediction for
$\MW$ within the MSSM is still significantly larger than in the SM,
where it is currently estimated to be about 4~MeV~\cite{Awramik:2003rn}.
We estimate the present~\cite{MWintrcurrent} and future intrinsic 
uncertainties to be
\BE
\De\MW^{\rm intr,current} = 10 \mev, \quad 
\De\MW^{\rm intr,future} = 2 \mev. 
\EE
The parametric uncertainties are dominated by the experimental error of
the top-quark mass
and the hadronic contribution to the shift in the
fine structure constant. The current errors induce the following
parametric uncertainties
\BEA
\de\mt^{\rm current} = 4.3 \gev &\Rightarrow&
\De\MW^{{\rm para},\mt, {\rm current}} \approx 26 \mev,  \\[.3em]
\de(\De\al_{\rm had}^{\rm current}) = 36 \times 10^{-5} &\Rightarrow&
\De\MW^{{\rm para},\De\al_{\rm had}, {\rm current}} \approx 6.5 \mev .
\EEA
At the ILC, the top-quark mass will be measured with an accuracy of
about 100~MeV~\cite{lctdrs}. The parametric uncertainties induced by the 
future experimental errors of $\mt$ and $\De\al_{\rm had}$~\cite{fredl}
will then be~\cite{deltamt}
\BEA
\de\mt^{\rm future} = 0.1 \gev &\Rightarrow&
\De\MW^{{\rm para},\mt, {\rm future}} \approx 1 \mev,  \\[.3em]
\de(\De\al_{\rm had}^{\rm future}) = 5 \times 10^{-5} &\Rightarrow&
\De\MW^{{\rm para},\De\al_{\rm had}, {\rm future}} \approx 1 \mev .
\EEA

\noindent
The present experimental value of $\MW$ is~\cite{lepewwg}
\BE
\MW^{\rm exp,current} = 80.425 \pm 0.034~{\rm GeV}.
\label{mwexp}
\EE
With the GigaZ option of the ILC (i.e.\ high-luminosity running at the
$Z$~resonance and the $WW$ threshold) the $W$-boson mass will be
determined with an accuracy of about~\cite{mwgigaz,blueband}
\BE
\de\MW^{\rm exp,future} = 7 \mev .
\label{mwexpfuture}
\EE

\begin{figure}[htb!]
\begin{center}
\includegraphics[width=.48\textwidth]{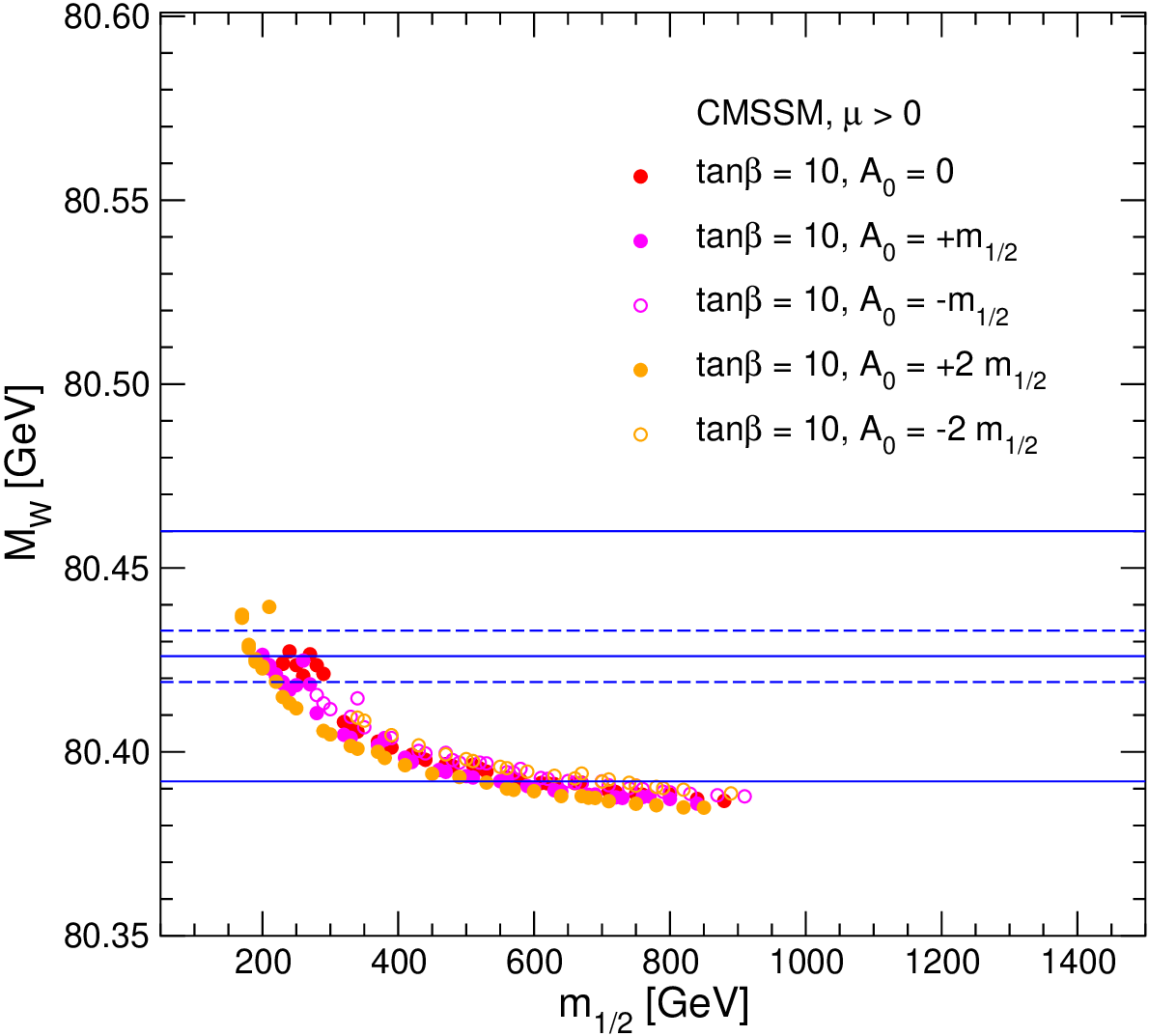}
\includegraphics[width=.48\textwidth]{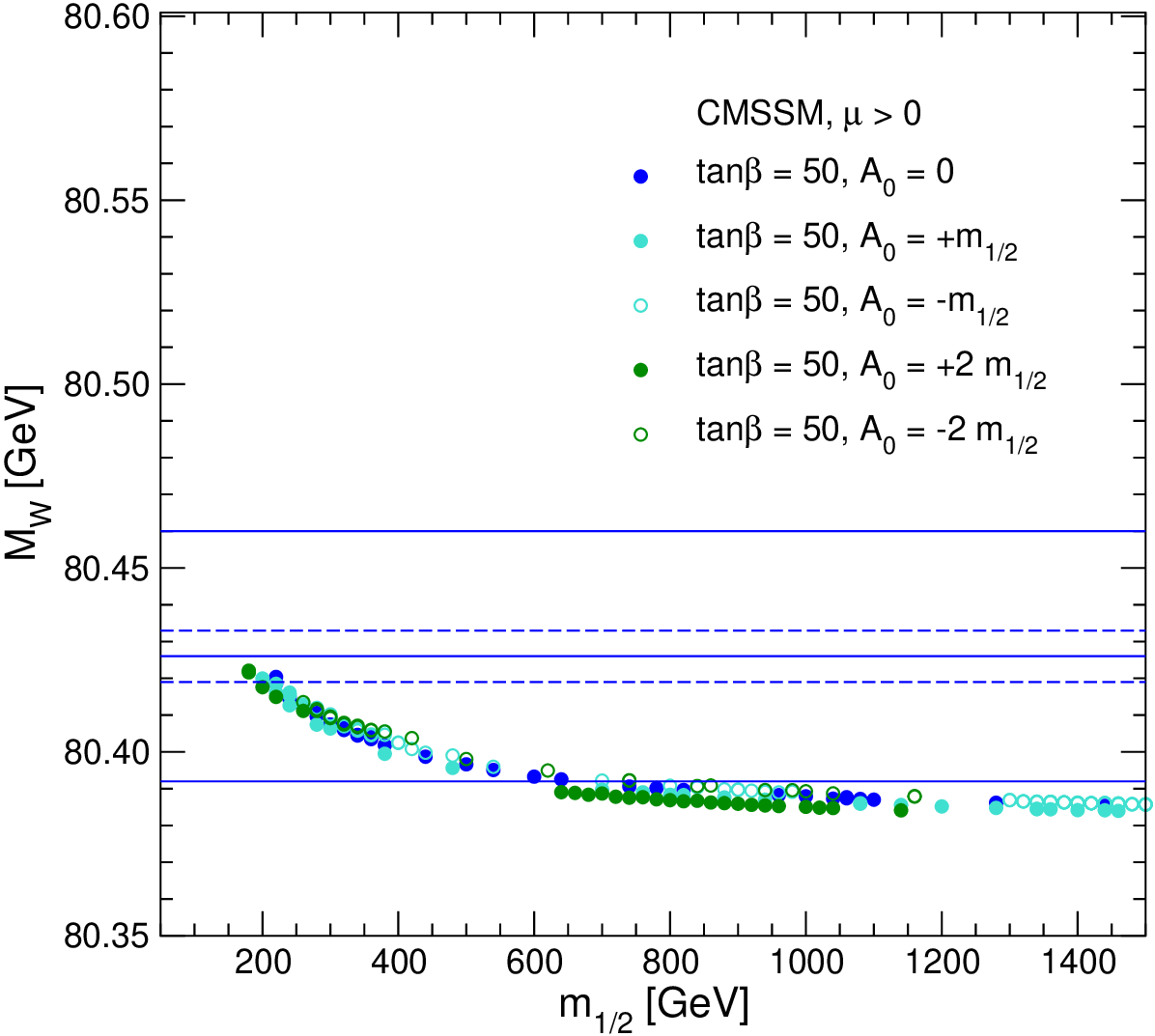}
\caption{%
The CMSSM prediction for $\MW$ as a function of $m_{1/2}$ along the 
WMAP strips for (a) $\tb = 
10$ and (b) $\tb = 50$ for various $A_0$ values. In each panel, the 
centre (solid) line is the present central
experimental value, and the (solid) outer lines show the current $\pm
1$-$\sigma$ range. The dashed lines correspond to the anticipated GigaZ
accuracy, assuming the same central value.
}
\label{fig:MW}
\end{center}
\end{figure}

In all plots of this section we show the theory predictions without
parametric and intrinsic theoretical uncertainties (using $\mt = 178$~GeV).
In the fits carried out in \refses{sec:combcurr} and \ref{sec:combfuture}
below we take both parametric and intrinsic theoretical uncertainties
into account.

We display in \reffi{fig:MW} the CMSSM prediction for $\MW$ and
compare it with the present measurement (solid lines) and a possible
future determination with GigaZ (dashed lines).
Panel (a) shows the values of $\MW$ obtained with $\tb = 10$ and 
$|A_0| \le 2$, and panel (b) shows the same for $\tb = 50$. It is striking 
that the present central value of $\MW$ (for both values of $\tb$)
favours low values of $m_{1/2} \sim 200$--$300$~GeV, though
values as large as 800~GeV are allowed at the 1-$\sigma$ level, and
essentially all values of $m_{1/2}$ are allowed at the $90 \%$ confidence
level. 
The GigaZ determination of $\MW$ might be able to determine
indirectly a low value of $m_{1/2}$ with
an accuracy of $\pm 50 \gev$, but even the GigaZ precision would still be
insufficient to determine $m_{1/2}$ accurately if $m_{1/2} \gsim 600 \gev$.


\subsection{The effective leptonic weak mixing angle}

The effective leptonic weak mixing angle at the $Z$~boson resonance 
can be written as
\BE
 \sweff = \frac{1}{4} \, \left( 1 - \re \frac{v_{\rm eff}}{a_{\rm eff}}  
\right) \ ,
\EE
where $v_{\rm eff}$ and $a_{\rm eff}$ 
denote the effective vector and axial couplings
of the $Z$~boson to charged leptons.
As in the case of $\MW$, the leading supersymmetric higher-order
corrections enter via the $\rho$~parameter, 
\BE
\de\sweff \approx - \frac{\cw^2 \sw^2}{\cw^2 - \sw^2} \De\rho .
\EE
Our theoretical prediction for $\sweff$ contains the same higher-order
corrections as described in \refse{subsec:mw}.

In the SM, the remaining intrinsic theoretical uncertainty in the prediction 
for $\sweff$ has been estimated to be about 
$5 \times 10^{-5}$~\cite{Awramik:2004ge}. For the MSSM, we use 
as present~\cite{MWintrcurrent} and future intrinsic uncertainties
\BE
\De\sweff^{\rm intr,current} = 12 \times 10^{-5}, \quad 
\De\sweff^{\rm intr,future} = 2 \times 10^{-5} . 
\EE
The current experimental errors of $\mt$ and $\De\al_{\rm had}$
induce the following parametric uncertainties
\BEA
\de\mt^{\rm current} = 4.3 \gev &\Rightarrow&
\De\sweff^{{\rm para},\mt, {\rm current}} \approx 14 \times 10^{-5},  \\[.3em]
\de(\De\al_{\rm had}^{\rm current}) = 36 \times 10^{-5} &\Rightarrow&
\De\sweff^{{\rm para},\De\al_{\rm had}, {\rm current}} \approx 
13 \times 10^{-5} .
\EEA
These should improve in the future to
\BEA
\de\mt^{\rm future} = 0.1 \gev &\Rightarrow&
\De\sweff^{{\rm para},\mt, {\rm future}} \approx 0.4 \times 10^{-5},  \\[.3em]
\de(\De\al_{\rm had}^{\rm future}) = 5 \times 10^{-5} &\Rightarrow&
\De\sweff^{{\rm para},\De\al_{\rm had}, {\rm future}} \approx 
1.8 \times 10^{-5} .
\EEA

It is well known that there is a 2.8-$\sigma$ discrepancy~\cite{lepewwg}
between the
leptonic and heavy-flavour determinations of the electroweak mixing angle,
with the leptonic measurement of $\sweff$ tending to pull down the value
of Higgs-boson mass
preferred in the SM fit, whereas the heavy-flavour measurements
favour a larger value of the Higgs mass.  
The Electroweak Working Group notes that
the overall quality of a global electroweak fit is quite acceptable, 
$\sim 26 \%$~\cite{lepewwg}, and we use their combination of the two sets 
of measurements:
\BE
\sweff^{\rm exp,current} = 0.23150 \pm 0.00016.
\label{swfit}
\EE
The experimental accuracy will improve to
about
\BE
\de\sweff^{\rm exp,future} = 1 \times 10^{-5} .
\label{swexpfuture}
\EE
at GigaZ~\cite{ewpo:gigaz2}.

\begin{figure}[htb!]
\begin{center}
\includegraphics[width=.48\textwidth]{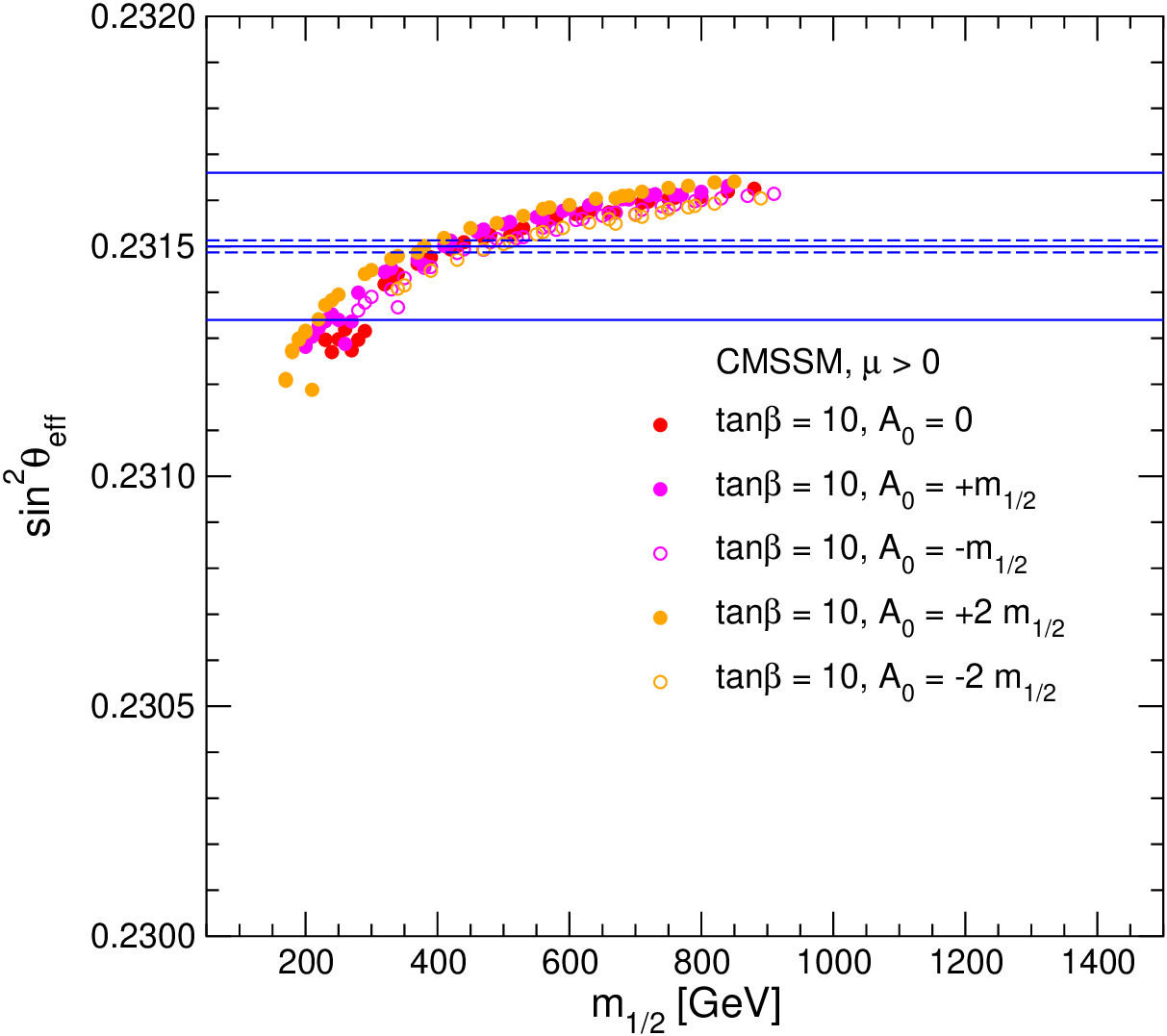}
\includegraphics[width=.48\textwidth]{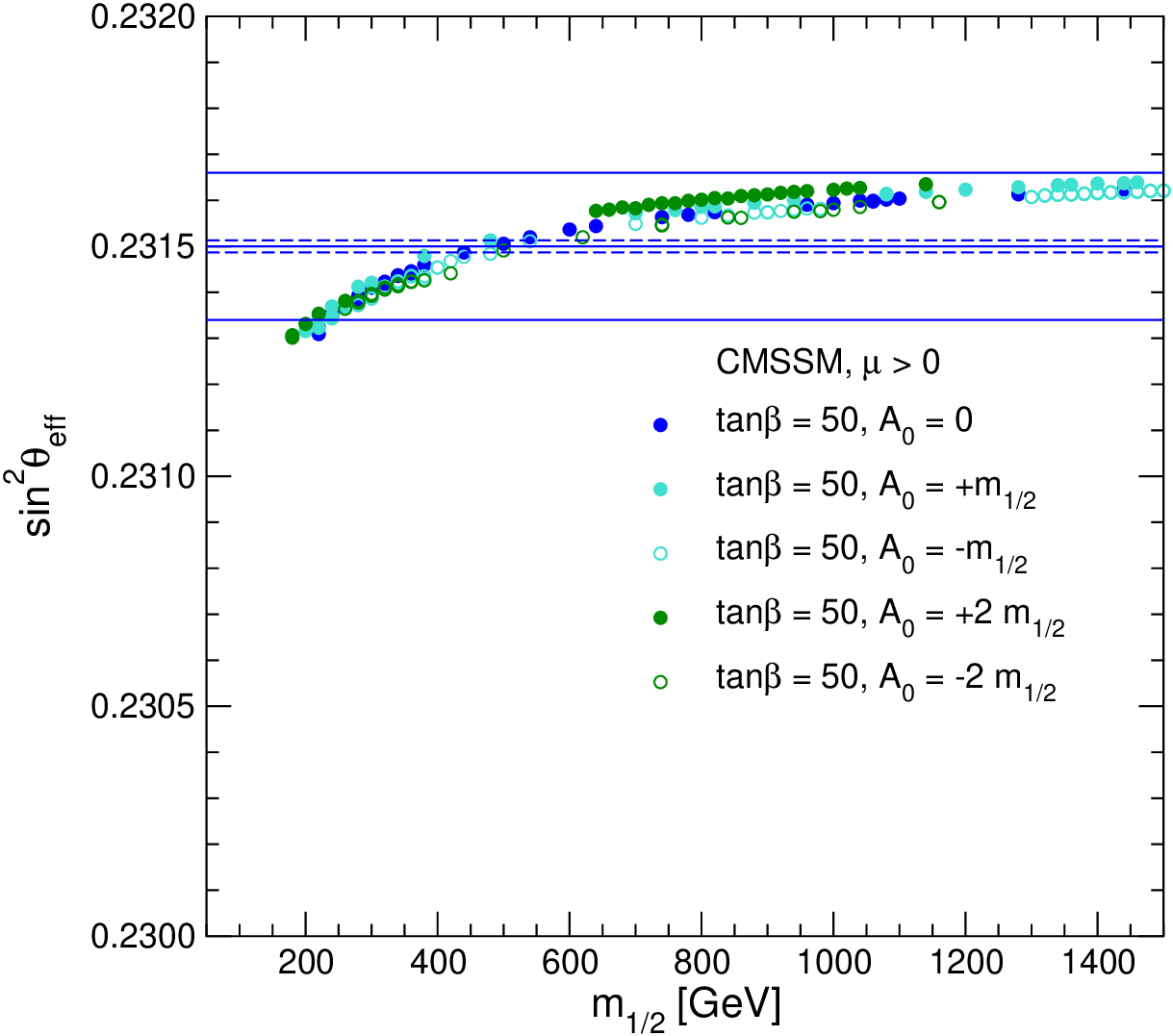}
\caption{%
The CMSSM prediction for $\sweff$ as a function of $m_{1/2}$ along the 
WMAP strips for (a) 
$\tb = 10$ and (b) $\tb = 50$ for various $A_0$ values. In each panel, 
the centre (solid) line is the present central
experimental value, and the (solid) outer lines show the current $\pm
1$-$\sigma$ range. The dashed lines correspond to the anticipated GigaZ
accuracy, assuming the same central value.
}
\label{fig:SW}
\end{center}
\end{figure}

\reffi{fig:SW} shows the prediction for $\sweff$ in the CMSSM compared
with the present and future experimental precision.
As in the case of $\MW$, low values of $m_{1/2}$ are also
favoured independently by $\sweff$. The present central value prefers 
$m_{1/2} = 300$--$500 \gev$, but the 1-$\sigma$ range extends beyond 1500~GeV
(depending on $A_0$), 
and all
values of $m_{1/2}$ are allowed at the $90 \%$ confidence level. 
The GigaZ precision on $\sweff$ would 
be able to determine $m_{1/2}$ indirectly with even greater accuracy 
than $\MW$ at low $m_{1/2}$, but would also be insufficient if 
$m_{1/2} \gsim 700 \gev$.


\subsection{The anomalous magnetic moment of the muon}

We now discuss the evaluation of the MSSM contributions to the anomalous
magnetic moment of the muon, $\amu \equiv (g-2)_\mu$. Since the possible
deviation of the SM prediction from the experimental result is crucial for
the interpretation of the $\amu$ results, we first review this aspect in 
the light of recent developments.

The SM prediction for the anomalous magnetic moment of 
the muon (see~\citeres{g-2review,g-2review2}
for reviews) depends on the evaluation of the
hadronic vacuum polarization and light-by-light (LBL) contributions. The
former have been evaluated in~\cite{DEHZ,g-2HMNT,Jegerlehner,Yndurain}
and the latter in~\cite{LBL,LBLnew}. The evaluations of the 
hadronic vacuum polarization contributions using $e^+ e^-$ and $\tau$ 
decay data give somewhat different results. Recently, new data have been 
published by the KLOE Collaboration~\cite{KLOE}, which agree well with 
the previous 
data from CMD-2. This, coupled with a greater respect for the 
uncertainties inherent in the isospin transformation from $\tau$ decay, 
has led to a proposal to use the $e^+ e^-$ alone and shelve the $\tau$ 
data, resulting in the estimate~\cite{Hocker:2004xc}
\BE
\amutheo = 
(11\, 659\, 182.8 \pm 6.3_{\rm had} \pm 3.5_{\rm LBL} \pm 0.3_{\rm QED+EW})
 \times 10^{-10},
\label{eq:amutheo}
\EE
where the source of each error is labelled~%
\footnote{
The updated QED result from~\cite{Kinoshita} is included.
}%
.

This result is to be compared with
the final result of the Brookhaven $(g-2)_\mu$ Experiment 
E821, namely~\cite{g-2exp}
\BE
\amuexp = (11\, 659\, 208.0 \pm 5.8) \times 10^{-10},
\label{eq:amuexp}
\EE
leading to an estimated discrepancy
\BE
\amuexp-\amutheo = (25.2 \pm 9.2) \times 10^{-10},
\label{delamu}
\EE
equivalent to a 2.7~$\sigma$ effect. In view of the chequered history of
the SM prediction, \refeq{eq:amutheo}, and the residual questions concerning
the use of the $\tau$ decay data, it would be premature to regard 
this discrepancy as firm evidence of new physics. We do note, on the
other hand,
that the $(g-2)_\mu$ measurement imposes an important constraint on 
supersymmetry, even if one uses the $\tau$ decay data. We use
\refeq{delamu} for our numerical discussion below.

The following MSSM contributions to the theoretical prediction for $\amu$
have been considered. We take fully into account the complete one-loop
contribution to $\amu$, which was evaluated nearly a decade ago 
in~\citere{g-2MSSMf1l}. We make no simplification in the sparticle mass
scales but, for illustrating the possible size of
corrections, a simplified formula can be used, in which relevant
supersymmetric mass scales are set to a common value,
$\msusy = m_{\cha{}} = m_{\neu{}} = m_{\Smu} = m_{\Sneum}$. The result
in this approximation is given by
\BE
\amu^{\SU,{\rm 1L}} = 13 \times 10^{-10}
             \KL \frac{100 \gev}{\msusy} \KR^2 \tb\;
 {\rm  sign}(\mu).
\label{susy1loop}
\EE
We see that supersymmetric effects can easily account for a
$(20\ldots30)\times10^{-10}$ deviation, if $\mu$ is positive and
$\msusy$ lies roughly between 100 GeV (for small $\tb$) and
600 GeV (for large $\tb$).
For this reason, in the rest of this paper, we restrict our attention 
to $\mu > 0$. Even in view of the possible size of experimental and
theoretical uncertainties, it is very difficult to 
reconcile $\mu < 0$ with the present data on $\amu$.

In addition to the full one-loop contributions, we also include several
two-loop corrections. The first class of corrections comprises
the leading $\log \KL m_\mu/\msusy\KR$ terms of supersymmetric 
one-loop diagrams with a photon
in the second loop, which are given by~\cite{g-2MSSMlog2l}:
\BE
\De\amu^{\SU,{\rm 2L,QED}} = \De\amuSUoL \times
\KL \frac{4\,\al}{\pi} \log\KL \frac{\msusy}{m_\mu} \KR \KR~.
\EE
These amount to about $-8\%$ of the supersymmetric one-loop contribution
for a supersymmetric mass scale $\msusy = 500 \gev$.

The second class of two-loop corrections comprises diagrams with a closed 
loop of SM
fermions or scalar fermions. These were calculated in~\citere{g-2FSf},
where it was demonstrated that these corrections may amount to $\sim 5
\times 10^{-10}$ in the general MSSM, if all experimental bounds are taken
into account. These corrections are included in the Fortran
code {\tt FeynHiggs}~\cite{feynhiggs,feynhiggs2.2}. We have furthermore
taken into account the 2-loop contributions to $\amu$ from 
diagrams containing a closed
chargino/neutralino loop, which have been evaluated in~\cite{g-2CNH}.
Here we use an approximate form for these corrections, which are typically
$\sim 1 \times 10^{-10}$. 

The current intrinsic uncertainties in the MSSM
contributions to $\amu$ can be estimated to be 
$\sim 6 \times 10^{-10}$~\cite{g-2CNH,PomssmRep}. In the more restricted CMSSM
parameter space the intrinsic uncertainties are smaller,
being about $1 \times 10^{-10}$. Once the full two-loop result in the
MSSM is available, this uncertainty will be further reduced. 
We assume that in the future the uncertainty in \refeq{delamu} will be
reduced by a factor two.

\begin{figure}[bht!]
\begin{center}
\epsfig{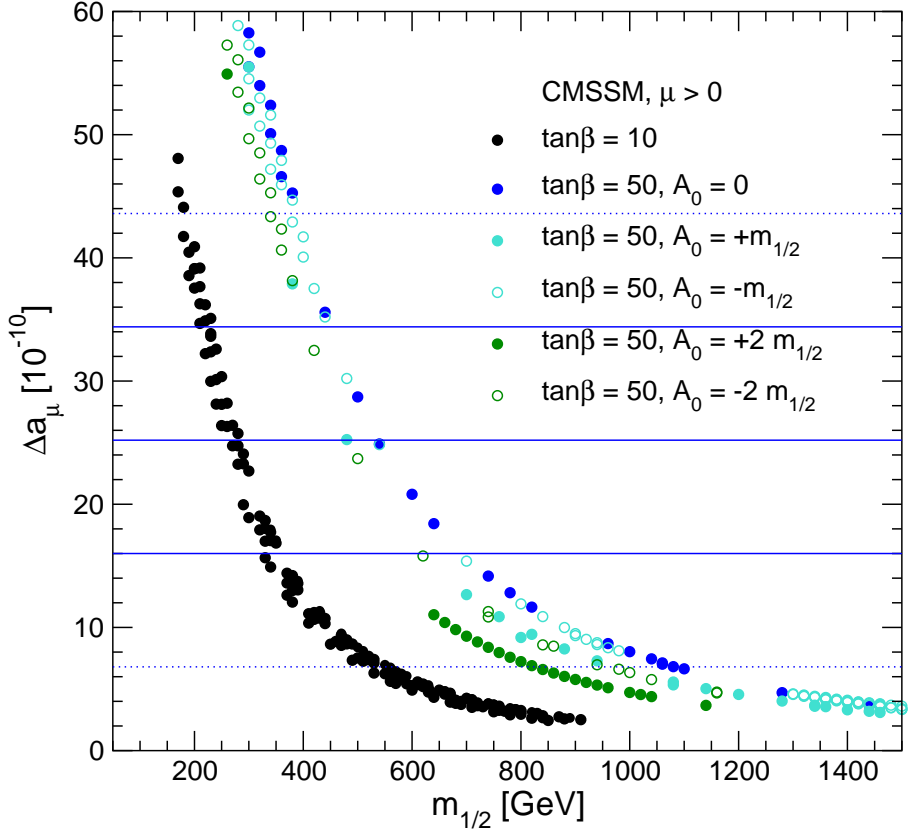}
\caption{%
The CMSSM prediction for $\De \amu$ as a function of $m_{1/2}$ along 
the WMAP strips for $\tb = 
10, 50$ and different $A_0$ values. The central (solid) line is the 
central value of the present discrepancy between experiment and the SM 
value evaluated using $e^+ e^-$ data (see text), and the other solid (dotted)
lines show the current $\pm 1(2)$-$\sigma$ ranges, \refeq{delamu}. 
}
\label{fig:AMU}
\end{center}
\end{figure}

As seen in \reffi{fig:AMU}, the CMSSM prediction for $\amu$ is almost 
independent of $A_0$ for $\tb = 10$, but substantial variations are possible 
for $\tb = 50$, except at very large $m_{1/2}$. In the case $\tb = 10$, 
$m_{1/2} \sim 200$--400~GeV is again favoured at the $\pm 1$-$\sigma$ 
level, but this preferred range shifts up to 400 to 800~GeV if $\tb = 
50$, depending on the value of $A_0$. At the 2-$\sigma$ level, there is 
nominally an upper bound $m_{1/2} \lsim 600 (1100) \gev$ for 
$\tb = 10 (50)$, but according to the discussion above it should be
interpreted with care. 
Nevertheless, the lower bound to $m_{1/2}$ for both
$\tb = 10$ and~50 should be regarded as relatively robust.
On the other hand, it is striking that $\MW$, $\sweff$ and $\amu$ all 
favour small $m_{1/2}$ for $\tb = 10$.
If $\tb = 50$, the consistency between the ranges 
preferred by the different observables is not so striking.


\subsection{The decay $b \to s \ga$}

Since this decay occurs at the loop level in the SM, the MSSM 
contribution might, {\it a priori}, be of similar magnitude. The most
up-to-date
theoretical estimate of the SM contribution to the branching ratio
is~\cite{ali}
\BE
\br( b \to s \ga ) = (3.70 \pm 0.30) \times 10^{-4},
\label{bsga}
\EE
where the calculations have been carried out completely to NLO in the 
\msbar\ renormalization scheme, and the error is dominated by 
higher-order QCD uncertainties. A complete NNLO QCD calculation is now 
underway, and will reduce significantly the uncertainty, once it is 
available. 

For comparison, the present experimental 
value estimated by the Heavy Flavour Averaging Group (HFAG)
is~\cite{bsgexp}
\BE
\br( b \to s \ga ) = (3.54^{+ 0.30}_{- 0.28}) \times 10^{-4},
\label{bsgaexp}
\EE
where the error includes an uncertainty due to the decay spectrum, as well 
as the statistical error. The very good agreement between \refeq{bsgaexp} 
and the SM calculation \refeq{bsga} imposes important constraints on the 
MSSM, as we see below.

Our numerical results have been derived and checked with three
different codes. The first is based on~\citeres{bsgKO1,bsgKO2}%
\footnote{
We are grateful to P.~Gambino and G.~Ganis for providing the
corresponding code.
}%
~and the second is based on \citeres{bsgKO2,bsgGH}%
\footnote{
We thank Gudrun Hiller for providing the corresponding Fortran code.
}%
. Results have been derived using the charm pole mass as well as the
charm running mass, giving an estimate of remaining higher-order
uncertainties. Finally, our results have been checked with the 
$\br(b \to s \ga)$ evaluation provided in~\citere{bsgMicro}, which
yielded very similar results to our two other approaches.
For the current theoretical uncertainty of the MSSM prediction for 
$\br(b \to s \ga)$ we use the value of \refeq{bsga}.
For the future uncertainty from the experimental as well as the
theoretical side we assume a reduction by a factor of~3.

\begin{figure}[htb!]
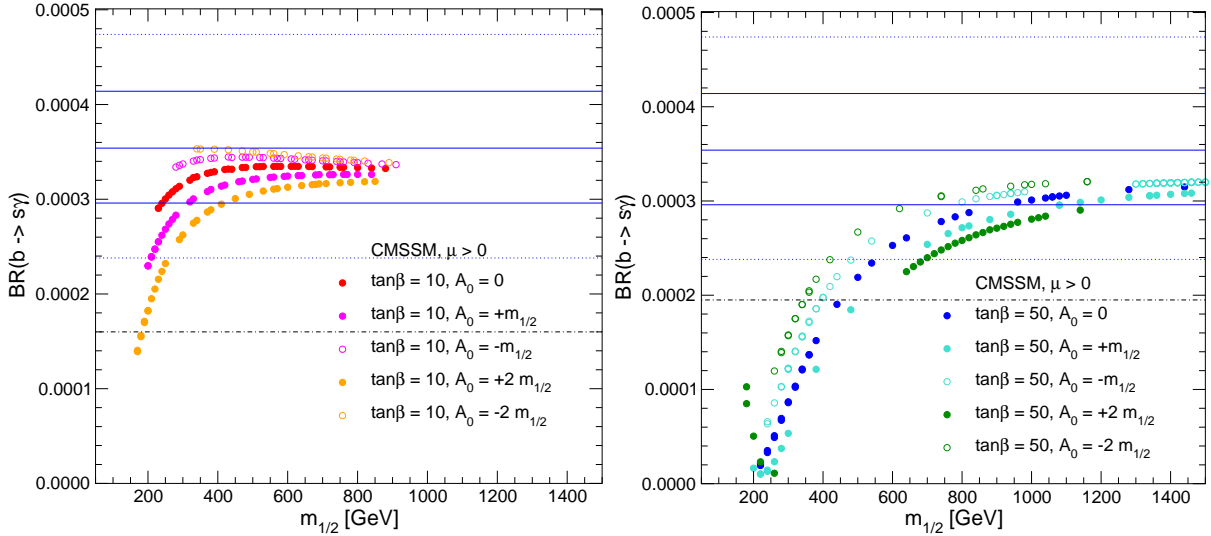

\begin{center}
\includegraphics[width=.48\textwidth]{ehow.BSG11a2.cl.eps}
\includegraphics[width=.48\textwidth]{ehow.BSG11b2.cl.eps}
\caption{%
The CMSSM predictions for $\br(b \to s \ga)$ as a function of $m_{1/2}$ 
along the WMAP strips for 
(a) $\tb = 10$ and (b) $\tb = 50$ and various choices of $A_0$. The 
uncertainty shown combines linearly the current experimental 
error and the present theoretical uncertainty in the SM prediction.
The central (solid) line indicates the current experimental
central value, and the other solid (dotted)
lines show the current $\pm 1(2)$-$\sigma$ ranges.
The dash-dotted line corresponds to a more conservative estimate of
intrinsic uncertainties (see text).
}
\label{fig:BSG}
\end{center}
\vspace{-1em}
\end{figure}

As already mentioned, the present central value of this branching ratio
agrees very well with the SM, implying that large values of $m_{1/2}$
cannot be excluded for any value of $\tb$. The uncertainty range 
shown in \reffi{fig:BSG} combines linearly the current 
experimental error and the present theoretical uncertainty in the SM 
prediction. Note however, that at present there is also an
uncertainty in the computed MSSM value (included in
obtaining the excluded regions in Figs. \ref{fig:mt} and \ref{fig:A})
from the uncertainty in the SUSY loop calculations.
Taking this conservatively into account results in a 95\% C.L.\
exclusion bound of 0.00016 in the case of $\tb = 10$, and of 0.000195
in the case of $\tb = 50$. These values are shown as dash-dotted lines
in \reffi{fig:BSG}. This allows a somewhat lower range in $m_{1/2}$
than depicted in \reffi{fig:BSG}. We assume that these uncertainties
can be significantly reduced in the future. We have checked that they
have no significant impact on the results presented below.

Since the CMSSM corrections are
generally smaller for smaller $\tb$, even values of $m_{1/2}$ as low as
$\sim 200 \gev$ would be allowed at the $90 \%$ confidence level if 
$\tb = 10$, whereas $m_{1/2} \gsim 450 \gev$ would be required if $\tb = 50$.
These limits are very sensitive to $A_0$, and, if the future error in 
$\br(b \to s \gamma)$ could indeed be reduced by a factor $\sim 3$, 
the combination of $\br(b \to s \gamma)$ 
with 
the other precision observables might be able, in principle,
to constrain $A_0$ significantly.


\subsection{The branching ratio $B_s \to \mu^+\mu^-$}
\label{subsec:bsmm}

The SM prediction for this branching ratio is $(3.4 \pm 0.5) \times
10^{-9}$~\cite{bsmmtheosm}, and 
the present experimental upper limit from the Fermilab Tevatron collider
is $3.4 \times 10^{-7}$ at the $95\%$ C.L.~\cite{bsmmexp}, providing ample 
room for the MSSM to dominate the SM contribution. The current Tevatron
sensitivity, being based on an integrated luminosity of about 410~\ipb\
summed over both detectors, is expected to improve significantly in the
future. A naive scaling of the present bound with the square root of the
luminosity yields a sensitivity at the end of Run~II of about 
$5.4 \times 10^{-8}$ assuming 8~\ifb\ collected with each detector. An
even bigger improvement may be possible with better signal acceptance
and more efficient background reduction. In \citere{heinemann} an
estimate of the future Tevatron sensitivity of $2 \times 10^{-8}$ at the 
90\%
C.L.\ has been given, and a sensitivity even down to the SM value can be
expected at the LHC. Assuming the SM value, i.e.\
$\br(B_s \to \mu^+ \mu^-) \approx 3.4 \times 10^{-9}$, it has been
estimated~\cite{lhcb} that LHCb can observe 33~signal events
over 10~background events within 3~years of low-luminosity
running. Therefore this process offers good prospects for probing the MSSM. 

For
the theoretical prediction we use
results from~\citere{bsmumu}%
\footnote{
We are grateful to A.\ Dedes for providing the corresponding code.
}%
, which include the full one-loop
evaluation and the leading two-loop QCD corrections. We are not aware of
a detailed estimate of the theoretical uncertainties from unknown
higher-order corrections.

\begin{figure}[tbh!]
\begin{center}
\epsfig{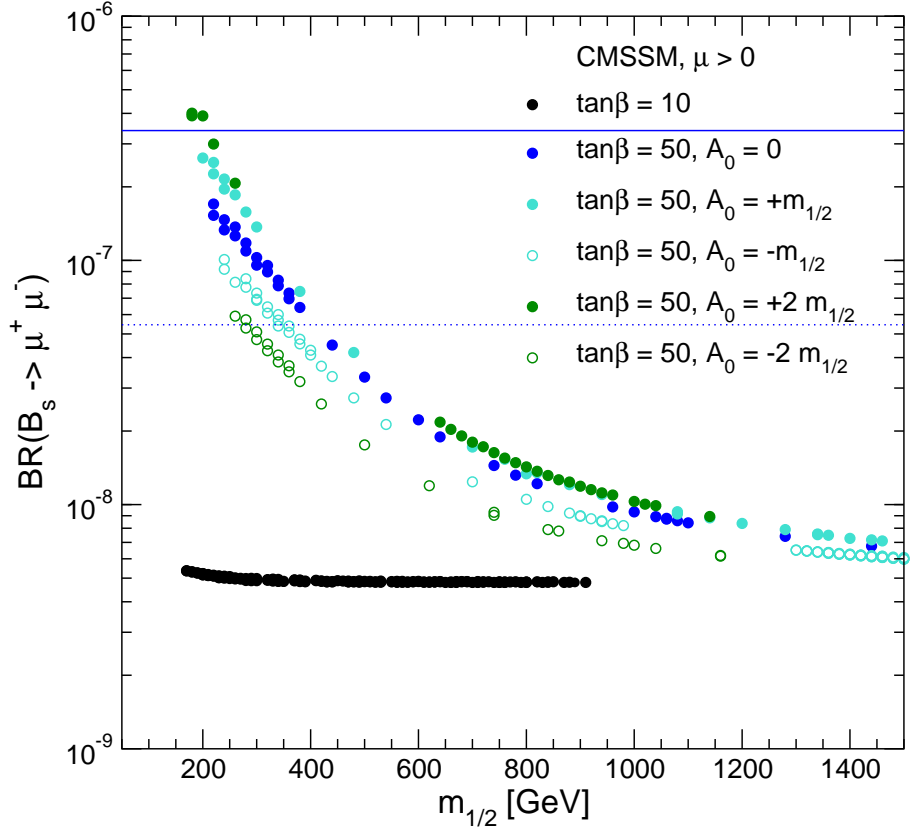}
\caption{%
The CMSSM prediction for $\br(B_s \to \mu^+\mu^-)$ as a function of 
$m_{1/2}$ along the WMAP strips for
$\tb = 10$ and all $A_0$ values, and for $\tb = 50$ with various 
values of $A_0$. The solid line shows the current Tevatron limit at the
95\% C.L., and the 
dotted line corresponds to an estimate for the
sensitivity of the Tevatron at
the end of Run~II.
}
\label{fig:BMM}
\end{center}
\end{figure}

In \reffi{fig:BMM} the CMSSM prediction for $\br(B_s \to \mu^+\mu^-)$ as a 
function of $m_{1/2}$ is compared with the present Tevatron limit and
our estimate for the sensitivity at the end of Run~II. For $\tb = 10$ 
the CMSSM prediction is significantly below the present and future Tevatron
sensitivity. With the current sensitivity, the Tevatron starts to probe
the CMSSM region with $\tb = 50$. The sensitivity at the end of Run~II 
will test the CMSSM parameter space with $\tb = 50$ and 
$m_{1/2} \lsim 600 \gev$, in particular for positive values of $A_0$. 
The LHC will be able to probe the whole CMSSM parameter space via this
rare decay.


\subsection{The lightest MSSM Higgs boson mass}

The mass of the lightest $\cp$-even MSSM Higgs boson can be predicted in 
terms of
the other CMSSM parameters. At the tree level, the two $\cp$-even Higgs 
boson masses are obtained as a function of $\MZ$, the $\cp$-odd Higgs
boson mass $\MA$, and $\tb$. 
In the Feynman-diagrammatic (FD) approach, which we employ here, 
the higher-order corrected 
Higgs boson masses are derived by finding the
poles of the $h,H$-propagator 
matrix. This is equivalent to solving 
\BE
\label{eq:proppole}
\left[p^2 - m_{h, \rm tree}^2 + \hSi_{hh}(p^2) \right] \times 
 \left[p^2 - m_{H, \rm tree}^2 + \hSi_{HH}(p^2) \right] 
- \left[\hSi_{hH}(p^2)\right]^2 = 0~,
\EE
where the $\hSi(p^2)$ denote the renormalized Higgs-boson self-energies,
and $p$ is the external momentum.

For the theoretical prediction of $\Mh$
we use the code {\tt FeynHiggs}~\cite{feynhiggs,feynhiggs2.2},
which includes all numerically relevant known higher-order corrections.
The status of the incorporated results for the self-energy contributions to
\refeq{eq:proppole} can be summarized as follows. For the
one-loop part, the complete result within the MSSM is 
known~\cite{ERZ,mhiggsf1lB,mhiggsf1lC}. 
Concerning the two-loop
effects, their computation is quite advanced, see \citere{mhiggsAEC} and
references therein. They include the strong corrections
at \order{\al_t\als} and Yukawa corrections at \order{\al_t^2},
as well as the dominant one-loop \order{\al_t} term, and the strong
corrections from the bottom/sbottom sector at \order{\al_b\als}. 
For the $b/\Sbot$~sector
corrections also an all-order resummation of the $\Tb$-enhanced terms,
\order{\al_b(\als\tb)^n}, is known~\cite{deltamb,deltamb1}.
Most recently, the \order{\al_t \al_b} and \order{\al_b^2} corrections
have been derived~\cite{mhiggsEP5}.
\footnote{Furthermore, a
two-loop effective potential calculation has been carried out in
\citere{fullEP2l}, but no public code based on this result is available.
}%

The current intrinsic error of $\Mh$ due to unknown higher-order
corrections and its prospective improvement in the future have
been estimated to be~\cite{mhiggsAEC,mhiggsFDalbals}
\BE
\De\Mh^{\rm intr,current} = 3 \gev , \quad 
\De\Mh^{\rm intr,future} = 0.5 \gev . 
\EE
The estimated future uncertainty assumes that a full two-loop result,
leading three-loop and possibly even higher-order corrections become available.

Concerning the parametric error on $\Mh$, the top-quark mass has the
largest impact, entering $\propto \mt^4$ at the one-loop level. As a
rule of thumb, an uncertainty of $\de\mt = 1\gev$ translates to an
induced parametric uncertainty in $\Mh$ of 
$\De\Mh^{\mt} \approx 1 \gev$~\cite{tbexcl}. We find for the parametric
uncertainties induced by the present 
experimental errors of $\mt$ and $\als$
\BEA
\de\mt^{\rm current} = 4.3 \gev &\Rightarrow&
\De\Mh^{{\rm para},\mt, {\rm current}} \approx 4 \gev ,  \\[.3em]
\de\als^{\rm current} = 0.002 &\Rightarrow&
\De\Mh^{{\rm para},\als, {\rm current}} \approx 0.3 \gev .
\EEA
These will improve in the future to
\BEA
\de\mt^{\rm future} = 0.1 \gev &\Rightarrow&
\De\Mh^{{\rm para},\mt, {\rm future}} \approx 0.1 \gev ,  \\[.3em]
\de\als^{\rm future} = 0.001 &\Rightarrow&
\De\Mh^{{\rm para},\als, {\rm future}} \approx 0.1 \gev .
\EEA
Thus, the intrinsic error would be the dominant source of uncertainty in
the future. On the other hand, a further reduction of the unknown
higher-order corrections to $\Mh$ is in principle possible.

\begin{figure}[hbt!]
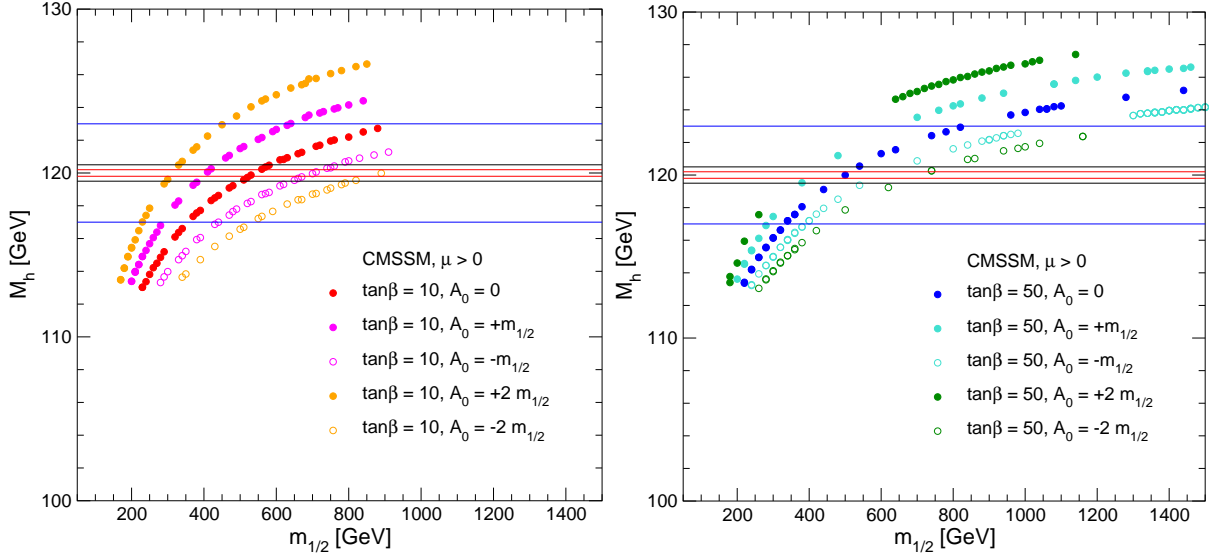

\begin{center}
\includegraphics[width=.48\textwidth]{ehow.Mh11a.cl.eps}
\includegraphics[width=.48\textwidth]{ehow.Mh11b.cl.eps}
\caption{%
The CMSSM predictions for $\Mh$ as functions of $m_{1/2}$ with (a) $\tb = 
10$ and (b) $\tb = 50$ for various $A_0$. 
A hypothetical experimental value is shown, namely $\Mh = 120 \gev$.
We display an optimistic anticipated theory uncertainty of $\pm 0.2 \gev$,
as well as a more realistic theory uncertainty of $\pm 0.5 \gev$ and 
the current theory uncertainty of $\pm 3 \gev$.
}
\label{fig:Mh}
\end{center}
\end{figure}

The experimental accuracy on $\Mh$ at the ILC~\cite{lctdrs}
will be even higher than
the prospective precision of the theory prediction,
\BE
\de\Mh^{\rm exp, future} = 0.05 \gev .
\EE
We show in \reffi{fig:Mh} we show the for $\Mh$, assuming a hypothetical
measurement at $\Mh = 120 \gev$. Since the experimental error
at the ILC will be smaller than the prospective theory uncertainties, we
display the effect of the current and future intrinsic uncertainties. In
addition, a more optimistic value of 200~MeV is also shown. The figure
clearly illustrates
the high sensitivity of this electroweak
precision observable to variations of the supersymmetric parameters 
(detailed results for Higgs boson phenomenology in
the CMSSM can be found in \citere{ehowasbs}).
The comparison between the measured value of $\Mh$ and a precise theory
prediction will allow one to set tight constraints on the allowed 
parameter space of $m_{1/2}$ and $A_0$.


\subsection{The Higgs boson branching ratios}

Within the CMSSM, various Higgs boson decay channels will be
accessible at the LHC and the
ILC. At the LHC, Higgs boson couplings~\cite{HcoupSM}
or ratios of them~\cite{lhctdrs,duehrssen} can in general be determined 
at the level of $\sim 10\%$ at best, depending on the Higgs-boson mass and
theoretical assumptions. Therefore we concentrate on ILC measurements
and accuracies.

It has been shown in \citere{deschi} that the observable combination
\begin{equation}
r \equiv \frac{\left[{\rm BR}(h \to b \bar b)/
                     {\rm BR}(h \to WW^*)\right]_{\rm MSSM}}
              {\left[{\rm BR}(h \to b \bar b)/
                     {\rm BR}(h \to WW^*)\right]_{\rm SM~~~}\,}
\label{eq:r}
\end{equation}
of Higgs boson decay rates is particularly sensitive
to deviations of the MSSM Higgs sector from the SM.
Even though the experimental error on the ratio of the two branching 
ratios is larger than that on the individual ones, the quantity $r$
has a stronger sensitivity to $M_A$ than any single branching ratio.

For the evaluation of $\br(\hbb)$, we use the results of~\citere{hff},
including the result of resumming the contributions of
\order{(\als\tb)^n}~\cite{deltamb,deltamb1}. The evaluation of
$\br(\hWW)$ is based on an effective-coupling approach, taking into
account off-shell effects. The corrections used for the effective-coupling 
calculation are the same as for the Higgs-boson mass
calculation, including the full one-loop and leading and subleading
two-loop contributions~\cite{feynhiggs,mhiggsAEC}. The evaluation has
been performed with {\tt FeynHiggs}~\cite{feynhiggs,feynhiggs2.2}. 

For the prospective accuracy at the ILC, we consider two cases. 
At the ILC with $\sqrt{s} = 500 \gev$ an accuracy of 4\% seems to be
feasible~\cite{lctdrs}, whilst at $\sqrt{s} = 1 \tev$ this accuracy could
be improved to~\cite{barklow}
\BE
\left(\frac{\de r}{r}\right)^{\rm exp,future} = 1.5\% .
\label{eq:rexp}
\EE
Since in this ratio of branching ratios many theoretical uncertainties
cancel, we assume that the future theoretical error can be neglected.
In the analysis in \refse{sec:combfuture} we use the accuracy of
\refeq{eq:rexp}.

\begin{figure}[htb!]
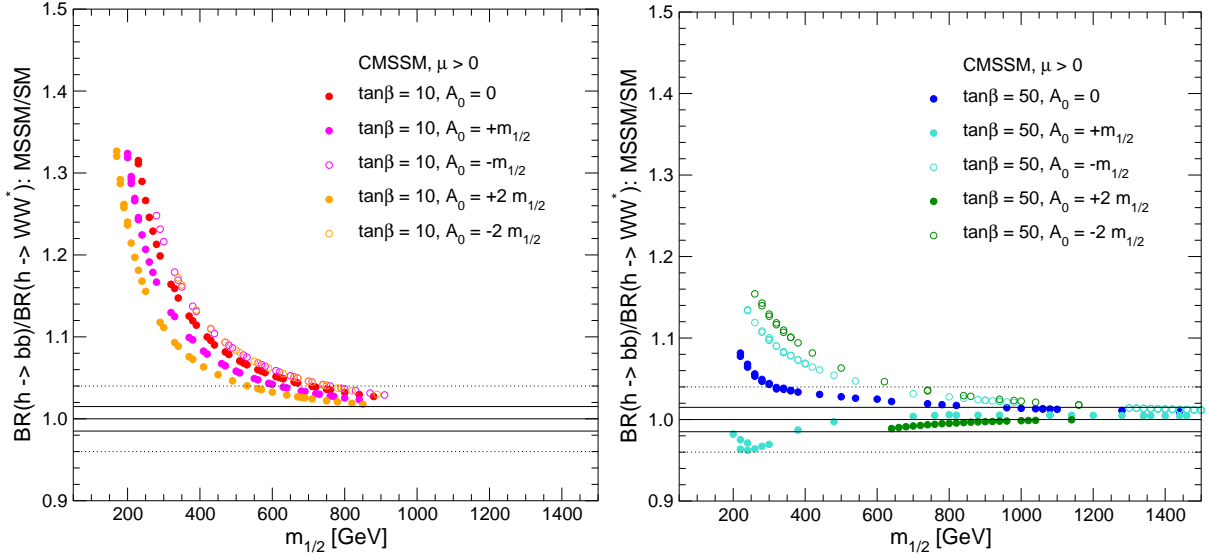

\begin{center}
\includegraphics[width=.48\textwidth]{ehow.BR11a.cl.eps}
\includegraphics[width=.48\textwidth]{ehow.BR11b.cl.eps}
\caption{%
The CMSSM predictions for 
$[\br(\hbb)/\br(\hWW)]_{\rm MSSM}/[\br(\hbb)/\br(\hWW)]_{\rm SM}$ 
as functions of $m_{1/2}$ for (a) $\tb = 10$ and (b) $\tb = 50$ with 
various values of $A_0$.
The central (solid) line corresponds to the SM expectation. The outer
(dotted) and inner (solid) lines indicate an ILC measurement with 4\%
and 1.5\% accuracy, respectively.
}
\label{fig:BR}
\end{center}
\end{figure}

In \reffi{fig:BR} the results for $r$ are shown as functions of
$m_{1/2}$ for $\tb = 10,50$. In the figure we indicate accuracies
of both 4\% and 1.5\%. For low $\tb$, the high ILC accuracy in $r$ will 
allow one to detect a deviation from the SM prediction for all CMSSM 
points.
For large $\tb$, the effects of the supersymmetric contributions to $r$ 
are in
general smaller. Deviations up to $m_{1/2} \approx 1$~TeV could be
visible, depending somewhat on $A_0$.


\section{Combined Sensitivity: Present Situation}
\label{sec:combcurr}

\subsection{Best fits for WMAP strips at fixed $A_0$}

We now investigate the combined sensitivity of the four low-energy
observables for which experimental measurements exist at present, namely
$\MW$, $\sweff$, $(g-2)_\mu$ and $\br(b \to s \ga)$. Since only an upper 
bound exists for $\br(B_s \to \mu^+ \mu^-)$, we discuss it separately 
below. We begin with an analysis of the sensitivity to $m_{1/2}$ moving 
along the WMAP
strips with fixed values of $A_0$ and $\tb$. The experimental central
values, the present experimental errors and theoretical uncertainties
are as described in \refse{sec:ewpo}. 
The experimental uncertainties, the intrinsic errors from unknown
higher-order corrections and the parametric uncertainties have been
added quadratically, except for $\br(b \to s \ga)$, where they have
been added linearly. Assuming that the four observables are
uncorrelated, a $\chi^2$ fit has been performed with
\BE
\chi^2 \equiv \sum_{n=1}^{N} \KL \frac{R_n^{\rm exp} - R_n^{\rm theo}}
                                 {\si_n} \KR^2~.
\EE
Here $R_n^{\rm exp}$ denotes the experimental central value of the
$n$th observable, so that $N = 4$ for the set of observables included in
this fit,
$R_n^{\rm theo}$ is the corresponding CMSSM prediction and $\si_n$
denotes the combined error, as specified above.
We have rejected all points of the CMSSM parameter space with either
$\Mh < 113 \gev$~\cite{LEPHiggsSM,LEPHiggsMSSM} or a chargino mass
lighter than $103 \gev$~\cite{pdg}.

\begin{figure}[thb!]
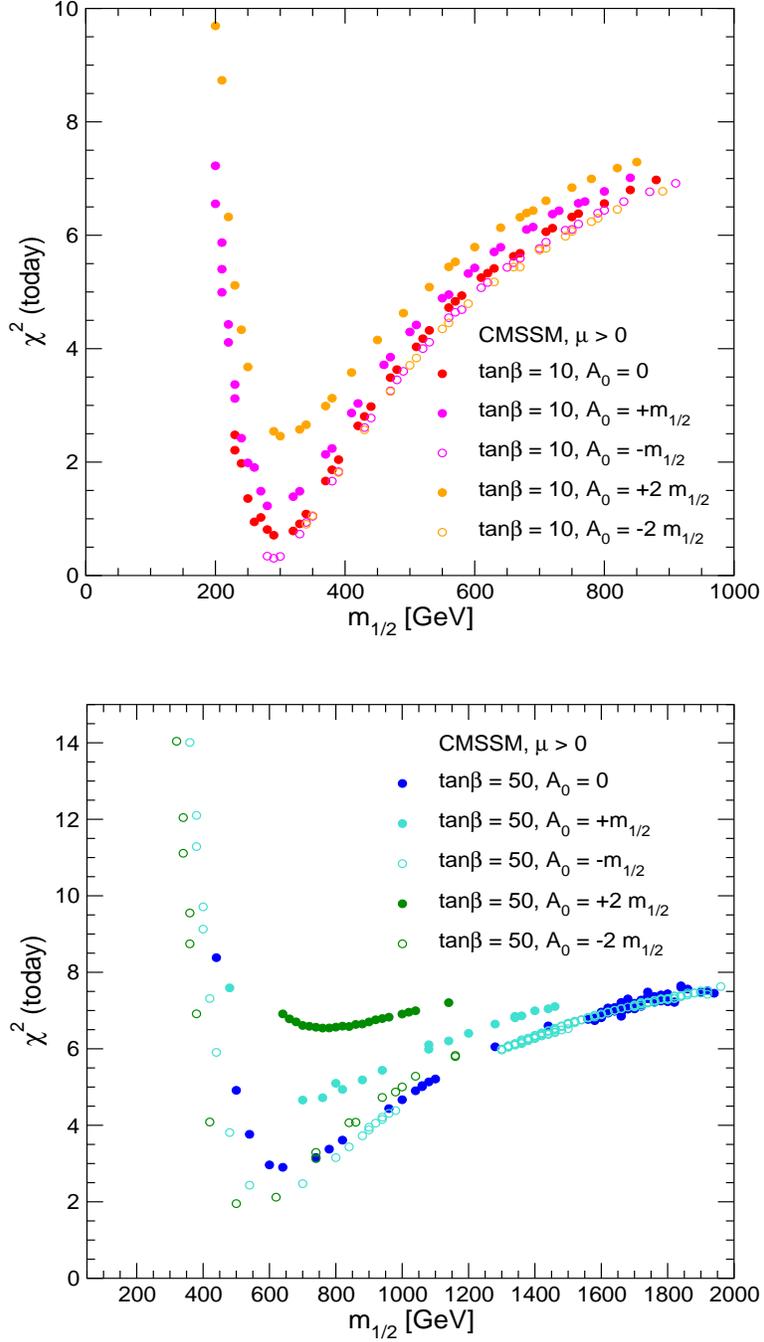

\begin{center}
\epsfig{figure=ehow.CHI11a.cl.eps,width=10cm,height=8.5cm}\\[2em]
\epsfig{figure=ehow.CHI11b.cl.eps,width=10cm,height=8.5cm}
\caption{%
The results of $\chi^2$ fits based on the current experimental
results for the precision observables $\MW$, $\sweff$, $(g-2)_\mu$ and
$\br(b \to s \ga)$ are shown as functions of $m_{1/2}$ in the CMSSM
parameter space with CDM constraints for different values of
$A_0$.
The upper plot shows the results for $\tb = 10$, and the lower plot
shows the case $\tb = 50$.
}
\label{fig:CHI}
\end{center}
\end{figure}

The results are shown in \reffi{fig:CHI} for $\tb = 10$ and $\tb = 50$.  
They indicate that, already at the present level of experimental
accuracies, the electroweak precision observables combined with the WMAP
constraint provide a sensitive probe of the CMSSM, yielding interesting
information about its parameter space. For $\tb = 10$, the CMSSM provides a
very good description of the data, resulting in a remarkably small minimum
$\chi^2$ value. The fit shows a clear preference for relatively small
values of $m_{1/2}$, with a best-fit value of about $m_{1/2} = 300 \gev$.
The best fit is obtained for $A_0 \leq 0$, while positive values of $A_0$
result in a somewhat lower fit quality. 
The fit yields an upper bound on $m_{1/2}$ of about 600~GeV at the
90\%~C.L. (corresponding to $\Delta \chi^2 \le 4.61$).

These results can easily be understood from the analysis in
\refse{sec:ewpo}. For $\tb = 10$, the CMSSM prediction with $m_{1/2}
\approx 300 \gev$ is very close to the experimental central values of $\MW$,
$\sweff$ and $(g-2)_{\mu}$ for all values of $A_0$, see
\reffis{fig:MW}--\ref{fig:AMU}. Also, $\br(b \to s \gamma)$ is well
described for $m_{1/2} \approx 300 \gev$ and $A_0 \leq 0$, while large
positive values of $A_0$ lead to a CMSSM prediction for $\br(b \to s
\gamma)$ which is significantly below the experimental value. Consequently,
in the case of $\tb = 10$, a very good fit quality is obtained for $m_{1/2}
\approx 300 \gev$ and $A_0 \leq 0$.%
\footnote{
A preference for relatively small values of $m_{1/2}$ within the CMSSM
has also been noticed in \citere{deboer1}, where only $(g-2)_{\mu}$ and
$\br(b \to s \gamma)$ had been analyzed.}
Some of the principal contributions to the increase in $\chi^2$ when 
$m_{1/2}$ increases for $\tb = 10$ are as follows.
For $A_0 = - m_{1/2}, m_{1/2} = 900$~GeV, we find that $(g - 2)_\mu$ 
contributes about 5 to $\Delta \chi^2$, $\MW$ nearly 1 and $\sweff$
about 0.2, whereas the contribution of $\br(b \to s \gamma)$ is 
negligible. On the other hand, for $A_0 = +2 m_{1/2}$, which is 
disfavoured for $\tb = 10$, the minimum in $\chi^2$ is due to a 
combination of the four observables, but $(g - 2)_\mu$ again gives the 
largest contribution for large $m_{1/2}$.

For $\tb = 50$ the overall fit quality is worse than for $\tb = 10$, and
the sensitivity to $m_{1/2}$ from the precision observables is lower. This
is related to the fact that, whereas $\MW$ and $\sweff$ prefer small values
of $m_{1/2}$ also for $\tb = 50$, as seen in \reffis{fig:MW} and
\ref{fig:SW}, the CMSSM predictions for $(g-2)_{\mu}$ and $\br(b \to s
\gamma)$ for high $\tb$ are in better agreement with the data for larger
$m_{1/2}$ values, as seen in \reffis{fig:AMU} and \ref{fig:BSG}.  Also in
this case the best fit is obtained for negative values of $A_0$, but the
preferred values for $m_{1/2}$ are 200--300~GeV higher than for $\tb = 10$.

\begin{figure}[htb!]
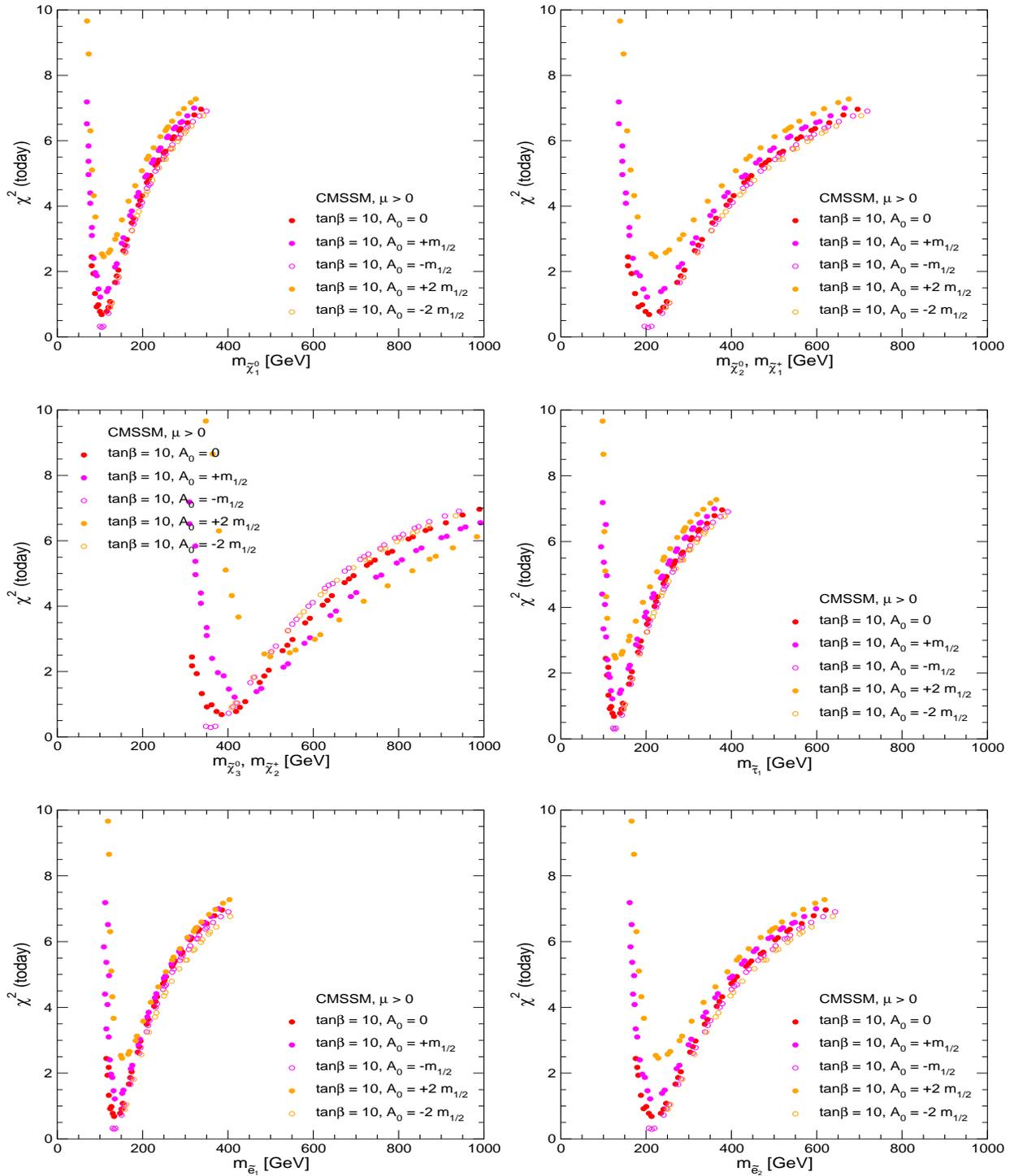

\mbox{}\vspace{-1cm}
\begin{center}
\epsfig{figure=ehow.mass11a.cl.eps,width=8.1cm,height=6.1cm}
\epsfig{figure=ehow.mass12a.cl.eps,width=8.1cm,height=6.1cm}\\[1em]
\epsfig{figure=ehow.mass14a.cl.eps,width=8.1cm,height=6.1cm}
\epsfig{figure=ehow.mass17a.cl.eps,width=8.1cm,height=6.1cm}\\[1em]
\epsfig{figure=ehow.mass15a.cl.eps,width=8.1cm,height=6.1cm}
\epsfig{figure=ehow.mass16a.cl.eps,width=8.1cm,height=6.1cm}
\caption{%
The $\chi^2$ contours in the CMSSM with $\tb = 10$ for different sparticle 
masses, based on the fits to the parameter space shown in \reffi{fig:CHI}.
The first row shows (left) the mass of the neutralino LSP,
$m_{\neone}$, and (right) the mass of the lighter chargino, 
$m_{\chone} \approx m_{\netwo}$.
The second row shows (left) the mass of the heavier chargino, 
$m_{\chtwo} \approx m_{\tilde{\chi}^0_3}$, and (right) the mass of the lighter
stau, $m_{\tilde{\tau}_1}$. The selectron masses are shown in the third
row.
}
\label{fig:massa}
\end{center}
\vspace{-0.8cm}
\end{figure}

\begin{figure}[htb!]
\mbox{}\vspace{-1cm}
\begin{center}
\epsfig{figure=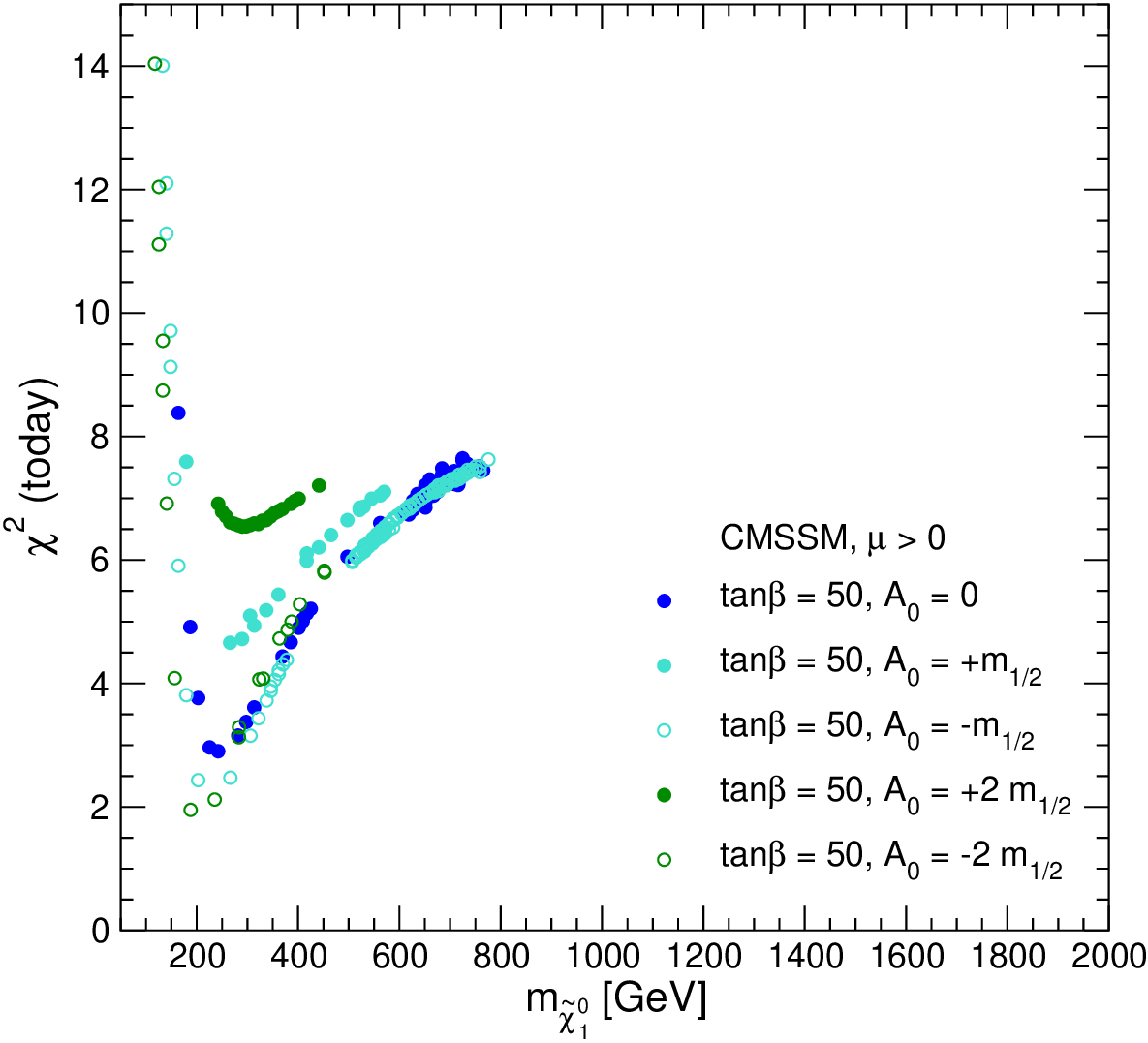,width=8.1cm,height=6.1cm}
\epsfig{figure=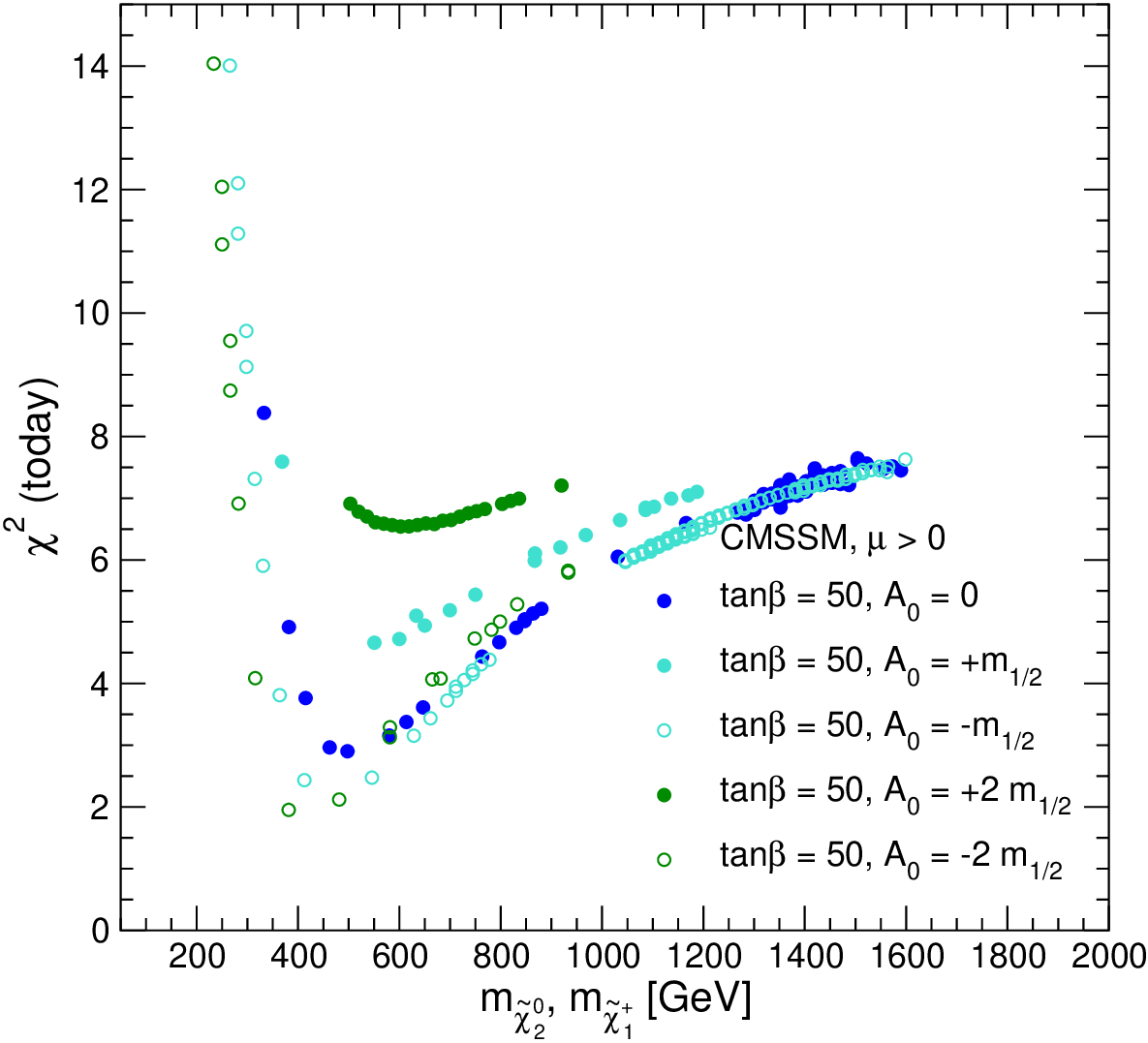,width=8.1cm,height=6.1cm}\\[1em]
\epsfig{figure=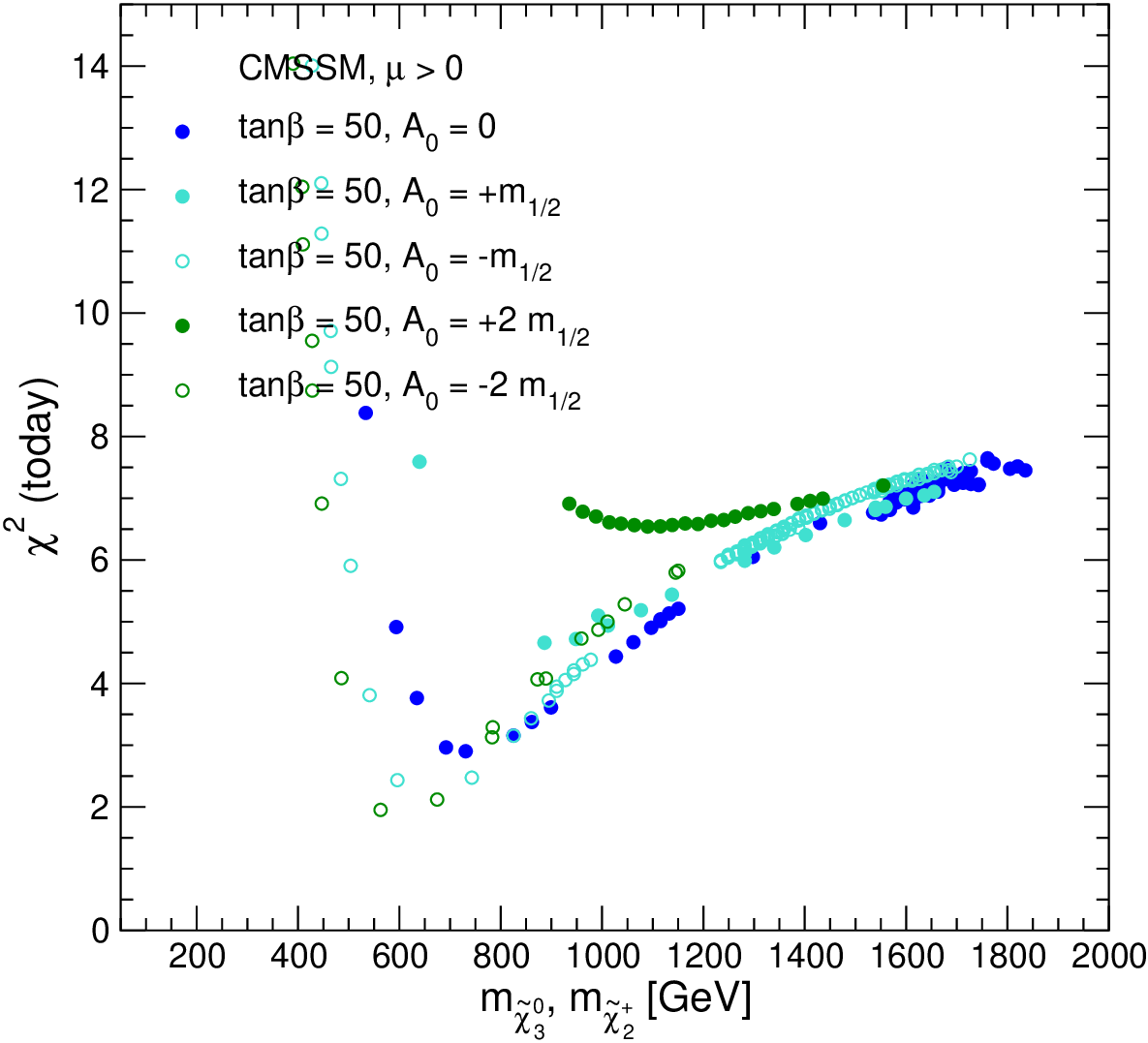,width=8.1cm,height=6.1cm}
\epsfig{figure=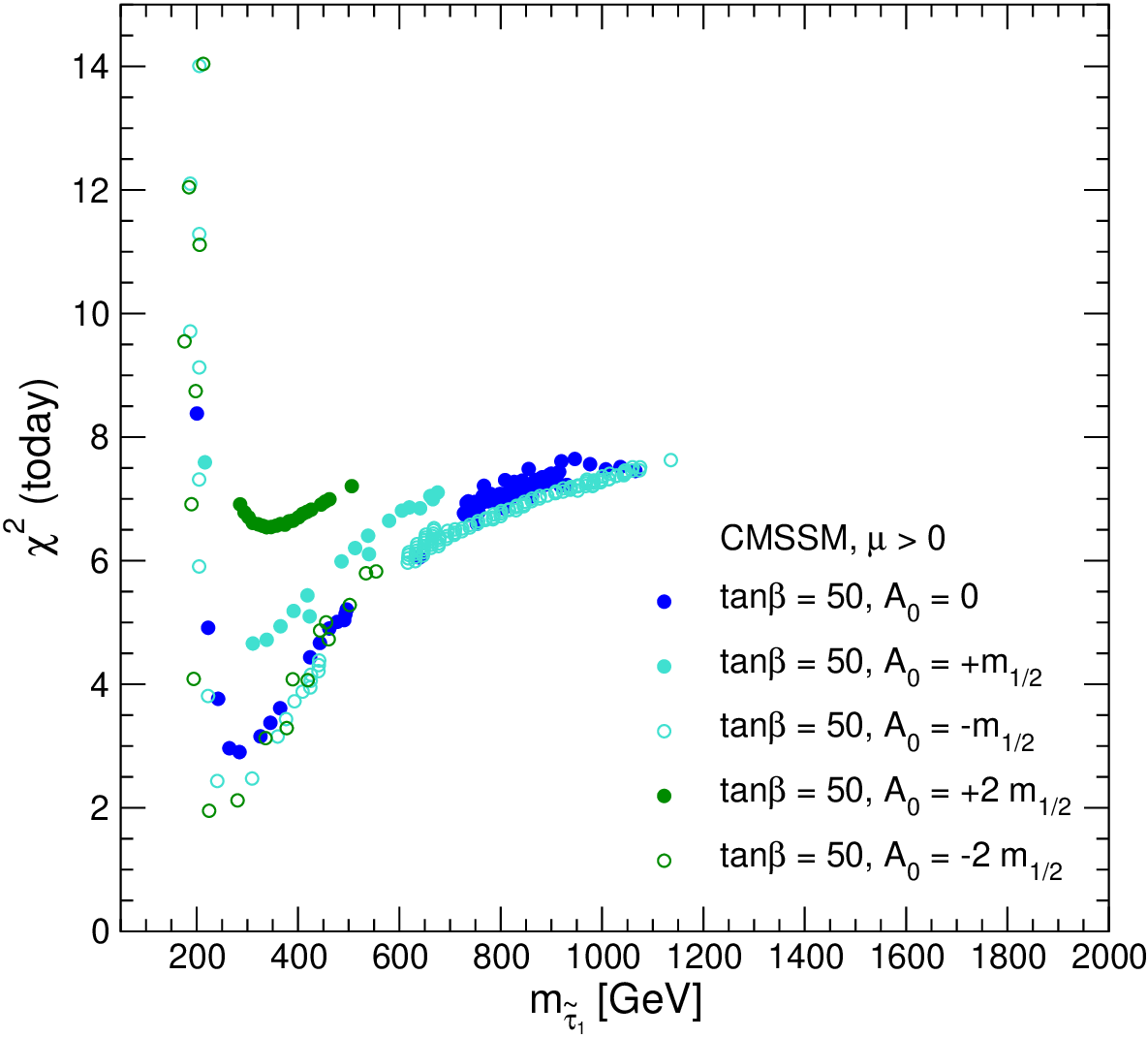,width=8.1cm,height=6.1cm}\\[1em]
\epsfig{figure=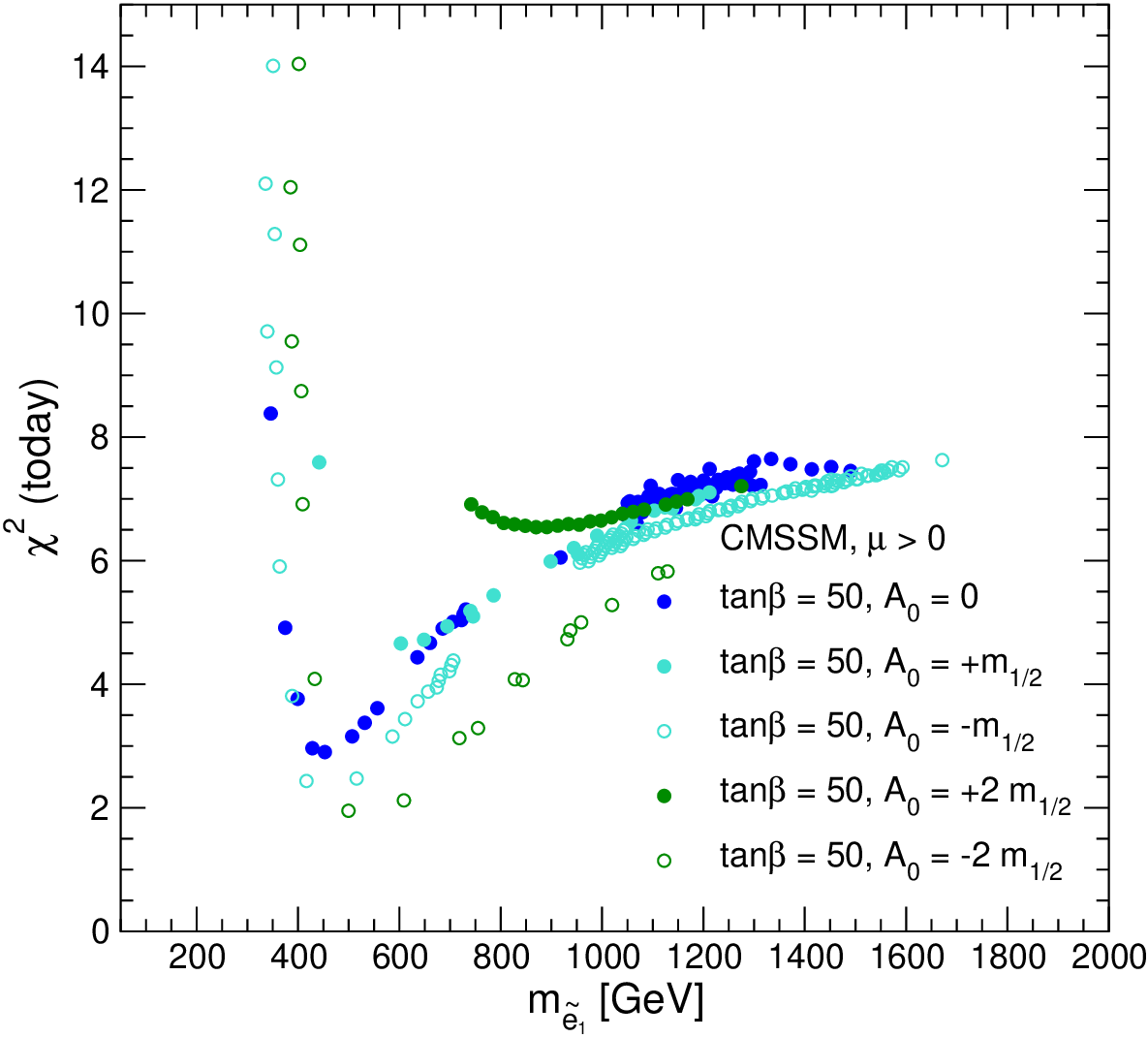,width=8.1cm,height=6.1cm}
\epsfig{figure=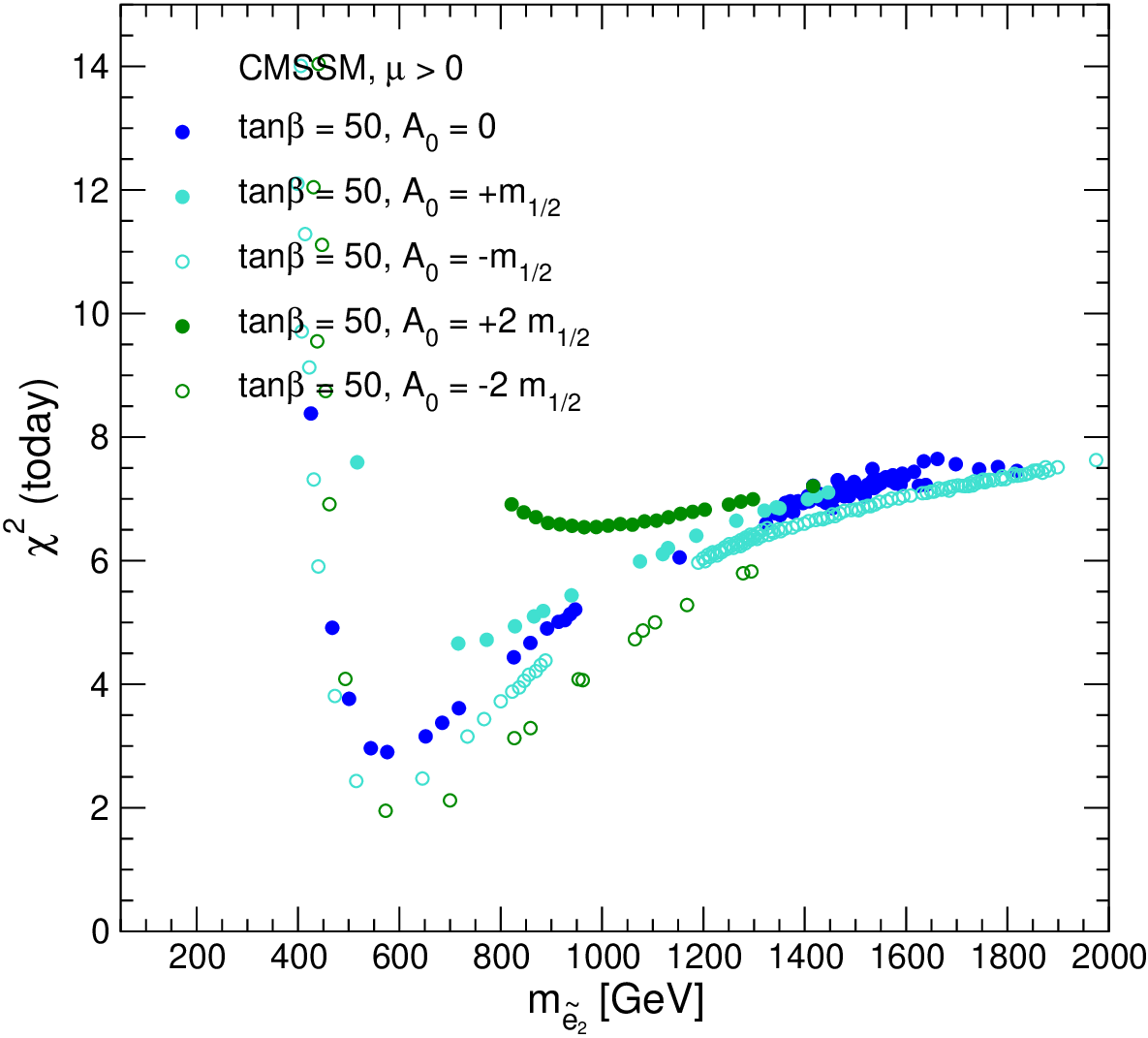,width=8.1cm,height=6.1cm}
\caption{%
The $\chi^2$ contours in the CMSSM with $\tb = 50$ for different sparticle
masses, based on the fits to the parameter space shown in \reffi{fig:CHI}.
The first row shows (left) the mass of the lightest neutralino,
$m_{\neone}$, and (right) the mass of the lighter chargino, 
$m_{\chone} \approx m_{\netwo}$.
The second row shows (left) the mass of the heavier chargino, 
$m_{\chtwo} \approx m_{\tilde{\chi}^0_3}$, and (right) the mass of the lighter
stau, $m_{\tilde{\tau}_1}$. The selectron masses are shown in the third
row.
}
\label{fig:massb}
\end{center}
\vspace{-0.8cm}
\end{figure}

\begin{figure}[htb!]
\mbox{}\vspace{-1cm}
\begin{center}
\epsfig{figure=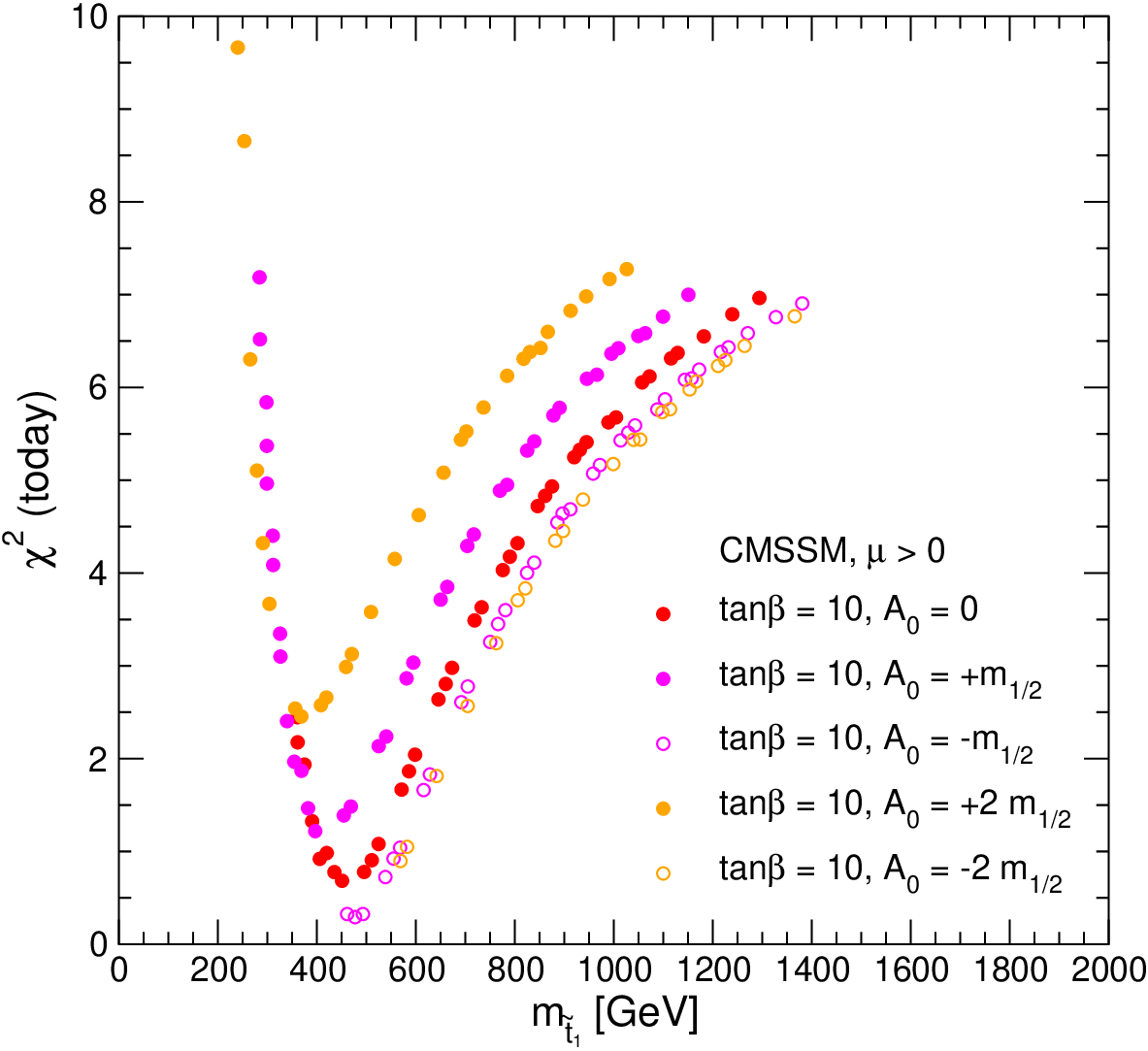,width=8.1cm,height=6.1cm}
\epsfig{figure=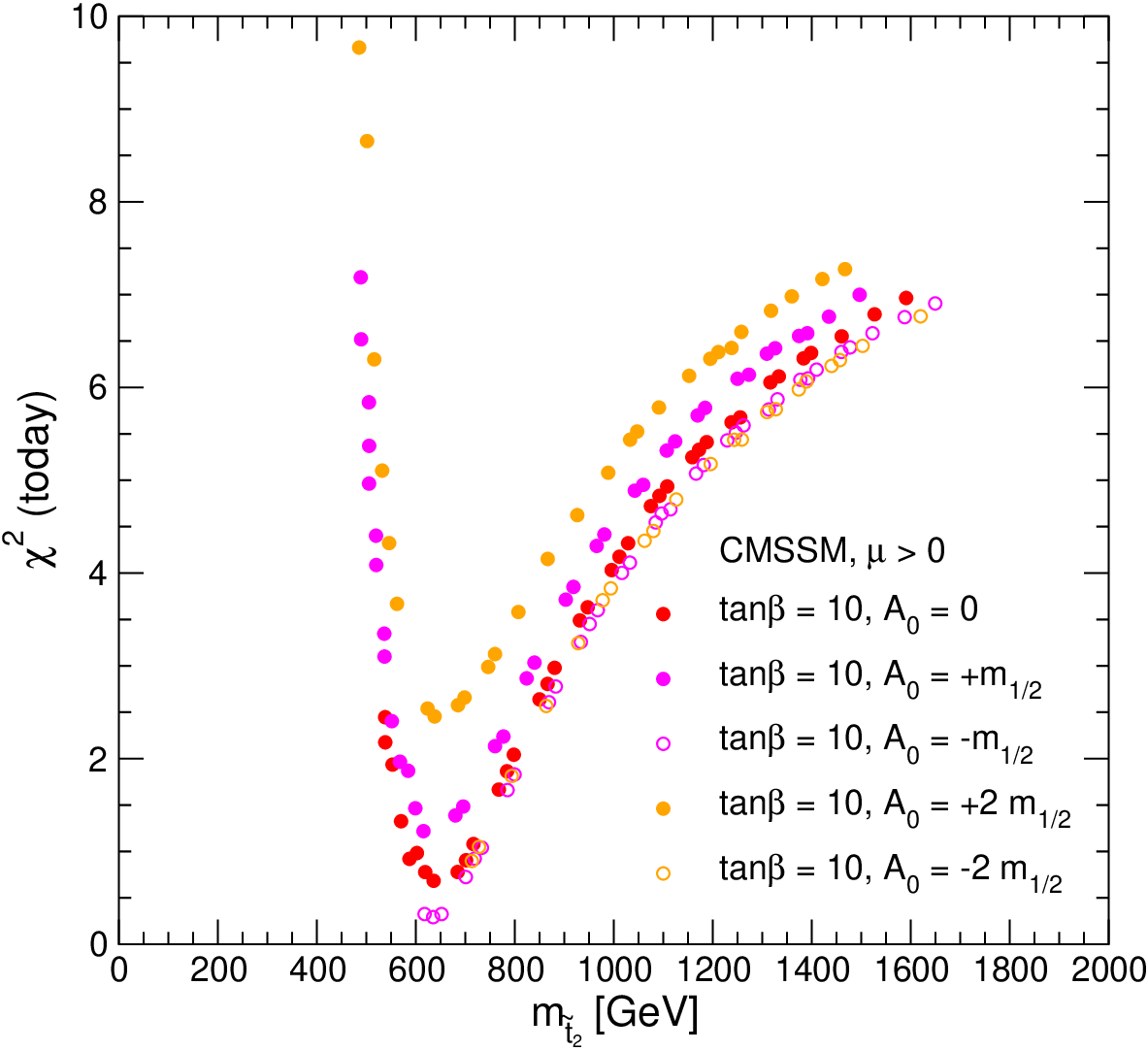,width=8.1cm,height=6.1cm}\\[1em]
\epsfig{figure=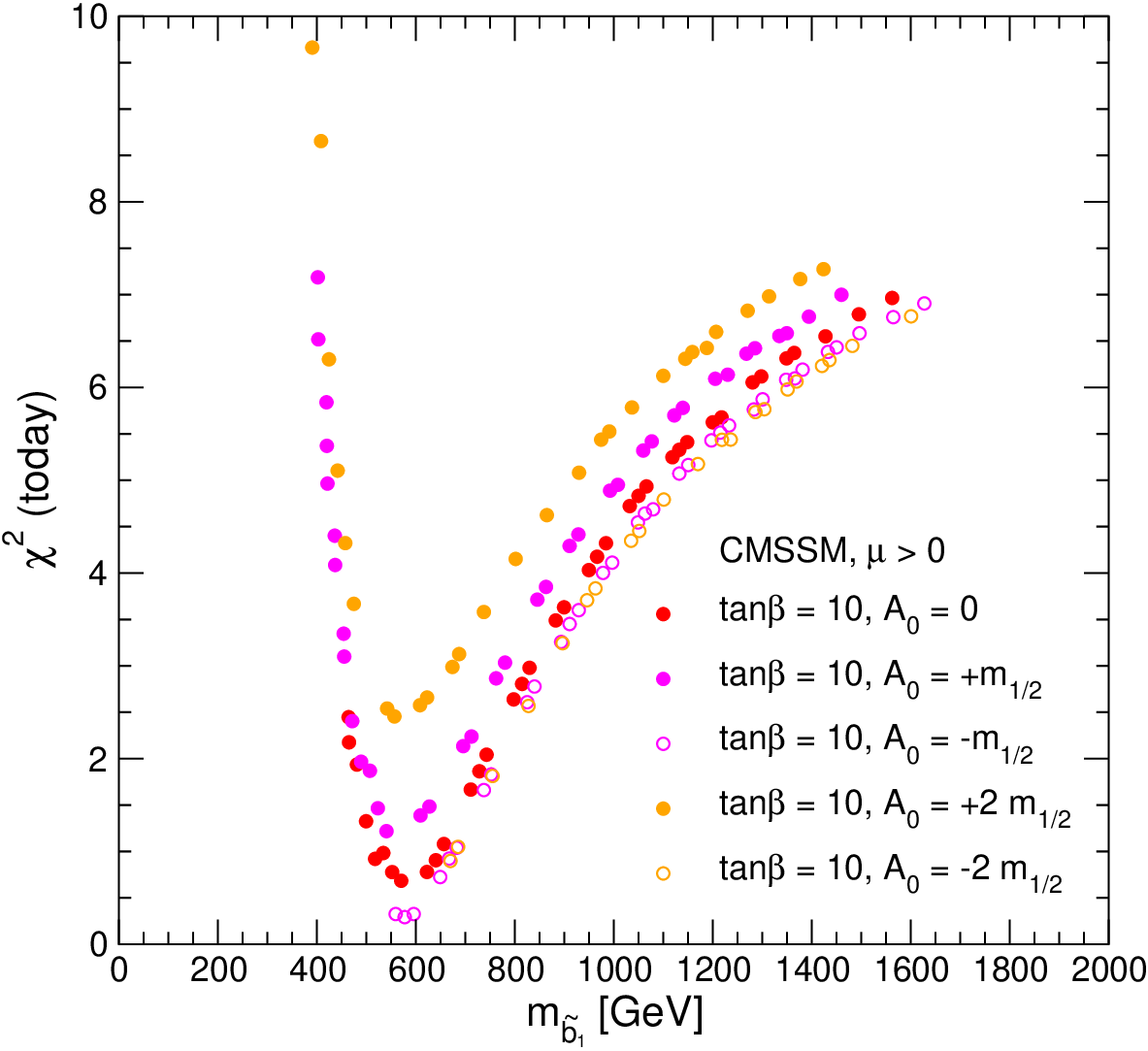,width=8.1cm,height=6.1cm}
\epsfig{figure=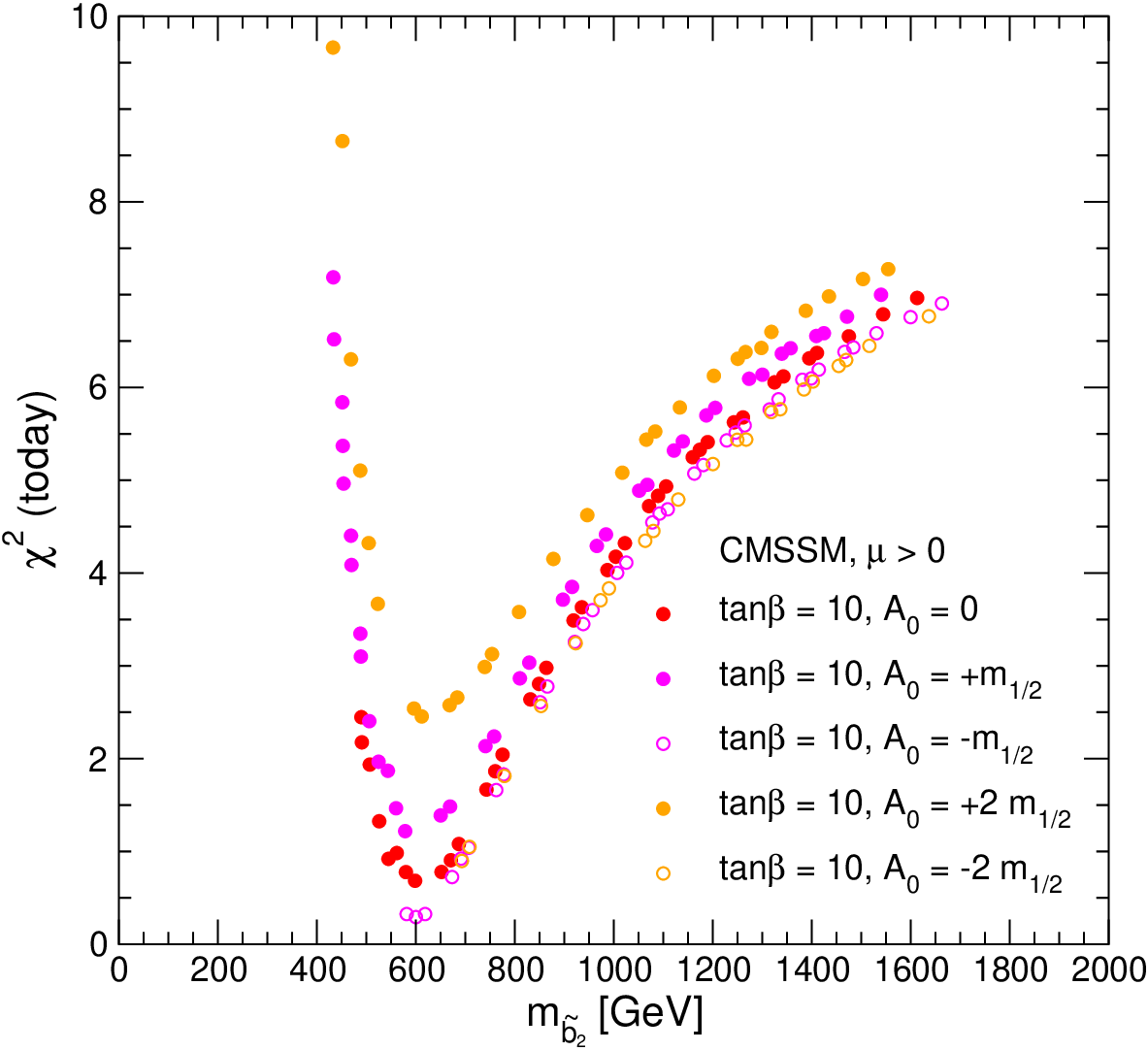,width=8.1cm,height=6.1cm}\\[1em]
\epsfig{figure=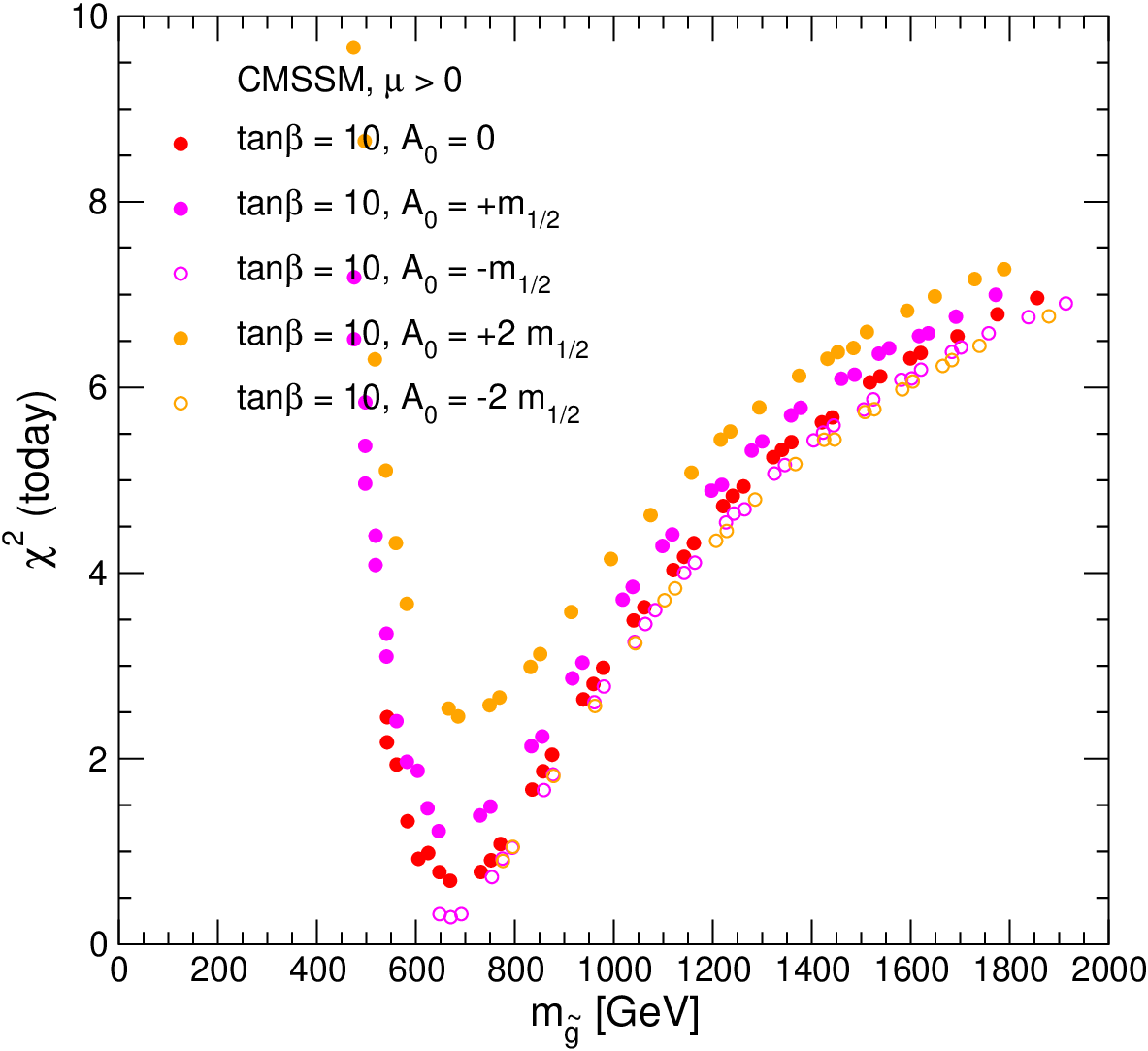,width=8.1cm,height=6.1cm}
\epsfig{figure=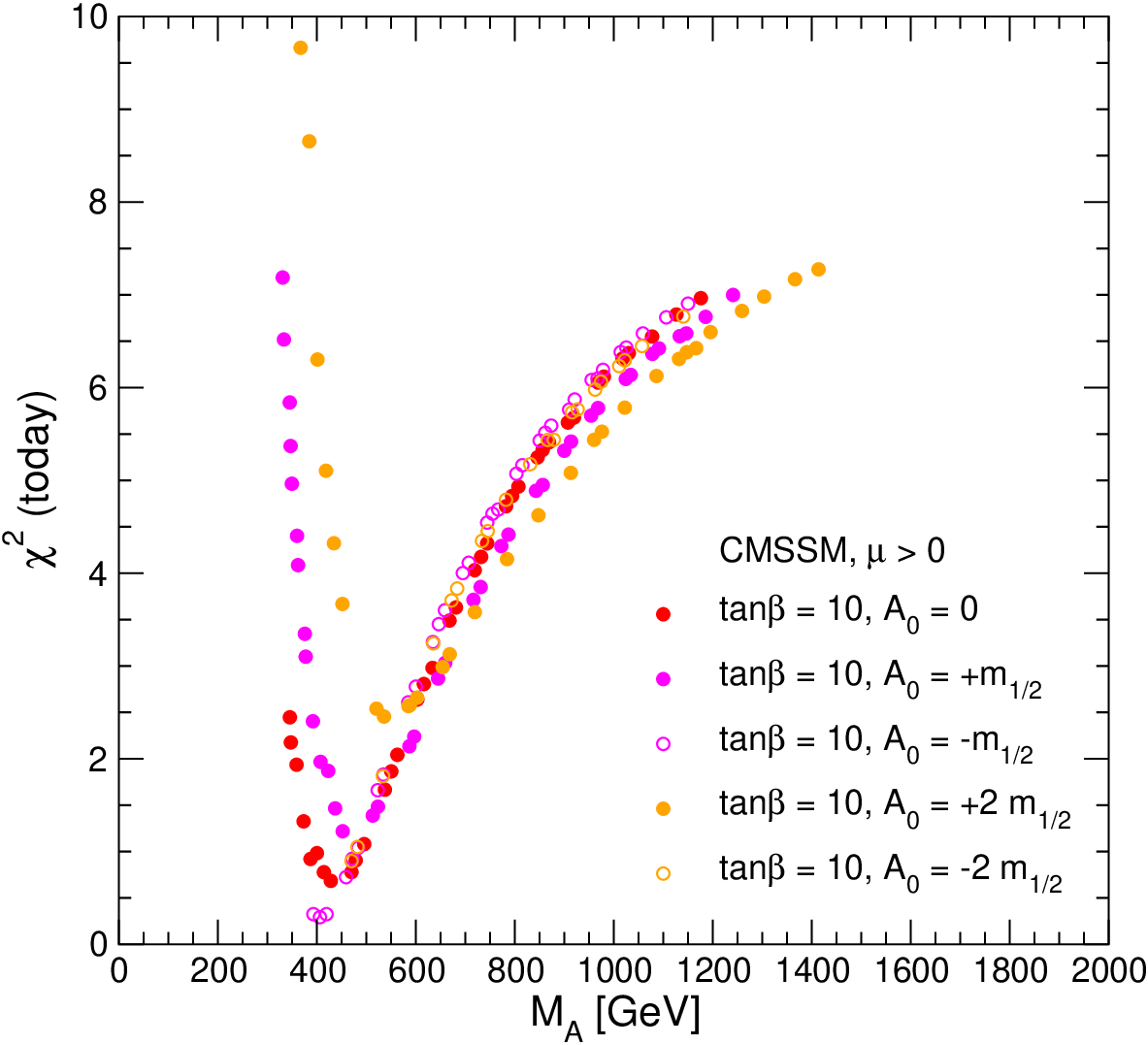,width=8.1cm,height=6.1cm}
\caption{%
The $\chi^2$ contours in the CMSSM with $\tb = 10$ for different sparticle
masses, based on the fits to the parameter space shown in \reffi{fig:CHI}.
The first row shows the scalar top masses, $\mste$, $\mstz$. 
The second row shows the scalar bottom masses, $\msbe$, $\msbz$.
The third row shows the gluino mass, $\mgl$, (left) and the mass of the
scalar Higgs boson, $\MA$ (right).
}
\label{fig:massa2}
\end{center}
\vspace{-0.8cm}
\end{figure}

\begin{figure}[htb!]
\mbox{}\vspace{-1cm}
\begin{center}
\epsfig{figure=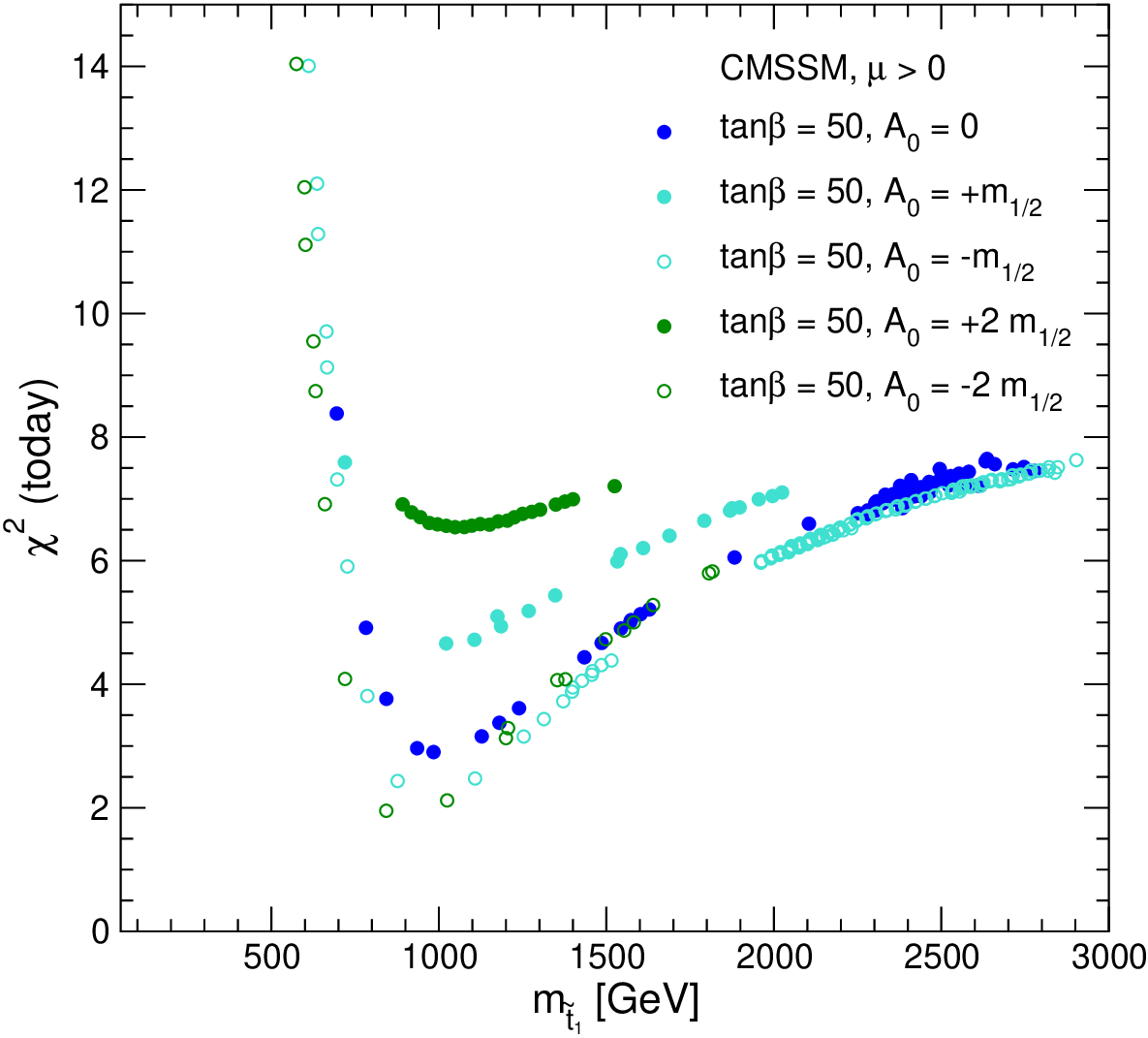,width=8.1cm,height=6.1cm}
\epsfig{figure=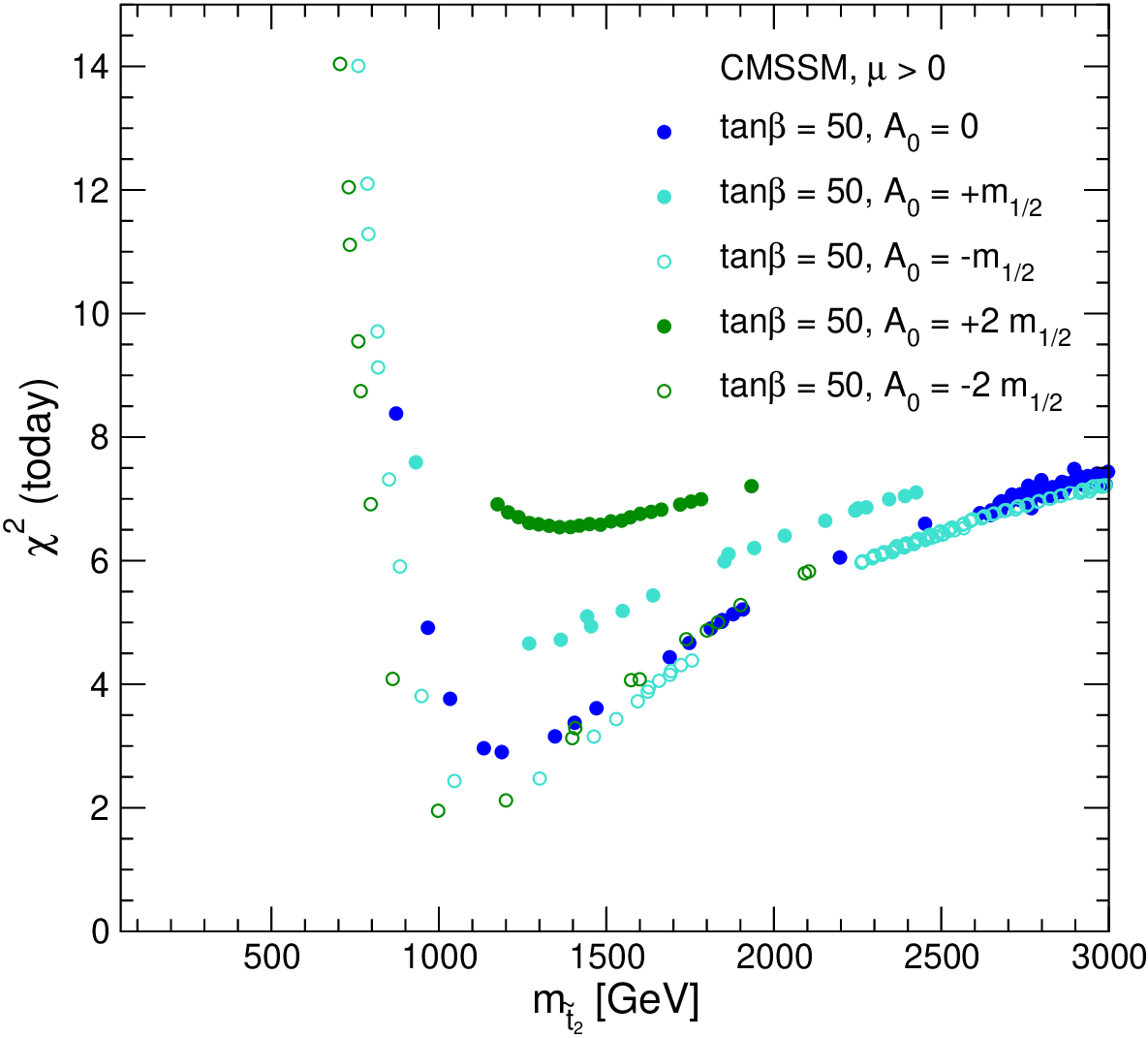,width=8.1cm,height=6.1cm}\\[1em]
\epsfig{figure=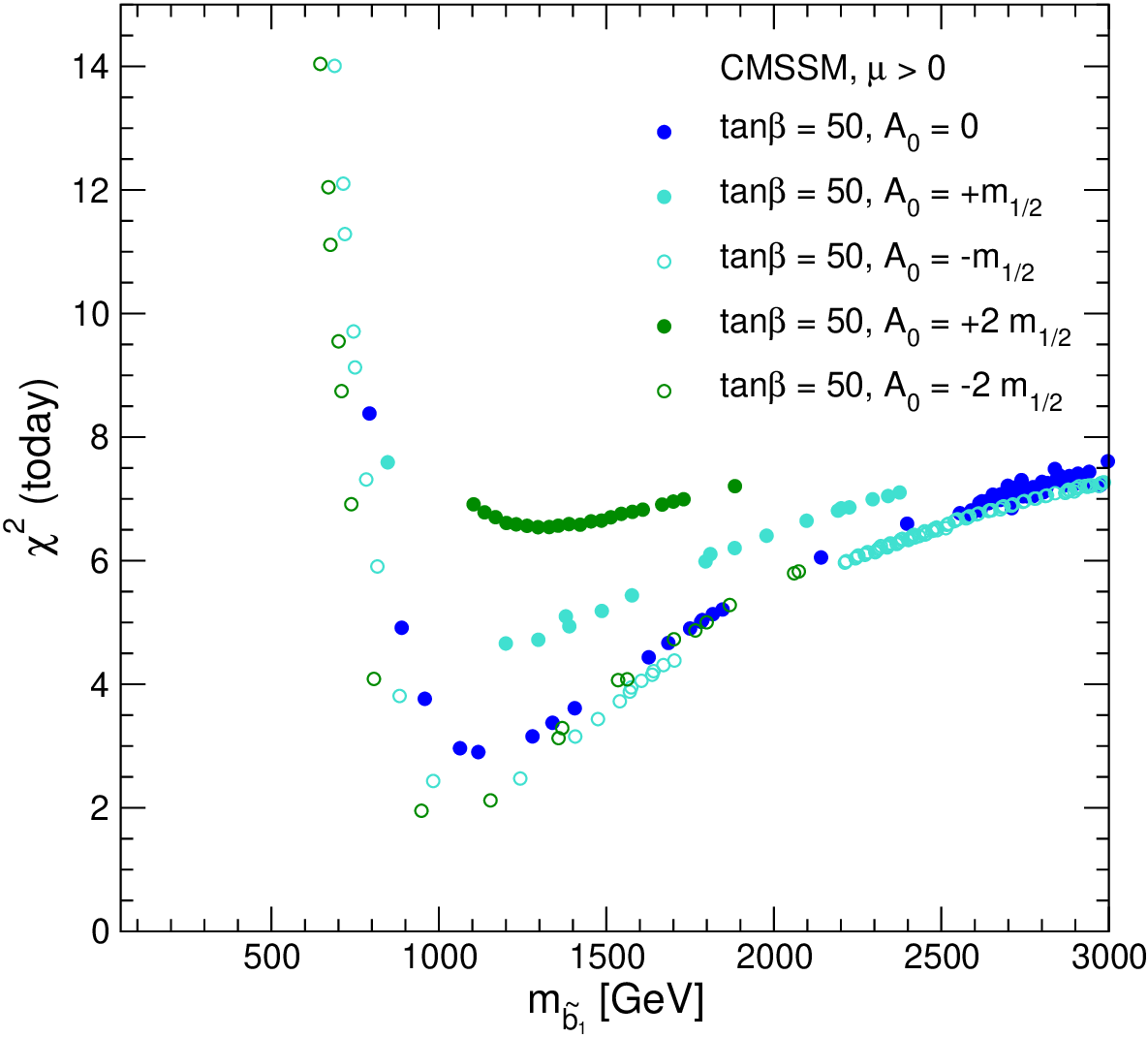,width=8.1cm,height=6.1cm}
\epsfig{figure=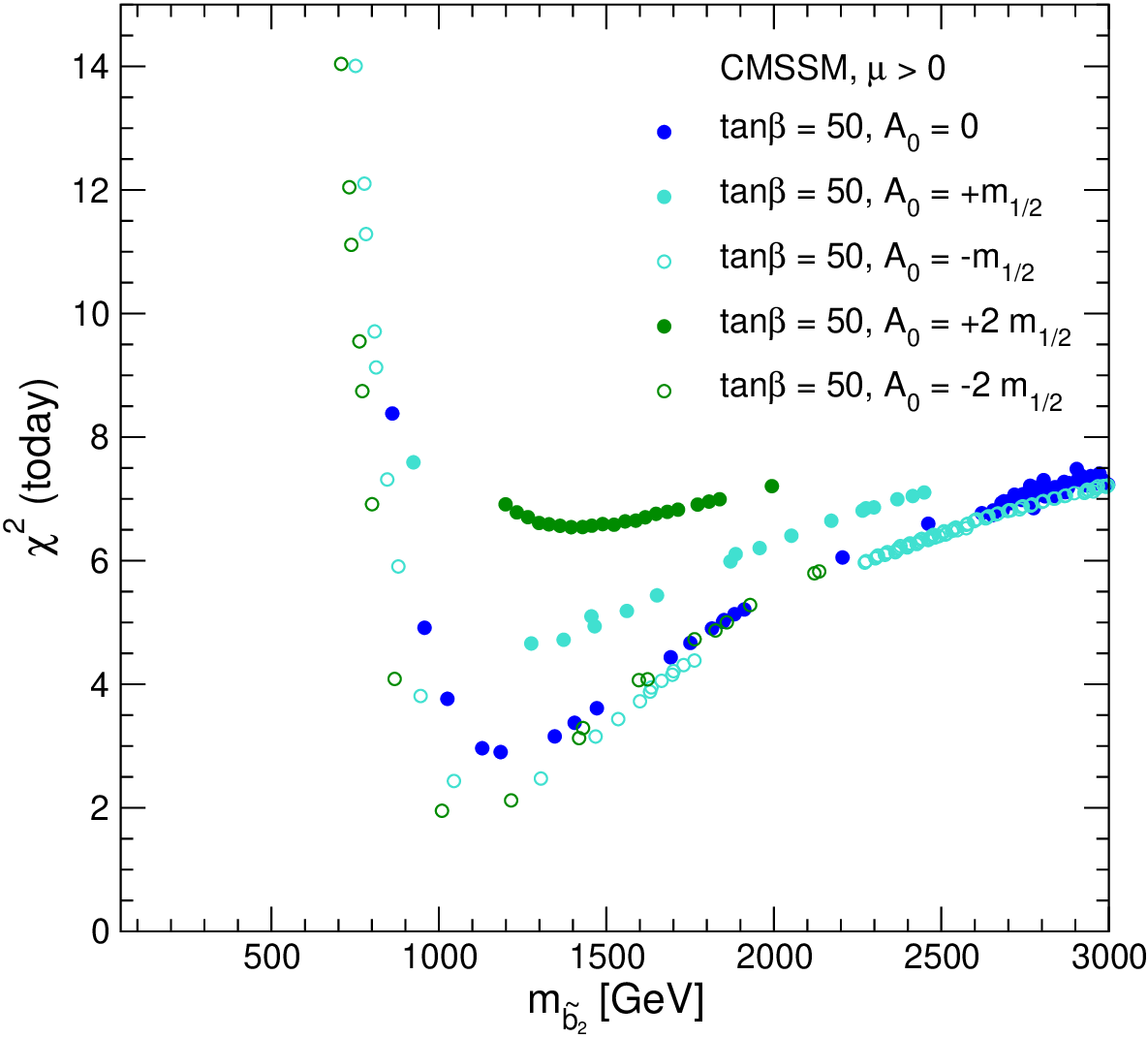,width=8.1cm,height=6.1cm}\\[1em]
\epsfig{figure=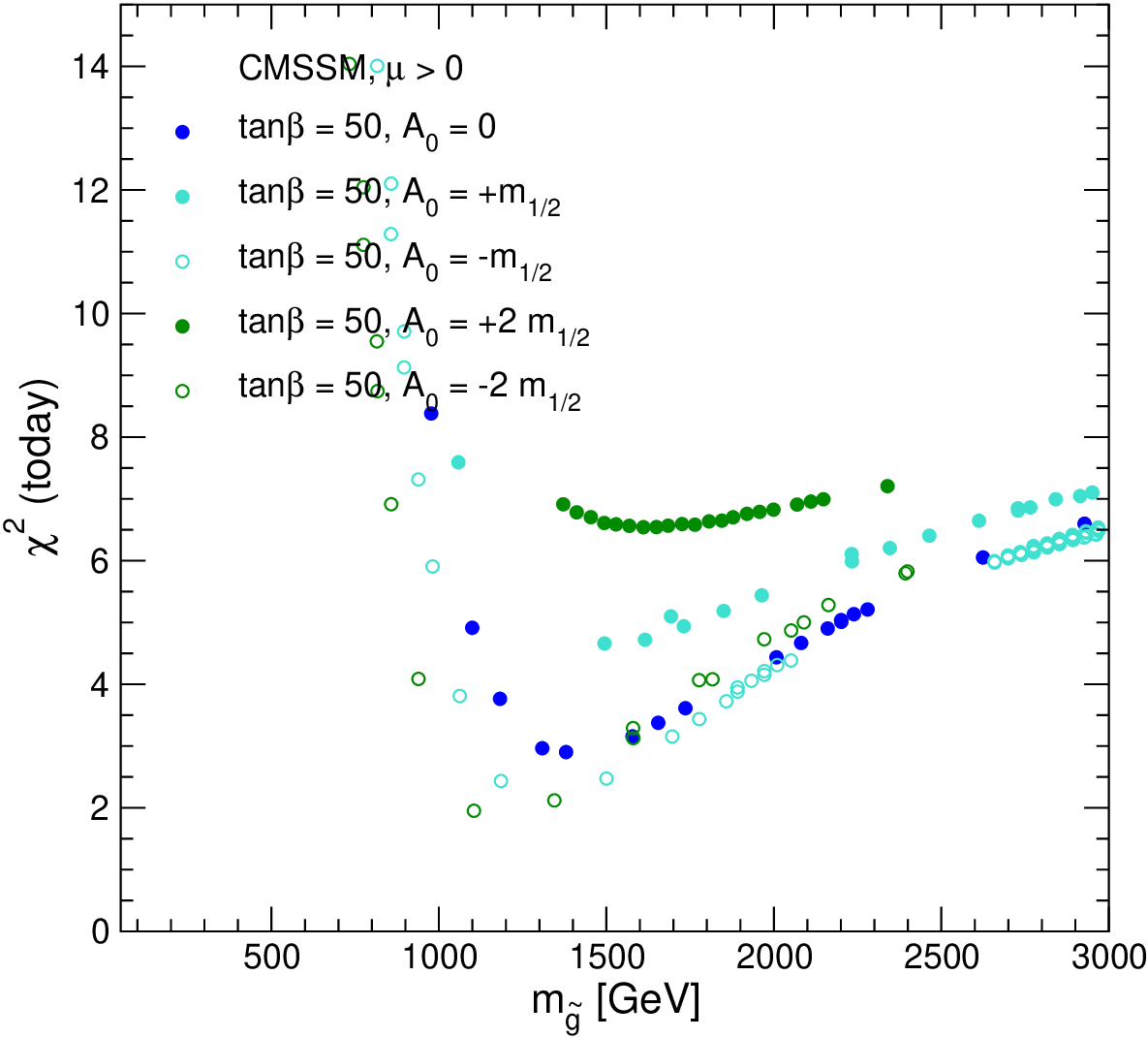,width=8.1cm,height=6.1cm}
\epsfig{figure=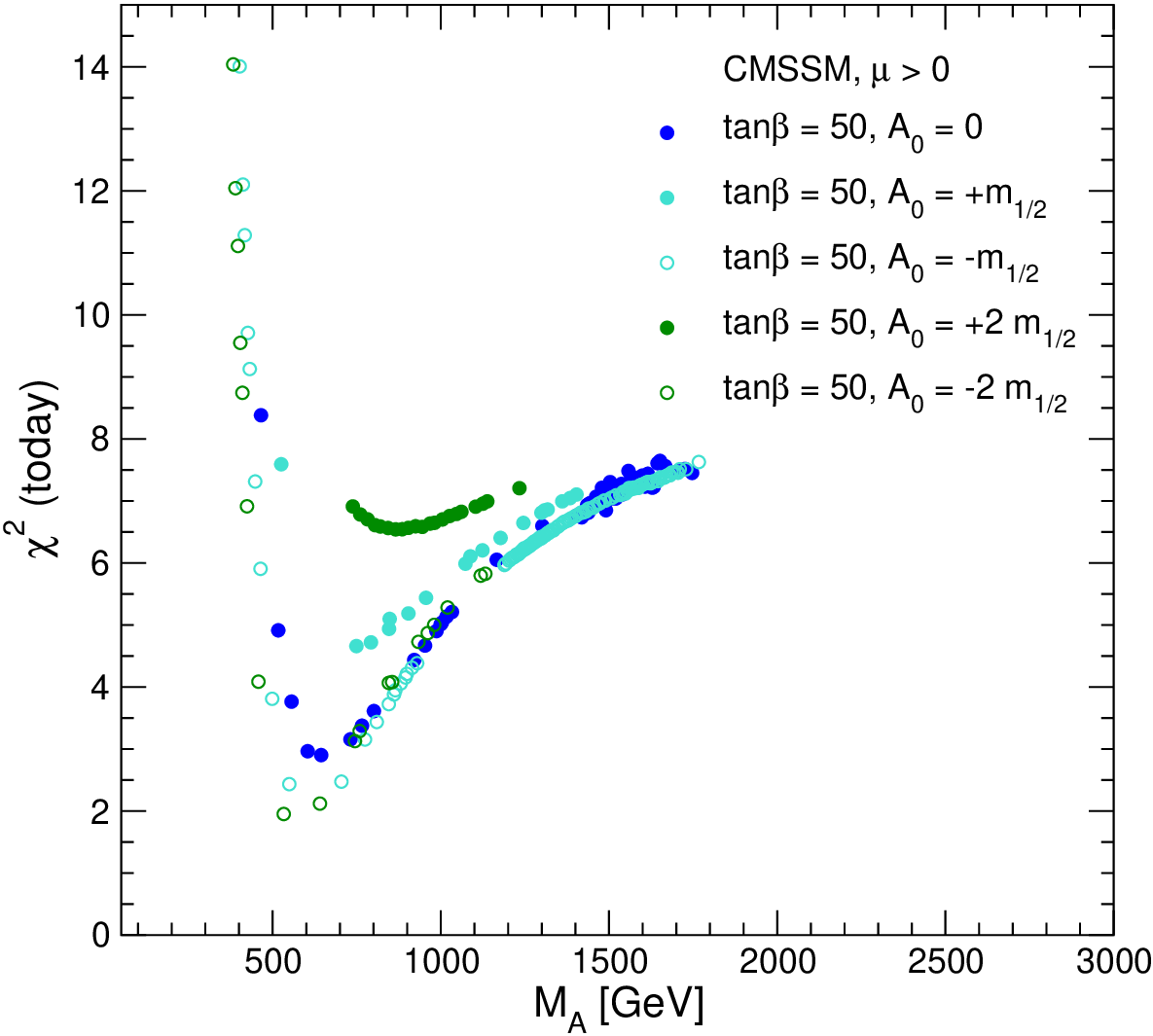,width=8.1cm,height=6.1cm}
\caption{%
The $\chi^2$ contours in the CMSSM with $\tb = 50$ for different sparticle
masses, based on the fits to the parameter space shown in \reffi{fig:CHI}.
The first row shows the scalar top masses, $\mste$, $\mstz$. 
The second row shows the scalar bottom masses, $\msbe$, $\msbz$.
The third row shows the gluino mass, $\mgl$, (left) and the mass of the
scalar Higgs boson, $\MA$ (right).
}
\label{fig:massb2}
\end{center}
\vspace{-0.8cm}
\end{figure}

In \reffis{fig:massa}--\ref{fig:massb2} the fit results of
\reffi{fig:CHI} are expressed in terms of the masses of different
supersymmetric particles. \reffi{fig:massa} shows that for $\tb = 10$ the
best fit is obtained if the lightest supersymmetric particle (LSP), which
within the CMSSM is the lightest neutralino, is lighter than about 200~GeV
(with a best-fit value $\sim 100$~GeV). The best-fit values for the
masses of the lighter chargino, the second-lightest neutralino 
(recall also that $m_{\chone}
\approx m_{\netwo}$), both sleptons and the lighter stau are all below
250~GeV, while the preferred region of the masses of the heavier chargino
and the heavier neutralinos is about 400~GeV. These masses offer good
prospects of direct sparticle detection at both the ILC and the LHC.
There are also some prospects for detecting the associated 
production of charginos and neutralinos at the Tevatron collider, via 
their trilepton decay signature, in particular. This is estimated to be 
sensitive to $m_{1/2} \lappeq 250$~GeV~\cite{BH}, covering much of the 
region below the best-fit value of $m_{1/2}$ that we find for $\tan \beta = 10$.

The same particle masses in the case $\tb = 50$ are shown in
\reffi{fig:massb}. Here the best-fit 
values for the LSP mass and the lighter stau are still below about 250~GeV.
The minimum $\chi^2$ for the other masses is shifted upwards compared to
the case with $\tb = 10$. The best-fit values are obtained in the region
400--600~GeV. Correspondingly, these sparticles would be harder to
detect. At the ILC with $\sqrt{s} \lsim 1 \tev$, the best prospects would 
be for the production of $\neu{1}\neu{2}$ or of $\Staue\AStaue$. Other
particles can only be produced if they turn out to be on the light
side of the $\chi^2$~function.

In \reffi{fig:massa2}, \ref{fig:massb2} we focus on the coloured part
of the supersymmetric spectrum and the Higgs mass scale. The case of $\tb 
= 10$
is shown in \reffi{fig:massa2}. The top row shows the two scalar top
masses, the middle row displays the two scalar bottom masses, and the bottom
row depicts the gluino mass and $\MA$. All the coloured particles should 
be accessible at the LHC. However, among them, only
${\tilde t}_1$ has a substantial part of its $\chi^2$-favoured spectrum 
below $500 \gev$, which would allow its detection at the ILC. The same
applies for the mass of the $A$~boson. The Tevatron collider has a 
sensitivity to $m_{{\tilde t}_1} \lappeq 450$GeV, which is not far 
below our best-fit value for $\tan \beta = 10$~\cite{BH}. 

Finally, in \reffi{fig:massb2} we show the same masses in the case of $\tb
= 50$. All the particles are mostly inaccessible at the ILC, though the
LHC has good prospects. However, at the 90\%~C.L.\ the coloured sparticle 
masses
might even exceed $\sim 3 \tev$, which would render their detection difficult.
Concerning the heavy Higgs bosons, their masses may well be below $\sim 1
\tev$. In the case of large $\tb$, this might allow their detection via
the process $b \bar b \to b \bar b H/A \to b \bar b \;
\tau^+\tau^-$~\cite{heavyHiggsLHC}. 


\subsection{Scan of the CMSSM Parameter Space}

Whereas in the previous section we presented fits keeping
$A_0/m_{1/2}$ fixed, we 
now analyse the combined sensitivity of the precision observables $\MW$,
$\sweff$, $\br(b \to s \ga)$ and $(g-2)_\mu$ in a scan over the $(m_{1/2},
A_0)$ parameter plane. In order to perform this scan, we have evaluated 
the
observables for a finite grid in the ($m_{1/2}$, $A_0$, $m_0$) parameter
space, fixing $m_0$ using the WMAP constraint. As before, we have
considered the two cases $\tb = 10$ and $\tb = 50$. Due to the finite grid
size, very thin lines in the \plane{m_{1/2}}{A_0} for $\tb = 50$, 
see \reffi{fig:A}, can
either be missed completely, or may be represented by only a few points. 

\begin{figure}[th!]
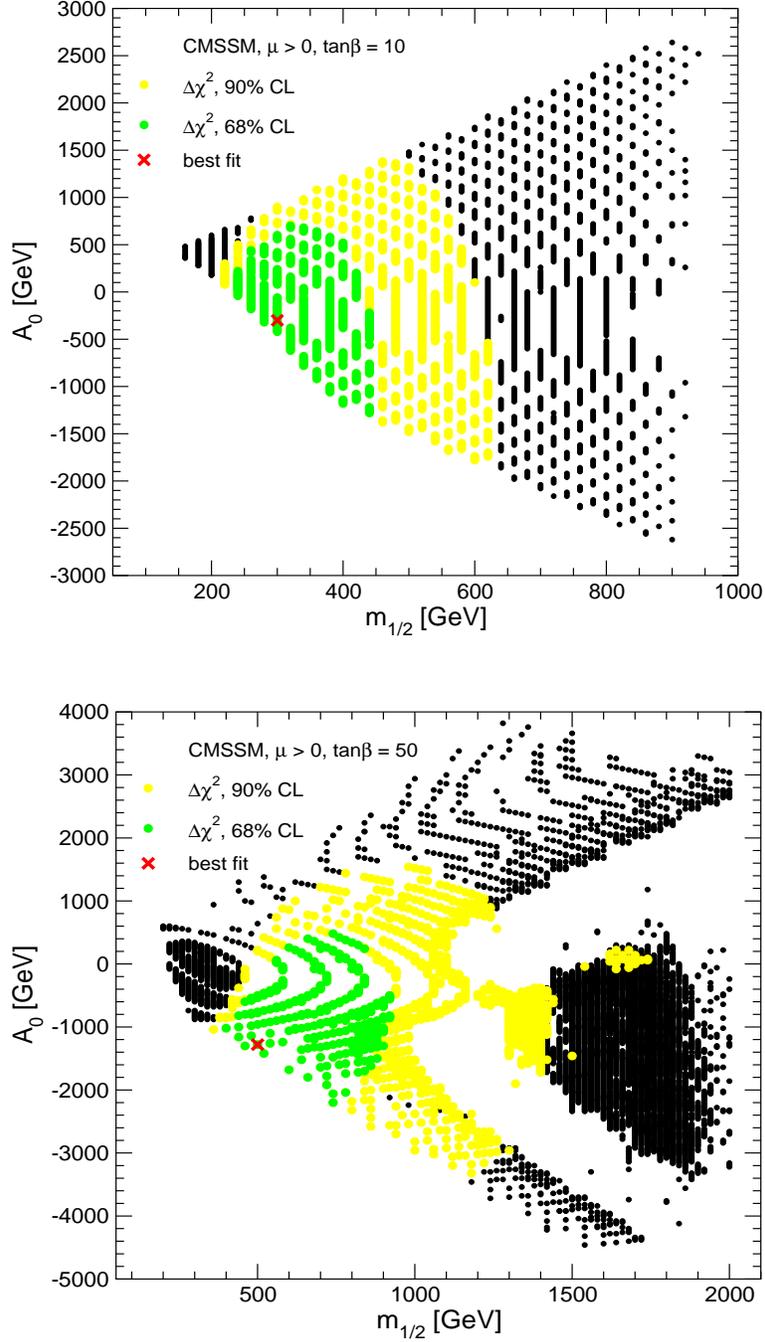

\begin{center}
\epsfig{figure=ehow.m12A04.bw.eps,width=10cm,height=8.5cm}\\[2em]
\epsfig{figure=ehow.m12A14.bw.eps,width=10cm,height=8.5cm}
\caption{%
The results of $\chi^2$ fits for $\tb = 10$ (upper plot) and $\tb = 50$
(lower plot) based on the current experimental
results for the precision observables $\MW$, $\sweff$, $(g-2)_\mu$ and
$\br(b \to s \ga)$ are shown in the \plane{m_{1/2}}{A_0}s of the CMSSM
with the WMAP constraint. The best-fit points are indicated, and 
the coloured regions correspond to the 68\% and 90\% C.L.\ regions,
respectively.
}
\label{fig:scancurrent}
\end{center}
\end{figure}

\reffi{fig:scancurrent} shows the WMAP-allowed regions in the
\plane{m_{1/2}}{A_0} for $\tb = 10$ and $\tb = 50$. 
The current best-fit values
obtained via $\chi^2$ fits for $\tb = 10$ and $\tb = 50$ are indicated. The
coloured regions around the best-fit values correspond to the 68\% and 90\%
C.L.\ regions (corresponding to $\Delta \chi^2 \le 2.30, 4.61$,
respectively). 

For $\tb = 10$ (upper plot of \reffi{fig:scancurrent}),
the precision data yield sensitive constraints on the available
parameter space for $m_{1/2}$ within the WMAP-allowed region. The
precision data are less sensitive to $A_0$.
The 90\% C.L.\ region contains all the WMAP-allowed $A_0$ values in
this region of $m_{1/2}$ values. As expected from the discussion above,
the best fit is obtained for negative $A_0$ and relatively small values
of $m_{1/2}$. At the 68\% C.L., the fit
yields an upper bound on $m_{1/2}$ of about 450~GeV. This bound is
weakened to about 600~GeV at the 90\% C.L.

As discussed above, the overall fit quality is worse for $\tb = 50$, and
the sensitivity to $m_{1/2}$ is less pronounced. This is demonstrated in
the lower plot of
\reffi{fig:scancurrent}, which shows the result of the fit in the 
\plane{m_{1/2}}{A_0} for $\tb = 50$. The best fit is obtained for
$m_{1/2} \approx 500 \gev$ and negative $A_0$. The upper bound on $m_{1/2}$
increases to nearly 1~TeV at the 68\% C.L. 

The holes in the coverage of the
\plane{m_{1/2}}{A_0} arise from the finite grid size of the scanning
procedure, as mentioned above. They would be filled if our scan would also
pick up the very thin lines, especially the wisps arising from
$\staustauH$. Thus, the holes correspond to
an extremely fine-tuned part of the parameter space, and are sparsely 
populated but not empty.

\begin{figure}[htb!]
\begin{center}
\epsfig{figure=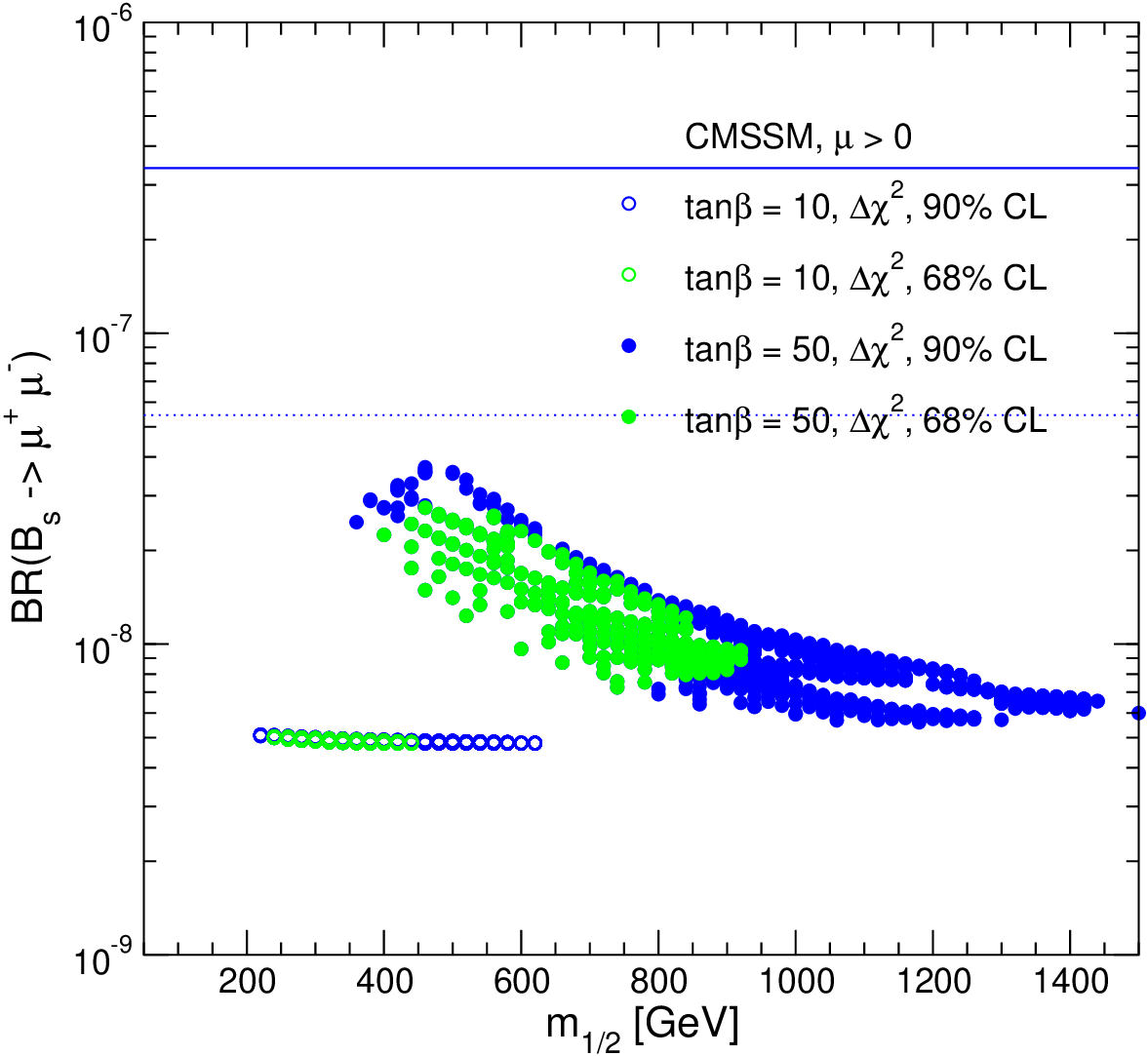,width=12cm}
\caption{%
Predictions for $\br(B_s \to \mu^+ \mu^-)$ within the CMSSM with WMAP
constraints are shown as functions of $m_{1/2}$, corresponding to the 
best-fit regions obtained by a $\chi^2$ fit (see
\reffi{fig:scancurrent}) based on the current experimental
results for the precision observables $\MW$, $\sweff$, $(g-2)_\mu$ and
$\br(b \to s \ga)$. The different
colours indicate the 68\% and 90\% C.L.\ regions. The present bound 
on $\br(B_s \to \mu^+ \mu^-)$ from the Tevatron (solid line) 
and our estimate for the prospective
sensitivity at the end of Run~II (dotted line) are also indicated (see text).
}
\label{fig:BMMTev}
\end{center}
\end{figure}

In \reffi{fig:BMMTev} we analyze the prospects for the Tevatron to observe
the process $B_s \to \mu^+ \mu^-$. We show the regions of the parameter
space that are favoured at the 68\% or 90\%~C.L., as a result of our fits
to the precision observables described above for $\tb = 10$ and $\tb = 50$.
The dotted line corresponds to our estimate of the final Tevatron
sensitivity at the 95\%~C.L.\ of $5.4 \times 10^{-8}$, see
\refse{subsec:bsmm}. It can be seen that, even for $\tb = 50$, all
parameter points result in a prediction for $\br(B_s \to \mu^+ \mu^-)$ that
is below our estimate of the future Tevatron sensitivity at the 95\% C.L.
Only with the more optimistic estimate of $2 \times 10^{-8}$ at the 90\%
C.L., discussed above, could a part of the favoured region for $\tb = 50$
be probed. The LHC, on the other hand, will cover the whole CMSSM parameter
space.


\section{Combined Sensitivity: ILC Precision}
\label{sec:combfuture}

\subsection{Best fits for WMAP strips at fixed $A_0$}

We now turn to the analysis of the future sensitivities of the precision
observables, based on the prospective experimental accuracies at the ILC
and the estimates of future theoretical uncertainties discussed in
\refse{sec:ewpo}. As before, we first display our results as functions of
$m_{1/2}$ moving along the WMAP strips with fixed values of $A_0$ and
$\tb$. We perform a $\chi^2$ fit for the combined sensitivity of the
observables $\MW$, $\sweff$, $(g-2)_\mu$, $\br(b \to s \ga)$, $\Mh$ and
$\br(\hbb) / \br(\hWW)$. We do not include $\br(B_s
\to \mu^+ \mu^-)$ into our fit. A measurement of this branching ratio at
the LHC could be used in combination with the above measurements at the
ILC.

\begin{figure}[htb!]
\mbox{}\vspace{-1cm}
\begin{center}
\epsfig{figure=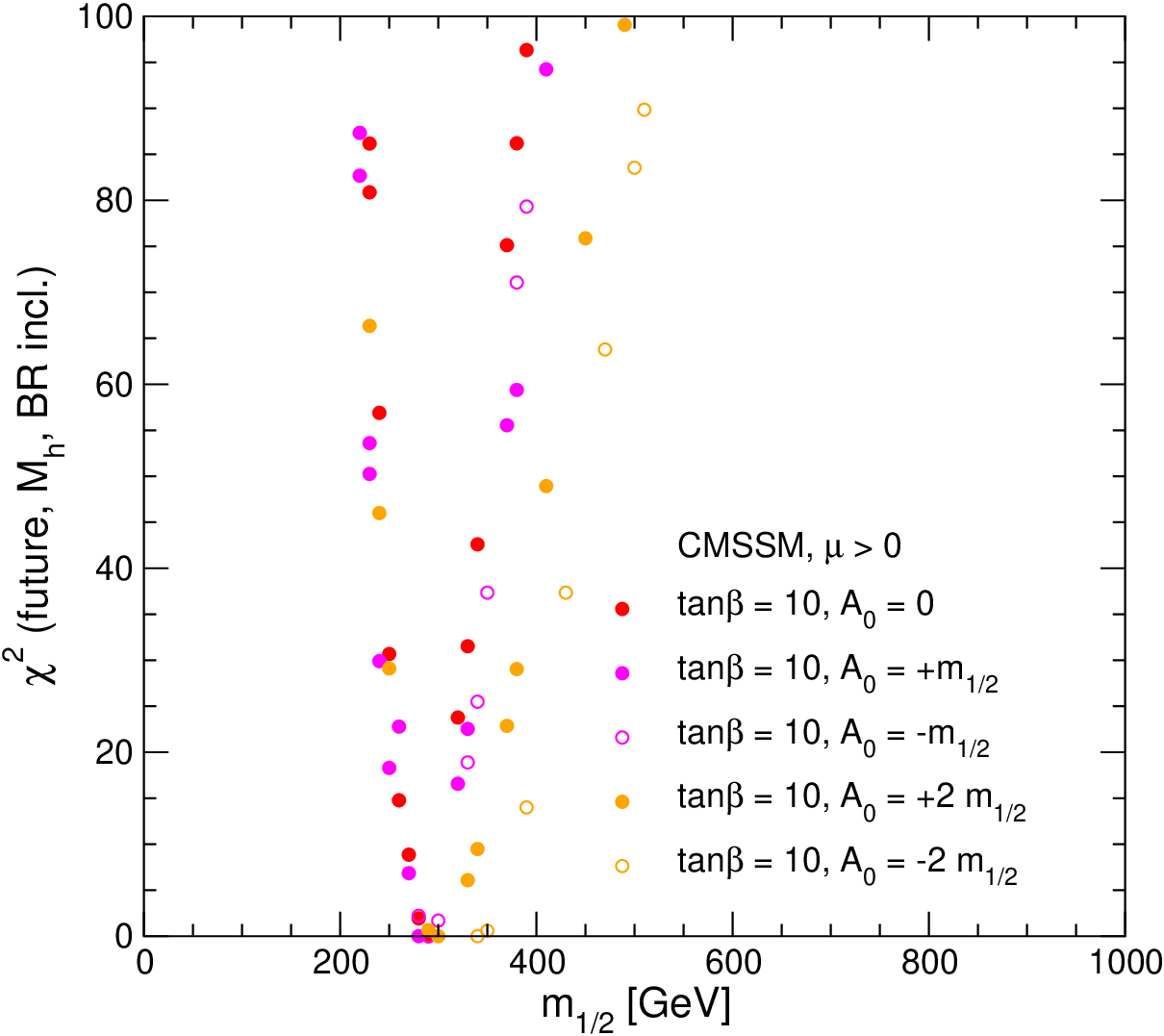,width=12cm,height=9.2cm}\\[1em]
\epsfig{figure=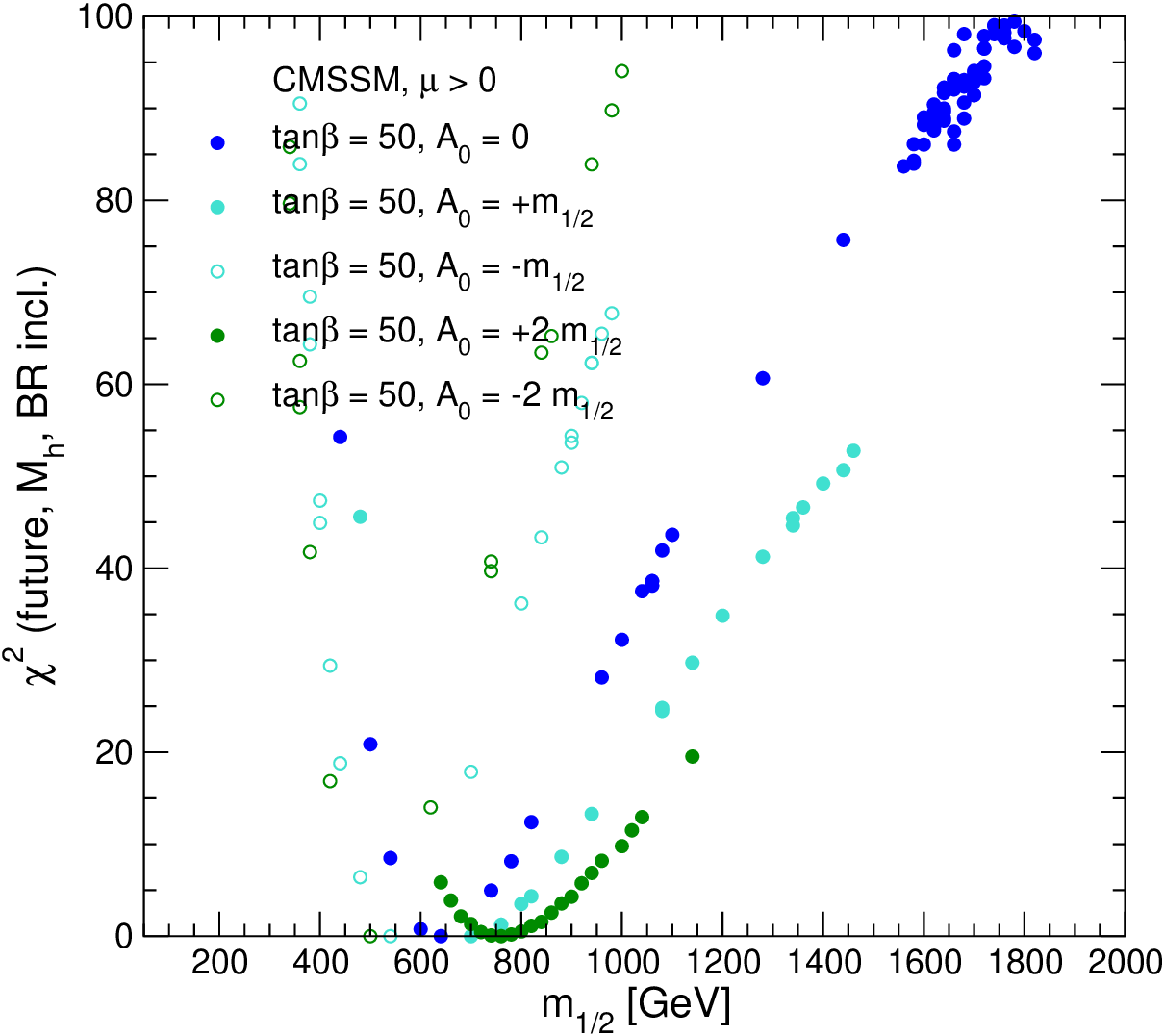,width=12cm,height=9.2cm}
\caption{%
The results of $\chi^2$ fits based on the prospective experimental
accuracies for the precision observables $\MW$, $\sweff$, $(g-2)_\mu$,
$\br(b \to s \ga)$, $\Mh$ and Higgs branching ratios at the ILC
are shown as functions of $m_{1/2}$ in the CMSSM
parameter space with the current WMAP constraints for $\tb = 10$ (upper
plot) and $\tb = 50$ (lower plot).
For each $A_0$ individually, the anticipated
future experimental central values are chosen according to the present 
best-fit point.  
}
\label{fig:CHIfut}
\end{center}
\vspace{-0.8cm}
\end{figure}
 
The results are shown in \reffi{fig:CHIfut} for $\tb = 10$ and $\tb = 50$.
The assumed future experimental central values of the observables have been
chosen such that they correspond to the best-fit value of $m_{1/2}$ in
\reffi{fig:CHI} for each individual value of $A_0$. Thus, the minimum of
the $\chi^2$ curve for each $A_0$ in \reffi{fig:CHIfut} occurs at $\chi^2 =
0$ by construction. The comparison of the prospective accuracies at the
ILC, \reffi{fig:CHIfut}, with the present situation, \reffi{fig:CHI}, shows
a big increase in the sensitivity to indirect effects of supersymmetric
particles within the CMSSM obeying the current WMAP constraints. For the
example shown here with best-fit values around $m_{1/2} = 300 \gev$ (upper
plot, $\tb = 10$), it is possible to constrain particle masses within about 
$\pm 10\%$
at the 95\% C.L.\ from the comparison of the precision data with the theory
predictions. We find a slightly higher sensitivity for $A_0 \leq 0$ than
for positive $A_0$ values. For the examples with best-fit values of
$m_{1/2}$ in excess of 500~GeV (lower plot, $\tb = 50$) the constraints 
obtained from the $\chi^2$ fit are weaker but still very significant.


\subsection{Scan of the CMSSM parameter space}

We now investigate the combined sensitivity of the precision
observables $\MW$, $\sweff$, $(g-2)_\mu$,
$\br(b \to s \ga)$, $\Mh$ and 
$\br(\hbb) / \br(\hWW)$
in the \plane{m_{1/2}}{A_0} of the CMSSM assuming ILC accuracies.
\reffi{fig:m12A0708} shows the fit results for $\tb = 10$, whilst
\reffi{fig:m12A1718} shows the $\tb = 50$ case.

\begin{figure}[htb!]
\mbox{}\vspace{-1cm}
\begin{center}
\epsfig{figure=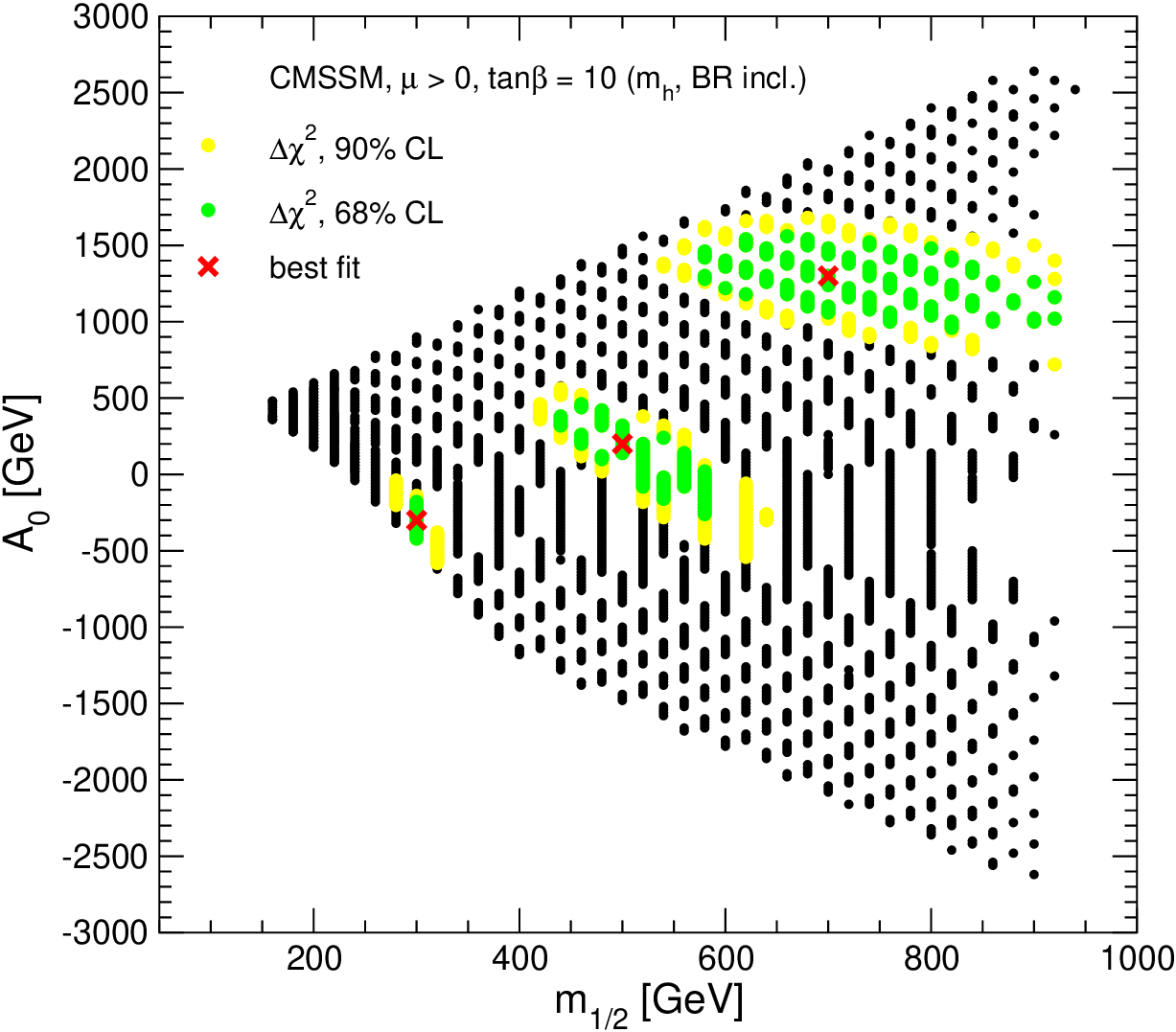,width=12cm,height=9.2cm}
\epsfig{figure=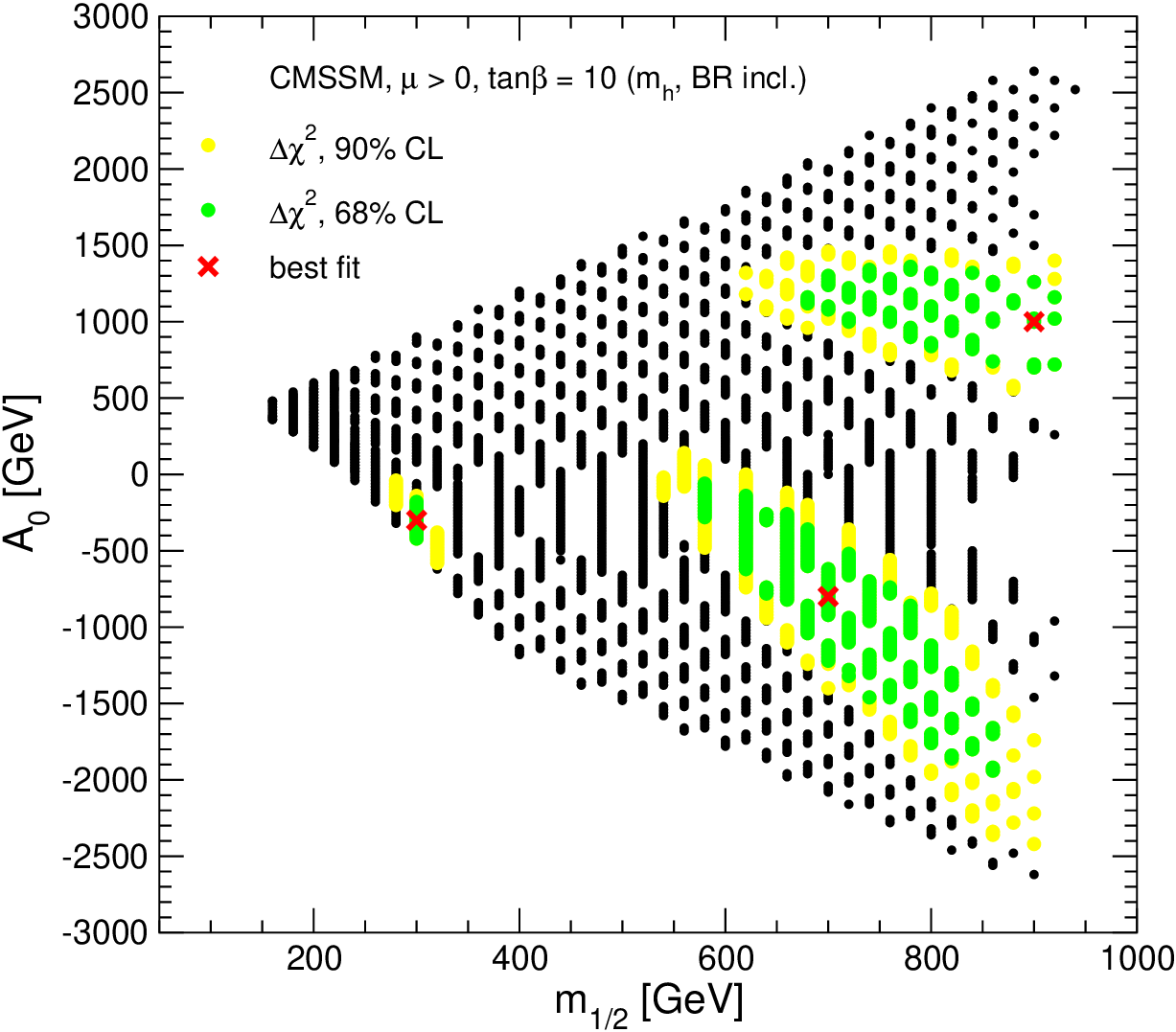,width=12cm,height=9.2cm}
\caption{%
The results of a $\chi^2$ fit based on the prospective experimental
accuracies for the precision observables $\MW$, $\sweff$, $(g-2)_\mu$,
$\br(b \to s \ga)$, $\Mh$ and Higgs branching ratios at the ILC
are shown in the \plane{m_{1/2}}{A_0} of the CMSSM
with WMAP constraints for $\tb = 10$. In both plots the 
WMAP-allowed region and the best-fit point 
according to the current situation (see \reffi{fig:scancurrent}) are
indicated. In both plots two further hypothetical future `best-fit' values 
have been chosen for illustration.
The coloured regions correspond to the 68\% and 90\% C.L.\ regions
according to the ILC accuracies.
}
\label{fig:m12A0708}
\end{center}
\vspace{-0.8cm}
\end{figure}

\begin{figure}[htb!]
\mbox{}\vspace{-1cm}
\begin{center}
\epsfig{figure=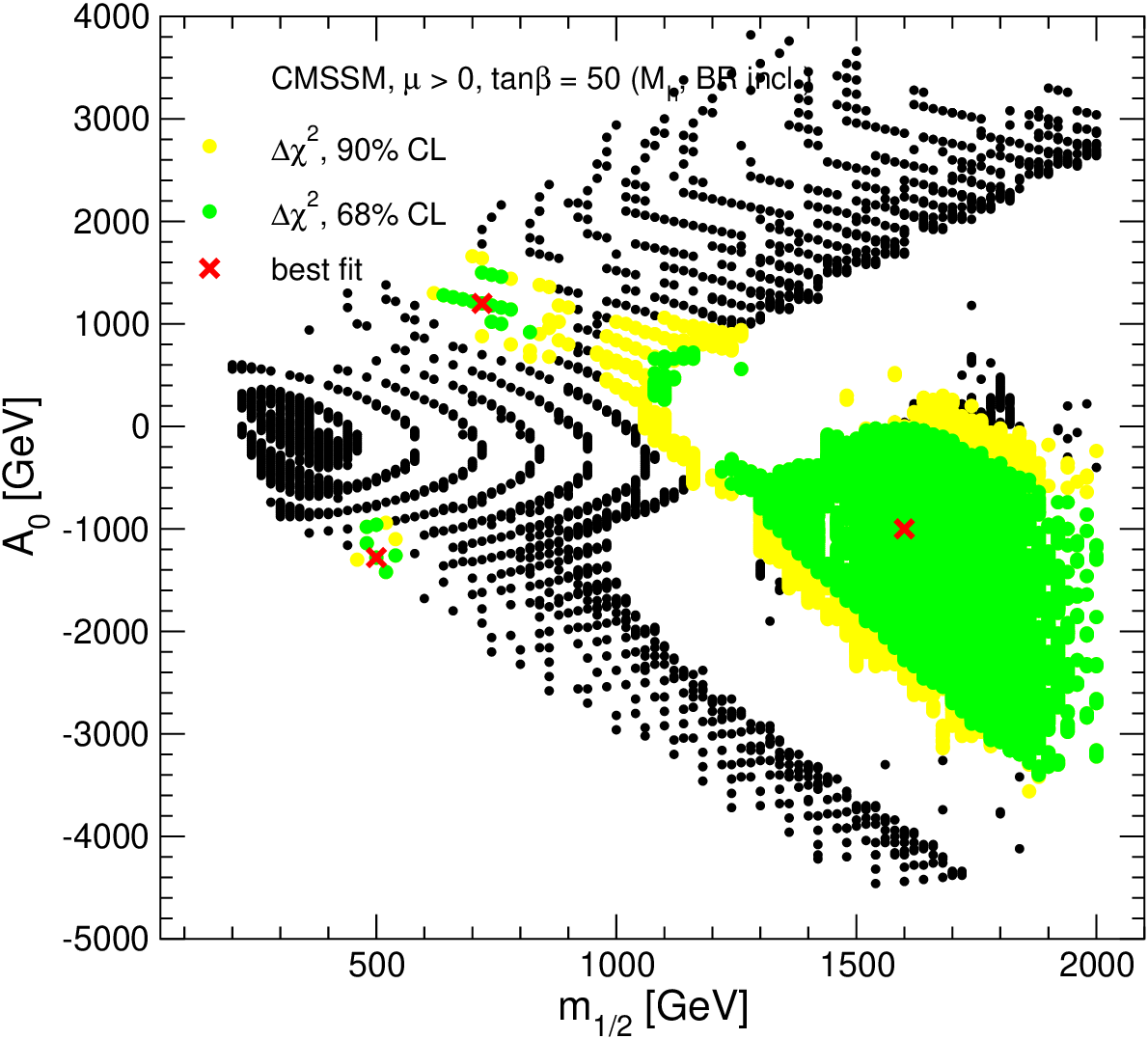,width=12cm,height=9.2cm}
\epsfig{figure=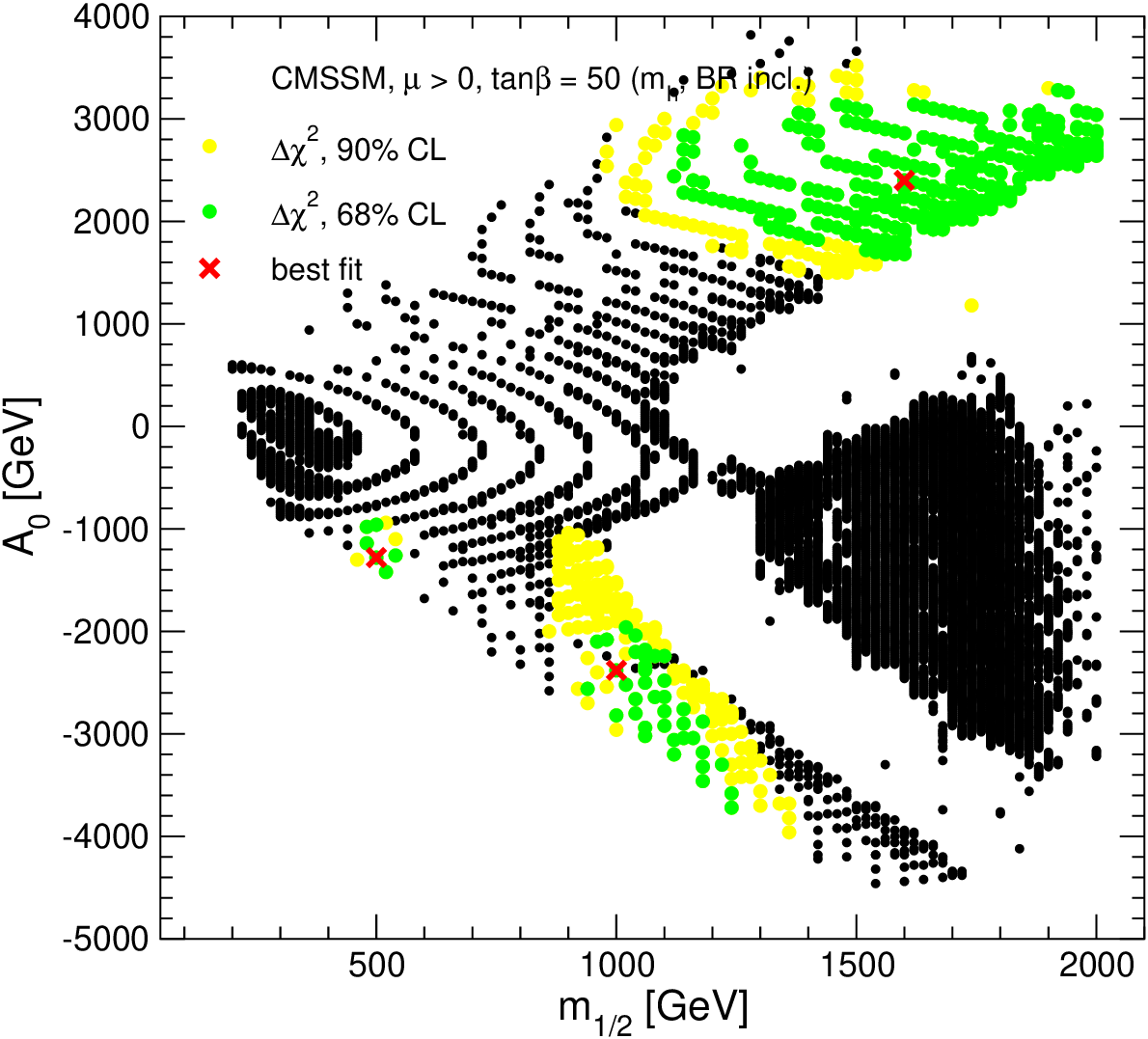,width=12cm,height=9.2cm}
\caption{%
The results of a $\chi^2$ fit based on the prospective experimental
accuracies for the precision observables $\MW$, $\sweff$, $(g-2)_\mu$,
$\br(b \to s \ga)$, $\Mh$ and Higgs branching ratios at the ILC
are shown in the \plane{m_{1/2}}{A_0} of the CMSSM
with WMAP constraints for $\tb = 50$. In both plots the 
WMAP-allowed region and the best-fit point for $\tb = 50$
according to the current situation (see \reffi{fig:scancurrent}) are
indicated. In both plots two further hypothetical future `best-fit' values 
have been chosen for illustration.
The coloured regions correspond to the 68\% and 90\% C.L.\ regions
according to the ILC accuracies.
}
\label{fig:m12A1718}
\end{center}
\vspace{-0.8cm}
\end{figure}

In each figure we show two plots, where the WMAP-allowed region and the
best-fit point according to the current situation (see
\reffi{fig:scancurrent}) are indicated. In both plots two further
hypothetical future `best-fit' points have been chosen for illustration.
For all the `best-fit' points, the assumed central experimental values of
the observables have been chosen such that they precisely coincide with the
`best-fit' points%
\footnote{
We have checked explicitly that assuming future experimental values of the 
observables with values distributed statistically around the present 
`best-fit' points with the estimated future errors does not degrade 
significantly the qualities of the fits.
}%
. The coloured regions correspond to the 68\% and 90\%
C.L.\ regions around each of the `best-fit' points according
to the ILC accuracies.

The comparison of \reffis{fig:m12A0708}, \ref{fig:m12A1718} with the result
of the current fit, \reffi{fig:scancurrent}, shows that the ILC
experimental precision will lead to a drastic improvement in the
sensitivity to $m_{1/2}$ and $A_0$ when comparing precision data with the
CMSSM predictions. For the best-fit values of the current fits for $\tb =
10$ and $\tb = 50$, the ILC precision would allow one to narrow down the
allowed CMSSM parameter space to very small regions in the 
\plane{m_{1/2}}{A_0}.
The comparison of these indirect predictions for $m_{1/2}$ and $A_0$
with the information from the direct detection of supersymmetric particles
would provide a stringent test of the CMSSM framework at the loop level. A
discrepancy could indicate that supersymmetry is realised in a more
complicated way than is assumed in the CMSSM.

Because of the decoupling
property of supersymmetric theories, the indirect constraints become
weaker for increasing $m_{1/2}$. 
The additional hypothetical `best-fit' points shown in
\reffis{fig:m12A0708}, \ref{fig:m12A1718} illustrate the indirect
sensitivity to the CMSSM parameters in scenarios where the precision
observables prefer larger values of $m_{1/2}$. 

For $\tb = 10$, we have investigated hypothetical `best-fit' values for
$m_{1/2}$ of 500~GeV, 700~GeV (for $A_0 > 0$ and $A_0 < 0$) and 900~GeV.
For $m_{1/2} = 500 \gev$, the 90\% C.L.\ region in the 
\plane{m_{1/2}}{A_0}
is significantly larger than for the current best-fit value of
$m_{1/2} \approx 300 \gev$, but interesting limits can still be set on both
$m_{1/2}$ and $A_0$. For $m_{1/2} = 700 \gev$ and $m_{1/2} = 900 \gev$, the
90\% C.L.\ region extends up to the boundary of the WMAP-allowed parameter
space for $m_{1/2}$. Even for these large values of $m_{1/2}$, however, the
precision observables (in particular the observables in the Higgs sector)
still allow one to constrain $A_0$.

For $\tb = 50$, where the WMAP-allowed region extends up to much higher
values of $m_{1/2}$~\footnote{We notice again the sparsely-populated 
`voids' due to our coarse sampling procedure.}, 
we find that for a `best-fit' value of $m_{1/2}$ 
as large as 1~TeV, which would lie close to the LHC limit and beyond the 
direct-detection reach of the 
ILC, the precision data would still allow one to establish an upper
bound on $m_{1/2}$ within the WMAP-allowed region. Thus, this indirect 
sensitivity
to $m_{1/2}$ could give important hints for supersymmetry searches at
higher-energy colliders. For `best-fit' values of $m_{1/2}$ in excess of
1.5~TeV, on the other hand, the indirect effects of heavy sparticles become
so small that they are difficult to resolve even with ILC accuracies.


\section{Conclusions}

We have investigated the sensitivity of precision observables, now and at
the ILC, to indirect effects of supersymmetry within the CMSSM. We have
taken into account the constraints from WMAP and other astrophysical and
cosmological data which effectively reduces the dimensionality of the CMSSM
parameter space.

We have performed a $\chi^2$ analysis based on the present experimental
results of the observables $\MW$, $\sweff$, $(g-2)_\mu$ and 
$\br(b \to s \ga)$ 
for two values of $\tb$, taking into account the current theoretical
uncertainties. For $\tb = 10$, we find that the CMSSM provides a very good
description of the data. A clear preference can be seen for relatively
small values of $m_{1/2}$, with a best-fit value of about 300~GeV and $A_0
\approx -m_{1/2}$. This result can be understood from the separate 
analyses of each of the observables, each of which is well described by
the CMSSM prediction for $m_{1/2} \approx 300 \gev$. At the 90\% C.L., we
find an upper bound on $m_{1/2}$ of about 600~GeV. The supersymmetric
particle spectrum corresponding to the best-fit region contains relatively
light states. There is a possibility that some sparticles might be 
detectable at the Tevatron collider, and many should be detectable at the 
LHC~\cite{lhctdrs} 
and the ILC~\cite{lctdrs}, allowing a detailed determination of their
properties~\cite{lhclc}.

For $\tb = 50$, the quality of the fit is worse than for the case with 
$\tb = 10$. While $\MW$ and $\sweff$ prefer small values of $m_{1/2}$
also for $\tb = 50$, $(g-2)_\mu$ and $\br(b \to s \ga)$ are better
described in this case by larger $m_{1/2}$ values. The indirect
constraints on $m_{1/2}$ are therefore less pronounced for 
$\tb = 50$. The best-fit value is obtained 
for $m_{1/2} \approx 500 \gev$ and negative $A_0$. The best-fit values for
the LSP mass and the lighter stau are still below about 250~GeV, while the
preferred mass values of the heavier neutralinos, the charginos and the
other sleptons are in the region of 500~GeV. The 90\% C.L.\ regions of
these masses extend beyond 1~TeV, but would be kinematically accessible at
a multi-TeV linear collider~\cite{clicreport}.
Coloured particles, such as the stops and sbottoms and the gluino
are likely to have masses within the reach of the LHC. However, at the
90\%~C.L.\ also masses beyond $\sim 3 \tev$ are possible. Heavy
Higgs bosons might also be accessible at the LHC in the
case of large $\tb$.

We have investigated the implications of our fit results for the prospects
for detecting a signal for $\br(B_s \to \mu^+ \mu^-)$. For both $\tb = 10$
and $\tb = 50$, we find that the 90\% C.L.\ region for $m_{1/2}$ and $A_0$
leads to predicted values of $\br(B_s \to \mu^+ \mu^-)$ that are below our
95\% C.L.\ estimate of the Tevatron sensitivity at the end of Run~II. With
a more optimistic estimate, the Tevatron could probe a part of the
parameter region for $\tb = 50$ at the \mbox{90\% C.L.} It seems more likely,
however, that detection of this process would have to await LHC data.

In the second part of our analysis, we have investigated the future
sensitivities of the precision observables to indirect effects of
supersymmetry, assuming the experimental accuracies achievable at the ILC
with a low-energy option running at the $Z$~resonance and the $WW$ threshold
and estimating the future theoretical uncertainties. As further precision
observables besides the ones discussed for the present situation, we have
included the mass of the lightest $\cp$-even Higgs boson and the ratio of
branching ratios $\br(\hbb) / \br(\hWW)$. We have
chosen several points in the \plane{m_{1/2}}{A_0} of the CMSSM with the
current WMAP constraints as examples for `best-fit' values, adjusting the
assumed future experimental central values of the precision observables to
coincide with the predictions of the `best-fit' values. With the
prospective ILC accuracies, the sensitivity to indirect effects of
supersymmetry improves very significantly compared to the present
situation. We find that for assumed `best-fit' values of $m_{1/2} \lsim
500 \gev$  the precision observables allow one to constrain tightly $m_{1/2}$
and $A_0$. Comparing these indirect predictions with the results from the
direct observation of supersymmetric particles will allow a stringent
consistency test of the model at the loop level.

Because of the decoupling property of supersymmetric theories, the indirect
constraints become weaker for larger $m_{1/2}$. Nevertheless, useful limits
on $m_{1/2}$ and $A_0$ can be obtained for `best-fit' values of $m_{1/2}$
as high as 1~TeV. Thus, the indirect sensitivity from the measurement of
precision observables at the ILC may even exceed the direct search reach of
the LHC and ILC.

Whilst this analysis has been restricted to the CMSSM, similar conclusions
are expected to apply if the assumption of universal soft
supersymmetry-breaking scalar masses is relaxed for the Higgs bosons, at
least for values of $\mu$ and $m_A$ not greatly different from those in
the CMSSM. The impact of the dark-matter constraint may well be rather
different if universality between the soft supersymmetry-breaking squark
and slepton masses is also relaxed, but we expect that the indication
found here for relatively light sparticle masses would be maintained.  
The investigation of these issues requires a more detailed study of models
beyond the CMSSM, which is in preparation.


\subsection*{Acknowledgements}

We thank R.~Clare, B.~Heinemann, G.~Hiller, T.~Kamon and C.~Weiser for
useful discussions. G.W. thanks the CERN Theory Division for kind
hospitality during the final stages of preparing this paper.


\end{document}